\documentclass[fleqn,usenatbib]{mnras}

\usepackage[T1]{fontenc}
\usepackage{ae,aecompl}

\usepackage{eso-pic}% http://ctan.org/pkg/eso-pic

\usepackage{graphicx}	% Including figure files
\usepackage{amsmath}	% Advanced maths commands
\usepackage{amssymb}	% Extra maths symbols
\usepackage[shortlabels]{enumitem}

\usepackage{multicol}
\title[MaNGA slow and fast rotator ellipticals]{Galaxy properties as revealed by MaNGA II. Differences in stellar populations of slow and fast rotator ellipticals and dependence on environment}

\author[Bernardi et al.]{
\parbox{\textwidth}{
M. Bernardi$^{1}$\thanks{E-mail: \texttt{\rm \texttt{bernardm@sas.upenn.edu}}}, H.~Dom\'{i}nguez S{\'a}nchez$^1$\thanks{\texttt{\rm \texttt{helenado@sas.upenn.edu}}}, J.~R.~Brownstein$^2$, N.~Drory$^3$ and R.~K.~Sheth$^{1}$}\\
 \vspace{0.cm}\\~\\
$^{1}$ Department of Physics and Astronomy, University of Pennsylvania, Philadelphia, PA 19104, USA\\
$^{2}$ Department of Physics and Astronomy, University of Utah, 115 S. 1400 E., Salt Lake City, UT 84112, USA\\
$^{3}$ McDonald Observatory, The University of Texas at Austin, 1 University Station, Austin, TX 78712, USA\\
}

\pubyear{2019}

\begin{document}
\label{firstpage}
\pagerange{\pageref{firstpage}--\pageref{lastpage}}
\maketitle

\begin{abstract}
We present estimates of stellar population (SP) gradients from stacked spectra of slow (SR) and fast (FR) rotator elliptical galaxies from the MaNGA-DR15 survey. We find that:
  1) FRs are $\sim 5$~Gyrs younger, more metal rich, less $\alpha$-enhanced and smaller than SRs of the same luminosity $L_r$ and central velocity dispersion $\sigma_0$. This explains why when one combines SRs and FRs, objects which are small for their $L_r$ and $\sigma_0$ tend to be younger. Their SP gradients are also different.
  2) Ignoring the FR/SR dichotomy leads one to conclude that compact galaxies are older than their larger counterparts of the same mass, even though almost the opposite is true for FRs and SRs individually.  
  3) SRs with $\sigma_0\le 250$~km~s$^{-1}$ are remarkably homogeneous within $\sim R_e$:  they are old, $\alpha$-enhanced and only slightly super-solar in metallicity. These SRs show no gradients in age and $M_*/L_r$, negative gradients in metallicity, and slightly positive gradients in [$\alpha$/Fe] (the latter are model dependent). SRs with $\sigma_0\ge 250$~km~$s^{-1}$ are slightly younger and more metal rich, contradicting previous work suggesting that age increases with $\sigma_0$. They also show larger $M_*/L_r$ gradients.
  4) Self-consistently accounting for $M_*/L$ gradients yields $M_{\rm dyn}\approx M_*$ because gradients reduce $M_{\rm dyn}$ by $\sim 0.2$ dex while only slightly increasing the $M_*$ inferred using a Kroupa (not Salpeter) IMF.
  5) The FR population all but disappears above $M_*\ge 3\times 10^{11}M_\odot$; this is the same scale at which the size-mass correlation and other scaling relations change. Our results support the finding that this is an important mass scale which correlates with the environment and above which mergers matter.

\end{abstract}


%%%%%%%%%%%%%%%%%%%%%%%%%%%%%%%%%%%%%%%%%%%%%%%%%%

%%%%%%%%%%%%%%%%% BODY OF PAPER %%%%%%%%%%%%%%%%%%

\section{Introduction}
Hierarchical models seek to understand galaxy formation within the cosmological context of large-scale structure formation \cite[e.g.][]{Mo2010}.  These models distinguish between the processes of star formation and its cessation, and those of the assembly of smaller stellar units into what becomes the final object.  In these models, there is a close connection between evolution and environment:  dense regions are like more evolved universes, so all the action in them happened longer ago.  Hence, they are expected to be populated by older, more massive galaxies.  If these galaxies acquired their stellar mass in situ rather than by a sequence of mergers, then they are expected to show strong gradients in their stellar populations \citep{Larson1974}; gradients which mergers are expected to erase, although residual central star formation can steepen them again \citep{White1980}. 

Spatially resolved spectroscopy of large galaxy samples allows one to test such predictions \cite[e.g.][]{Davies1993}.  Early work used long slit-spectroscopy to study gradients in small samples of galaxies, finding strong color gradients mostly driven by metallicity \cite[e.g.][and references therein]{SanchezBlazquez2007, Spolaor2009, Spolaor2010, Koleva2011}.  The advent of Integral Field Units (IFUs) has driven a revolution in this field \citep[see][for a recent review]{C16}.  The SAURON \citep{Emsellem2004} and ATLAS$^{\rm 3D}$ \citep{Cappellari2011} surveys of a decade ago, the more recent CALIFA \citep{CALIFA} and SAMI \citep{SAMI} surveys, and the ongoing and substantially larger MaNGA survey \citep{Bundy2015, Law2015, Wake2017, Westfall2019}, provide estimates of kinematic gradients (i.e., rotation curves and velocity dispersion profiles) and stellar population gradients in tens to hundreds of early-type galaxies, each sampled by tens to hundreds of spaxels.  There is now general agreement that metallicity increases towards the central regions of early-type galaxies \cite[e.g.][]{Scott2009, GonzalezDelgado2014, McDermid2015, GonzalezDelgado2015, Greene2015, Boardman2017, vdSande2018, Li2018, Parikh2019,Zhuang2019, Zibetti2019, Ferreras2019}, whereas age gradients are less pronounced.  In addition, recent studies of IMF gradients \citep{MN2015, MN2015b, MN2015c, LaBarbera2016, Vaughan2018a, Vaughan2018b, Sarzi2018} consistently find that the central regions of galaxies favour a bottom-heavy IMF.

Most of these studies are based on samples that are significantly smaller than what MaNGA provides.  The larger sample size, and the availability of more precise morphological classifications \citep{Fischer2019}, motivated us (\citealt[hereafter Paper~I]{DS2019}) to divide MaNGA early-types into Ellipticals (Es) and S0s, and study gradients in the Es. (The S0s are studied in Paper III -- in prep.)  In particular, in Paper~I we constructed stacked spectra of Es at $z < 0.08$ binned in $\sigma_0$ and luminosity $L_r$ using the multiple spectra provided by the MaNGA survey (\citealt{Bundy2015}) which is a component of the Sloan Digital Sky Survey IV (\citealt{Blanton2017}). We estimated absorption line strengths in these high S/N spectra, and then used a variety of single stellar population synthesis models to estimate stellar population parameters. We focused on the effects of the IMF variations inside a galaxy and across the galaxy population. We found that for the ellipticals with the largest $L_r$ and $\sigma_0$, the results are consistent with those associated with the commonly used Salpeter IMF in the central regions, approaching values from a Kroupa-like IMF by $\sim 1 R_e$ (assuming [X/Fe] enhancement variations are limited). For these galaxies we find that the stellar mass-to-light ratio decreases at most by a factor of 2 from the central regions to $R_e$. In contrast, for lower $L_r$ and $\sigma_0$ galaxies, the IMF is shallower and the $M_*/L_r$ of central regions is similar to the outskirts. That gradients become less important at lower masses is also consistent with previous work \citep{Parikh2018}. Although a factor of 2 is smaller than previous reports based on a handful of galaxies \cite[e.g.][]{vD2017}, it is still large enough to matter for dynamical mass estimates \citep{Bernardi2018b}. Our results show that accounting self-consistently for the $M_*/L_r$ gradients when estimating both $M_*$ and $M_{\rm dyn}$ brings the two into good agreement: gradients reduce $M_{\rm dyn}$ by $\sim 0.2$ dex while only slightly increasing the $M_*$ inferred using a Kroupa IMF. This is a different resolution of the $M_*$-$M_{\rm dyn}$ discrepancy than has been followed in the recent literature where $M_*$ of massive galaxies is increased by adopting a Salpeter IMF while leaving $M_{\rm dyn}$ unchanged \cite[e.g.][]{Cappellari2013b, Li17}. Our results are consistent with previous work on reconciling estimates of the stellar and dynamical mass functions \citep{Bernardi2018b}.

Paper~I also found that galaxies with larger $\sigma_0$ or $L_r$ tend to be older and more metal rich, and that, within a galaxy, age and metallicity tend to increase with $\sigma$:  i.e., they are all larger in the center.  These findings are in qualitative agreement with previous work \cite[e.g.][]{LaBarbera2013, MN2015, LaBarbera2016, vD2017, TW2017, Parikh2018}. However, in Paper~I we also found that Es with $200 \le \sigma_0 \le 250$~km~s$^{-1}$ and $-21.5\le M_r \le -22.5$ tend to be the oldest in the sample, even though they are neither the largest $\sigma$ nor the largest $L_r$.  One of the goals of the present study is to investigate why.

In addition, recent work has emphasized the fact that, if early-type galaxies are classified as being slow or fast rotators \cite[following][]{Emsellem2007}, then the slow rotators are much more likely to have had merger-dominated assembly histories \citep{C16}.  \cite{Fischer2019} showed the value of using the \cite{DS2018} morphological classificiations to separate the Es from S0s in MaNGA when studying the slow and fast rotator dichotomy.  This motivates us to further subdivide our sample of Es on this basis.  Additional motivation comes from the fact that \cite{McDermid2015} do not report a significant difference in the age of the stellar populations of slow and fast rotating early-type galaxies in the ATLAS$^{\rm 3D}$ survey (they only find a significant difference in the metallicity, with fast rotators being more metal rich), whereas \cite{vdSande2018} find that fast rotators in the SAMI survey tend to be younger.  Our MaNGA sample should allow us to determine whether or not rotation plays a key role and, if it does, whether the age-rotation anti-correlation is driven by morphology (e.g. S0s are fast and Es are slow rotators) or is present even in a sample of Es alone.\footnote{If most fast rotating Es are simply mis-classified S0s \cite[e.g.][]{Emsellem2011,C16} then the issue of morphology dependence is less interesting.  However, because there exists substantial previous work on S0s as opposed to Es in MaNGA \cite[][and references therein]{mangaS0}, we thought it useful to keep the two populations separate.  Moreover, it may be that stellar population gradients, which were not the focus of the ATLAS$^{\rm 3D}$ or SAMI analyses, are able to discriminate between S0s and fast rotators Es.}
\cite{McDermid2015} do report another correlation with age:  At a given mass, more compact galaxies are older.  We study if this anti-correlation between age and size is present in our sample, although it spans a much smaller range of $L$ and $\sigma_0$, and how it relates to the slow/fast rotator dichotomy.

In Section~\ref{sec:split} we first subdivide each $L_r$ and $\sigma_0$ bin on the basis of half-light radius $R_e$, and then still further on the basis of rotation.
%Since $L_r$ and $R_e\sigma_0^2$ are both well-correlated with galaxy mass, this allows us to show how stellar populations depend on mass and density.
%In Section~\ref{fastSlow} we subdivide still further into slow and fast rotators.
Section~\ref{sec:ssp} provides estimates of the stellar populations (age, metallicity, $\alpha$-enhancement, IMF and stellar mass-to-light ratio) and stellar population gradients in these bins. It also discusses the ages and sizes at a given $M_*$ accounting for the slow/fast rotator dichotomy and compares stellar and dynamical mass estimates of slow and fast rotators when one self-consistently accounts for $M_*/L$ gradients. Section~\ref{Sec:env} discusses how the observed properties correlate with environment. 
%Section~\ref{sec:B10} studies the Es with $\sigma\approx 230$~km~s$^{-1}$ and stellar masses $\approx 10^{11}M_\odot$, and presents a plausible explanation for why they are anomalous.
A final section summarizes.  Appendix~\ref{sec:SN} demonstrates that, even after subdividing by size and rotation, our stacked spectra have sufficient S/N to provide reliable measurements.  It also discusses how, when studing gradients, one must be careful about the fact that, if the same galaxy is located closer to us, then its inner regions will be sampled by more MANGA spaxels, but its outer regions may not be sampled at all.  Appendix~\ref{sec:compare} contrasts our results with previous work.

\section{Binning in $\sigma_0$, $L_r$, $R_e$ and rotation}\label{sec:split}
In Paper~I we defined a sample of MaNGA Es with redshift $z\le 0.08$ to cover only a small range of lookback times. We constructed stacked spectra of these MaNGA Es, binned in $\sigma_0$ and luminosity $L_r$.  Here we further subdivide each bin into objects with above and below average sizes $R_e$ within the bin, before making stacked spectra. (In what follows $R_e$ is the truncated semimajor axis, i.e. $R_e = R_{e,{\rm maj}}$, from the best-fit indicated by FLAG$\_$FIT in the MaNGA PyMorph Photometric Value Added Catalogue, hereafter MPP-VAC, see \citealt{Fischer2019}).  We then separate slow and fast rotators in each $\sigma_0$, $L_r$ and $R_e$ bin. 

\begin{figure}
  \centering
  \includegraphics[width=0.99\linewidth]{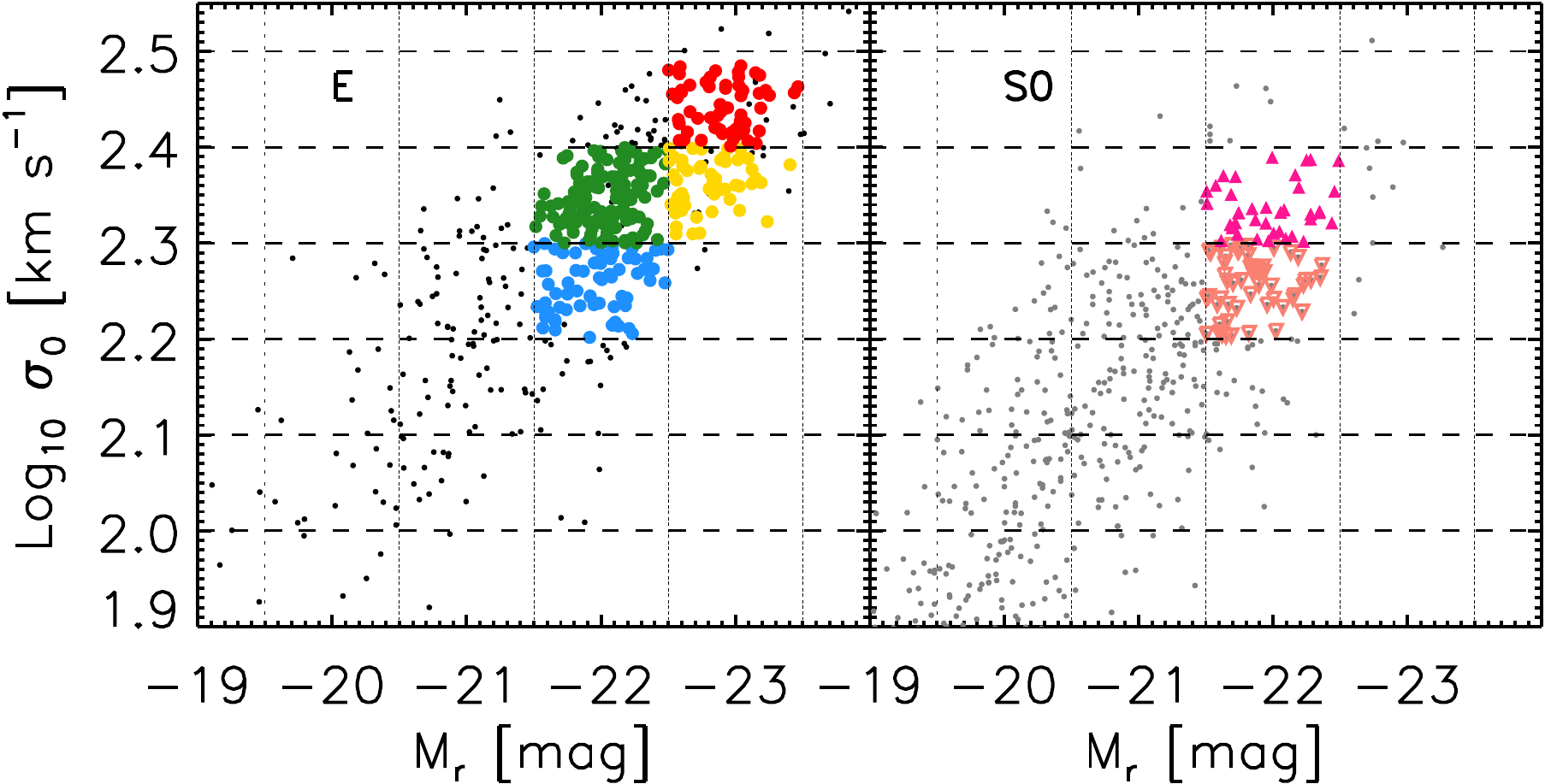}
   \vspace{-0.5cm}
  \caption{Correlation between $\sigma_0$ and $L_r$ for Es (left) and S0s (right) at $z\le 0.08$ selected as described in Paper I.  Blue, green, yellow and red symbols in the left-hand panel show Es in bins B00, B10, B11 and B21 (see Table~\ref{tab:bin}).  Open inverted and filled upright triangles in the right-hand panel show S0s in bins B00 and B10; there are essentially no S0s in bins B11 and B21.}
  \label{fig:VL}
\end{figure}

\begin{figure}
 \centering
  \includegraphics[width=0.9\linewidth]{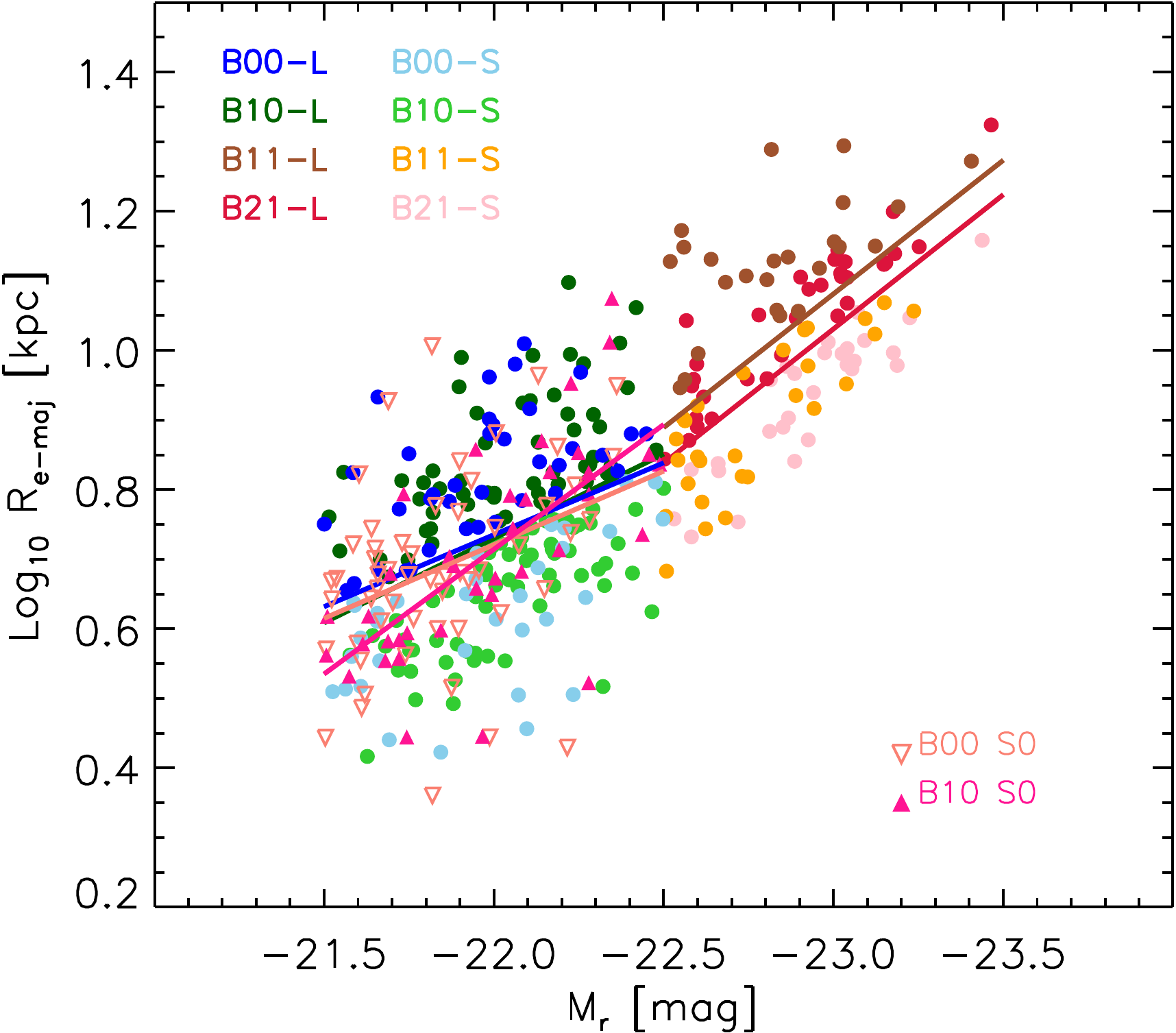}
  \caption{Size-luminosity relation of MaNGA elliptical galaxies in the four bins described in Table~\ref{tab:bin}. These were further subdivided into two approximately equal bins in size using the four straight lines shown.  Lighter shades (light blue, light green, orange and pink) represent galaxies with smaller sizes given their $\sigma$ and $L_r$; darker shades (dark blue, dark green, brown, red) represent galaxies with larger sizes.  Triangles show S0s in the bin B00 and B10 (i.e., compare with the blue and green Es, respectively).}
  \label{fig:split}
\end{figure}

Although the results which follow are all based on splitting the sample based on the distribution in the $R_e$-$L_r$ plane at each $\sigma_0$, we have also studied what happens if we split based on the $R_e$-$\sigma_0$ plane at each $L_r$.  This results in similar, but noisier trends, compared to those we show below.  We comment on splitting based on Fundamental Plane parameters $R_e$, $\sigma_0$ and surface-brightness in Appendix~\ref{sec:compare}.

\begin{table}%[htb]
\centering
BINNING OF GALAXIES\\
\begin{tabular}{lcccccc}
  \hline
  Bin &  M$_r$ & Log$_{10}$ $\sigma_0$ & All & SR & FR & FR \\
  &  [mag] & [km s$^{-1}$]         & & & $ \lambda_e < 0.2$ & $\lambda_e > 0.2$ \\
 \hline
 B00-S  & $-21.5, -22.5$  &  2.20, 2.30  & 33 & 5 & 6 & 22\\
 B00-L  & $-21.5, -22.5$  &  2.20, 2.30  & 37 & 20 & 9 & 8\\ 
 B10-S  & $-21.5, -22.5$  &  2.30, 2.40  & 64 & 25 & 18 & 21\\
 B10-L  & $-21.5, -22.5$  &  2.30, 2.40  & 57 & 27 & 14 & 16\\
 B11-S  & $-22.5, -23.5$  &  2.30, 2.40  & 28 & 13 & 2 & 13\\
 B11-L  & $-22.5, -23.5$  &  2.30, 2.40  & 24 & 13 & 7 & 4\\
 B21-S  & $-22.5, -23.5$  &  2.40, 2.50  & 28 & 16 & 9 & 3\\
 B21-L  & $-22.5, -23.5$  &  2.40, 2.50  & 32 & 18 & 8 & 6\\ 
\hline
\hline
\end{tabular}
\caption{Number of galaxies in each bin with large $R_e$ (B**-L) and small $R_e$ (B**-S) with respect to the mean $R_e$-$L_r$ relation. Slow (SR) and fast (FR) rotators are galaxies inside and outside the box in the bottom corner of Figure~\ref{fig:lambda}, respectively.}
\label{tab:bin}
\end{table}

\subsection{Splitting on size at fixed $\sigma_0$ and $L_r$}\label{sec:RLV}
The left hand panel of Figure~\ref{fig:VL} shows how Paper~I defined the $\sigma_0$ and $L_r$ bins we use in our study of Es. Table~\ref{tab:bin} describes the bin limits, and the number of objects in each bin. Note that bins B00 and B10 have the same $L_r$ but different $\sigma_0$, as do bins B11 and B21; bins B10 and B11 have the same $\sigma_0$ but different $L_r$.
%Later in this paper we will study the Es in bin B10 in more detail (see Section~\ref{sec:B10}).
One of our tests will compare these Es with S0s. Therefore, the right hand panel shows the S0 distribution in the corresponding bins.  Notice that there are essentially no S0s brighter than $M_r\le -22.5$.  It is worth noting that this is the luminosity scale (mass scale $M_*\sim 2\times 10^{11}M_\odot$ assuming a Chabrier IMF) at which many early-type galaxy scaling relations change slope \citep{Bernardi2011}, and where the population becomes dominated by slow rotators \citep{C16}. In Section~\ref{sec:Mdyn} we show that the IMF is not Chabrier, so we provide a better estimate of this critical stellar mass scale in \ref{sec:ageSize}.

Figure~\ref{fig:split} illustrates how the galaxies in each $\sigma_0$ and $L_r$ bin were subdivided based on $R_e$.  Note that the division is not a straight cut on $R_e$; rather we separate the objects in a bin based on whether they are larger or smaller than the mean $R_e$-$L_r$ relation for the bin.  Hereafter, we use B00-L and B00-S to refer to the objects in bin B00 that are larger or smaller than the average for their $L_r$ (and $\sigma_0$), and similarly for the other bins.  On average, the objects in B**-L have smaller surface brightnesses than those in B**-S. In the next subsection (Section~\ref{fastSlow}) we subdivide even further based on angular momentum:  Table~\ref{tab:bin} gives the relevant numbers for this as well.

\begin{figure}
  \centering
  \includegraphics[width=0.75\linewidth]{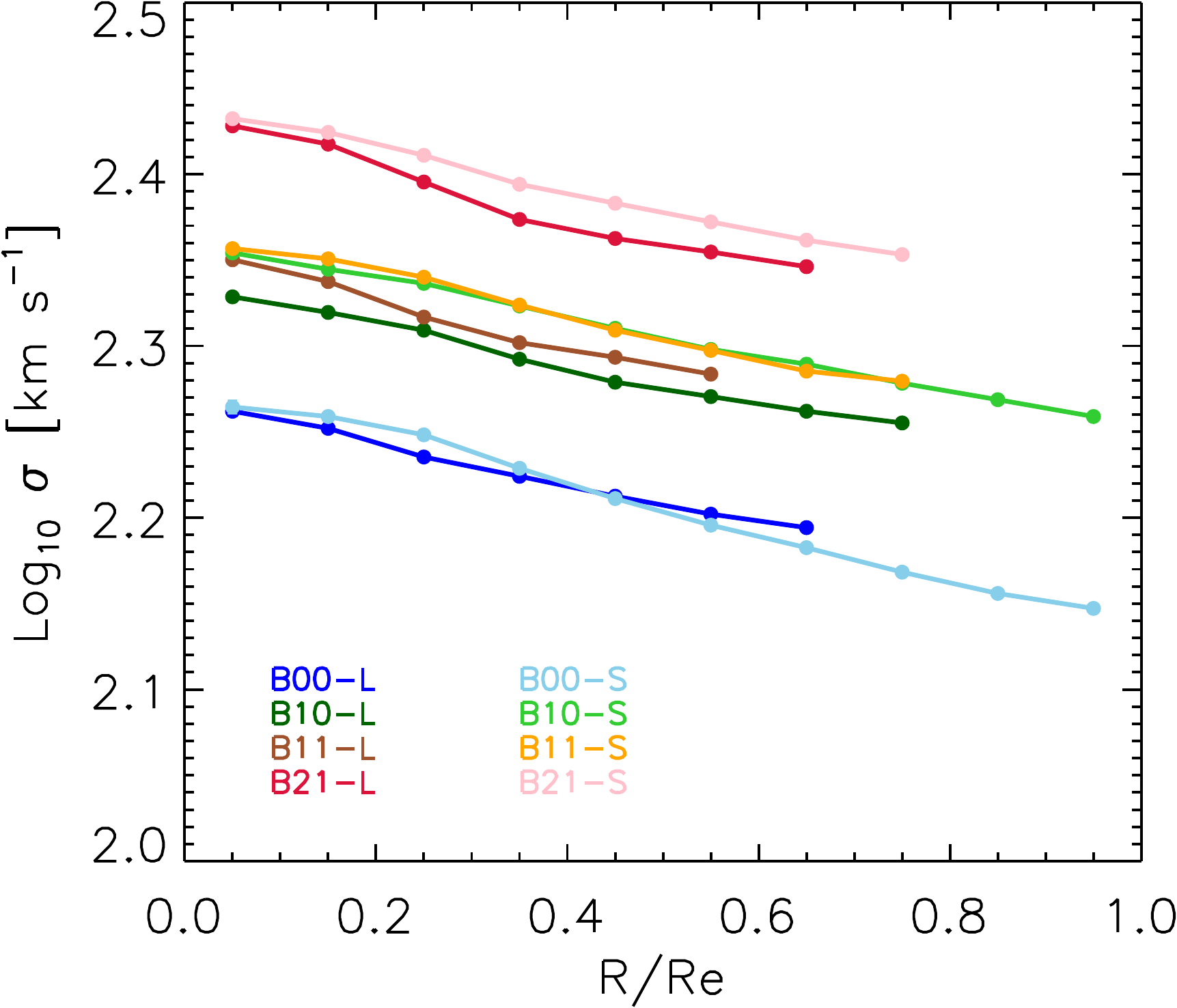}
  \caption{Velocity dispersion profiles for the eight bins described in Table~\ref{tab:bin}. Except for bin B00 (blue), the objects with larger sizes (for their central velocity dispersion and $L_r$) in a bin tend to have $\sigma$ at $R/R_e\sim 0.5$ smaller by about 0.02~dex.}
  \label{fig:sigmaR}
\end{figure}

We cannot subdivide into arbitrarily fine bins, since we require spectra with signal-to-noise $\sim 100$ and this is larger than that in a typical spaxel (Figure~\ref{fig:SNsplitindiv} in Appendix~\ref{sec:SN}).  To produce high signal-to-noise spectra, spaxels of the galaxies in each of the bins defined in Table~\ref{tab:bin} were stacked based on their values of $R/R_e$ (see Paper~I for details).  The number of spaxels which contribute, and the resulting S/N ratios are provided in Appendix~\ref{sec:SN}.  In particular, Figure~\ref{fig:Nspx} shows the number of spaxels in each radial bin which contribute to our results for the eight bins in $L_r$, $\sigma_0$ and $R_e$. Some of the curves decrease at large $R$ because, for the largest galaxies, the spaxels may not cover the entire region within $R_e$.  Figure~\ref{fig:Nspx} shows that this problem is most severe for bin B11-L.  In what follows we only show those radial bins which are well represented by all galaxies (e.g. for bin B11-L (brown) we show our results out to $R/R_e \sim 0.6$ and for B21-L (dark red) out to $R/R_e\sim 0.7$).
Moreover, within a bin, we expect some galaxy-to-galaxy variation over which we would like to average so as to obtain representative values.  Therefore, we also require more than $\sim 10$ galaxies per stack.  While this is not an issue for the analysis of this Section, we must take care when we subdivide based on rotation in the next Section.

%\begin{figure}
%  \centering
%  \includegraphics[width=0.9\linewidth]{FIGURES/MANGA_rmsLR_S0.pdf}  
%  \caption{}
%  \label{fig:rms}
%\end{figure}

Figure~\ref{fig:sigmaR} shows the velocity dispersion profiles for these bins.  While the binning was based on $L_r$ and the central $\sigma_0$, the bins with larger $R_e$ in each bin tend to have slightly smaller velocity dispersions at larger $R$.  Nevertheless, it is apparent that $\sigma_0$ of galaxies in bin B00 is approximately the same as $\sigma_e$ at the outer regions ($\sim R_e$) of bins B10 and B11.  Similarly, $\sigma_0$ of B10 and B11 galaxies is similar to $\sigma_e$ of B21 galaxies.  This will allow us to test if stellar population parameters which vary with $\sigma$ across the population are really a function of the local $\sigma$.

\subsection{Separating fast and slow rotators}\label{fastSlow}
%We now study how the stellar populations in our subsamples are correlated with stellar kinematics.  Specifically,
We noted in the introduction that it is interesting to further subdivide our sample on the basis of rotation, in part because slow rotators are expected to have merger-dominated histories.

\begin{figure}
  \centering
  \includegraphics[width=0.99\linewidth]{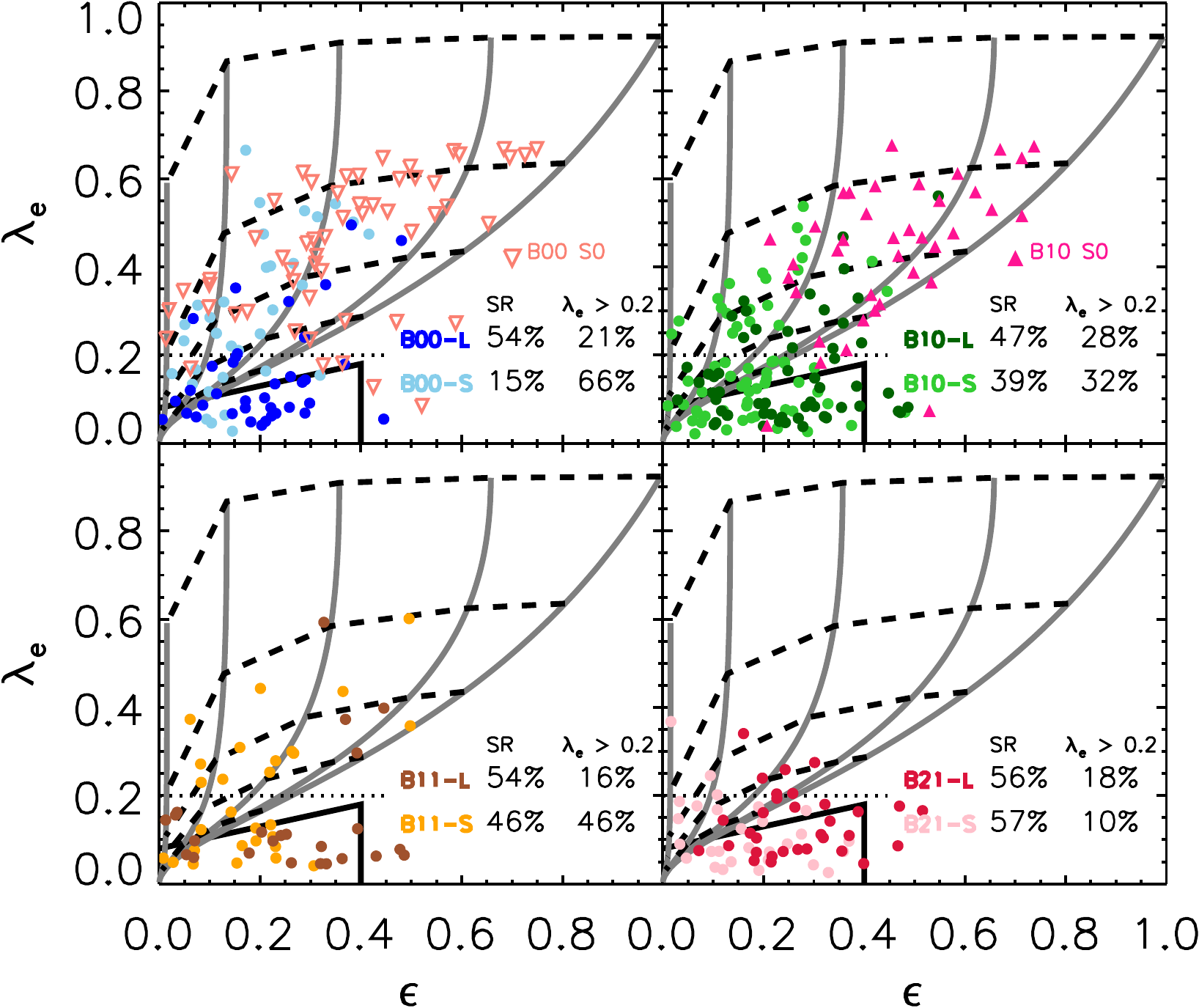}
    \vspace{-0.5cm}
    \caption{Joint distribution of shape $\epsilon\equiv 1-b/a$ and spin $\lambda_e$ (defined in the main text) for the objects (filled circles) in our four bins. The open and filled triangles show the S0s (there are essentially no S0s in bins B11 and B21). Right-most solid grey line shows an isotropic rotator viewed edge-on; the others show different inclinations.  Dashed lines show how an object of intrinsic $\epsilon$ (given by the right-most value) moves across the diagram as inclination changes.  Black box in bottom left corner of each panel shows the region which defines `slow rotators'.  In each bin, the objects with larger $R_e$ (B**-L) always have a significantly larger fraction of `slow rotators' (approximately the same for each bin), while for galaxies with smaller $R_e$ (B**-S) large $\lambda_e$ values are more common in the top left panel (bin B00).}
  \label{fig:lambda}
\end{figure}

Figure~\ref{fig:lambda} shows the joint distribution of photometric ellipticity $\epsilon\equiv 1-b/a$ and spin $\lambda_e\equiv \sum_i F_i R_i |V_i|/\sum_i F_i R_i \sqrt{V_i^2 + \sigma^2_i}$, where $F$ is the flux, $V_i$ and $\sigma_i$ are the rotation speed and velocity dispersion in pixel $i$ which is at distance $R_i$ from the image center, and the sum is over all spaxels within $R_e$ \citep{Emsellem2007}.  (We have corrected this quantity for seeing following \citealt{Graham2018}.)  The box in the bottom corner shows the region which is used to identify slow rotators \cite[hereafter SR; see, e.g.,][]{C16}.  Objects outside this box are usually called fast rotators (FR).  In what follows, we will sometimes select objects with $\lambda\ge 0.2$ as a simple way of decreasing contamination from SRs, and will often simply refer to them as FRs.  The text in each panel shows the fraction of FRs and those with $\lambda_e>0.2$ in each bin, and Table~\ref{tab:bin} gives the actual number of SRs, FRs with $\lambda_e<0.2$ and FRs with $\lambda_e>0.2$ in each bin.  

\cite{Graham2018} and \cite{Fischer2019} show how MaNGA galaxies populate this plane as a function of morphology.  Here, we mainly concentrate on Es, and further subdivide the Es based on $L_r$, $\sigma_0$ and $R_e$. It is clear that large $\lambda_e$ values are more common in the top left panel (bin B00) than bottom right (B21). It is well known that SRs only begin to dominate at large $L_r$ and that the typical $\lambda_e$ for Es decreases as $L_r$ increases \citep{C16, Fischer2019}.  Note however that the SR fraction is approximately independent of $L_r$ for the larger objects (B**-L) in each $\sigma_0$ and $L_r$ bin.

Finally, consider the open inverted and filled upright triangles in the two top panels.  These show S0s with the same $L_r$ and $\sigma_0$ in each bin. (It is worth noting that the S0 vs E morphological classification was based entirely on imaging:  i.e. it did not use $\lambda_e$ at all.) Clearly, the S0s are generally rotating faster than the Es. In addition, while we see S0s with $\epsilon<0.2$ in bin B00 (triangles in the top left panel) there are no S0s with $\epsilon\le 0.2$ in bin B10 (filled triangles in top right panel). We will discuss this in the next Section. 

\subsection{Separating Es from S0s}
\begin{figure}
  \centering
  \includegraphics[width=0.99\linewidth]{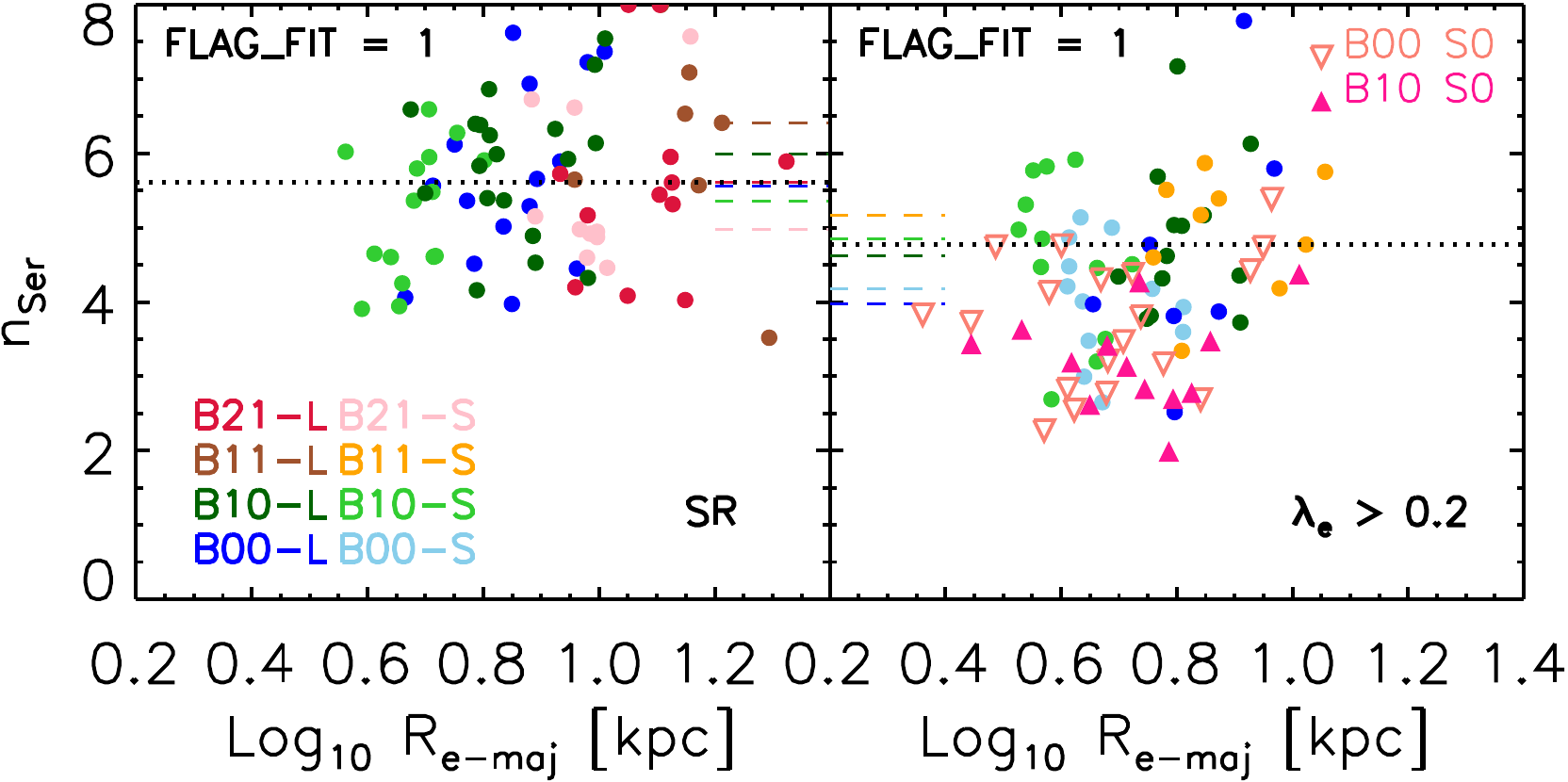}
   \vspace{-0.5cm}
   \caption{Correlation between Sersic index $n$ and size $R_e$ in our eight bins, for SRs (left) and FRs (right) with {\tt FLAG$\_$FIT=1} (i.e. objects whose photometry is better described by a single Sersic profile:  $\sim 64 \%$ of our Es). Dotted horizontal lines show the median $n$ for the Es in each panel. Dashed colored horizontal lines show the median $n$ in each bin:  SRs tend to have larger $n$ than their FR counterparts.  Triangles in the right hand panel show {\tt FLAG$\_$FIT=1} S0s ($\sim 35 \%$ of the S0s) with the same $L_r$ and $\sigma_0$ as the Es in bins B00 and B10; they have substantially lower $n$.}
  \label{fig:lambdanR}
\end{figure}

Figure~\ref{fig:lambdanR} shows additional evidence of an important structural difference between Es and S0s:  the correlation between Sersic index $n$ and size $R_e$ in our eight bins, for SRs (left) and FRs (right). Here we show galaxies with {\tt FLAG$\_$FIT=1} ($\sim 64 \%$ of our E sample), i.e. whose photometry is better described by a single Sersic profile (see \citealt{Fischer2019} for details). We also only show those bins which include at least five galaxies. The dotted horizontal lines show the median $n$ for the Es in each panel, and dashed lines show the median for the Es in each bin.  This shows that SR Es tend to have larger $n$ than their FR counterparts.  The open and filled triangles in the FR panel show that the B00 and B10 S0s ($\sim 35$\% of the S0s have {\tt FLAG$\_$FIT=1}) tend to have the smallest $n$ at each $R_e$ than their fast rotating E counterparts.  
%% E flag1 = 64%, 28% SR, 20% lambda>0.2
%%   flag 0 or 2 = 36%, 15% SR, 10% lambda>0.2

%% S0 flag1 = 35%
%%    flag 0 or 2 = 65%

% eint(eint-2) = e(e-2)/sini^2

\begin{figure}
  \centering
  \includegraphics[width=0.99\linewidth]{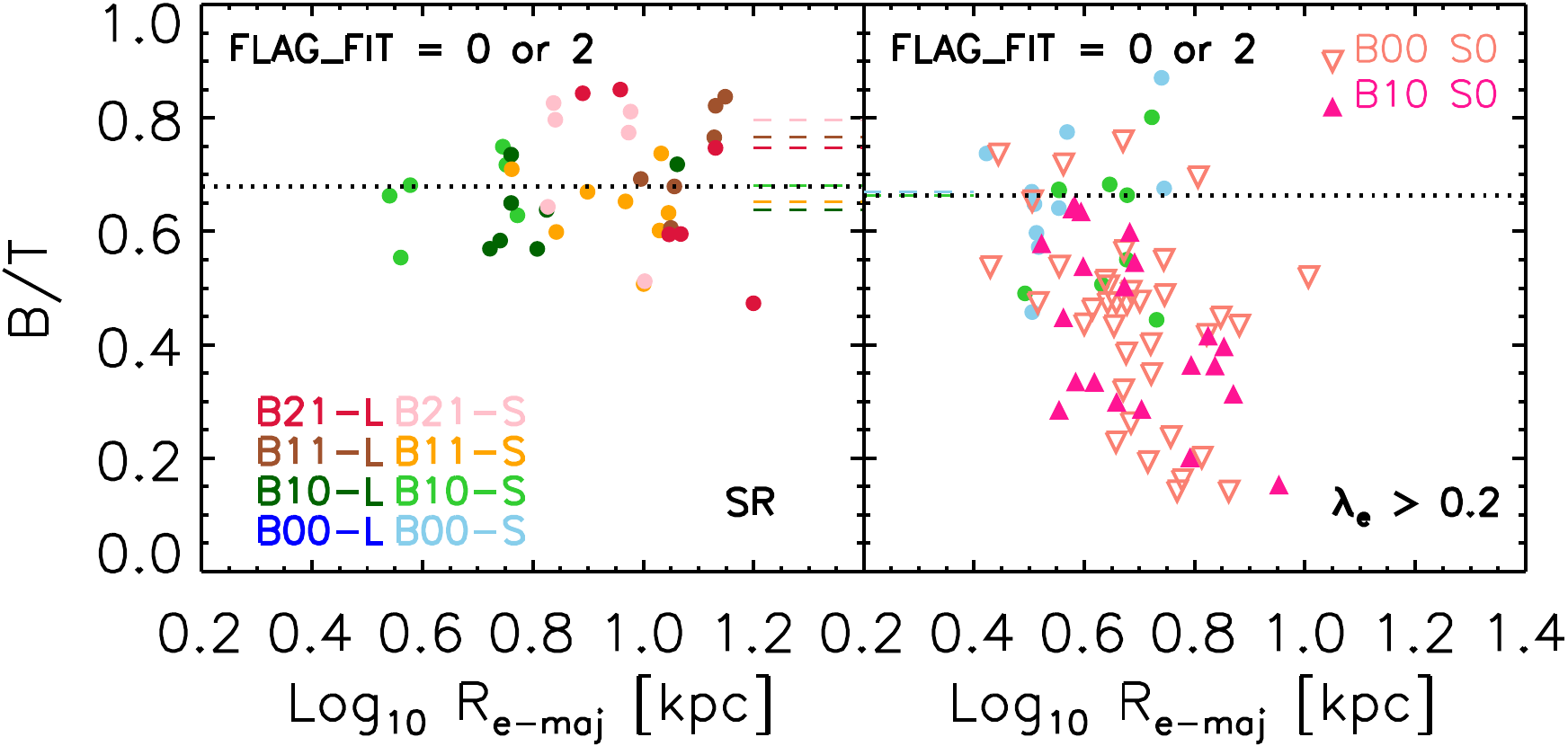}
   \vspace{-0.5cm}
   \caption{Correlation between bulge to total fraction B/T and size $R_e$ in our eight bins, for SRs (left) and FRs (right) with {\tt FLAG$\_$FIT=0} or 2 (i.e. objects whose photometry is as well or better-fit by a two-component Sersic+Exponential profile:  $\sim 36 \%$ of our Es). Dotted horizontal lines show the median B/T for the Es in each panel. Dashed colored horizontal lines show the median B/T in each bin.  Triangles in the right hand panel show {\tt FLAG$\_$FIT=0} or 2 S0s  ($\sim 65 \%$ of the S0s) with the same $L_r$ and $\sigma_0$ as the Es in bins B00 and B10; they have lower B/T.}
  \label{fig:lambdaBT}
\end{figure}

To study the other 36\% of our E sample ($\sim 65$\% for S0s), Figure~\ref{fig:lambdaBT} shows the objects classified as having {\tt FLAG$\_$FIT=0} or 2 (as for the previous Figure, we only show those bins which include at least five galaxies). These are objects for which a two-component Sersic+Exponential fit was as good as or preferred to the single-Sersic fit.  In this case, rather than showing $n$ of the Sersic bulge component we show B/T, the fraction of the light that is associated with the bulge.  The panel on the left shows that although some of the slow rotator Es may have two components, they always have B/T $> 0.6$ or so.  Note that there are no dark or light blue symbols in this panel: almost none of the SR Es in bin B00 are 2-component systems.  The panel on the right shows fast rotators.  Here, there are no dark blue or green symbols -- i.e., most of the larger galaxies in a bin are well described by a single component.  While some of the more compact Es in a bin (light blue and light green) are better fit by two components, these tend to have B/T values that are similar to the SRs.  In particular, these B/T values are larger than for the S0s of similar $L_r$ and $\sigma_0$.

Although S0s tend to have the smallest $n$ and B/T at each $R_e$ compared to their fast rotating E counterparts, in neither case is there an obvious gap between the FRs Es and S0s. Also there are no S0s with ellipticity $\epsilon\le 0.2$ in bin B10 of Figure~\ref{fig:lambda} (filled triangles in top right panel).  Since the dashed lines trace the expected locus in the $\lambda_e$-$\epsilon$ plane as the inclination with respect to the line of sight changes, seeing objects at large $\epsilon$ but not at small values may indicate that our Deep Learning morphologies have a tendency to label face-on S0s as Es.  This is less of an issue for bin B00 where we do see S0s with $\epsilon<0.2$ (triangles in the top left panel of Figure~\ref{fig:lambda}), so if there is an inclination-dependent bias, it is $\sigma_0$-dependent.  \cite{Fischer2019} provide more discussion of the $\epsilon$ distribution. In addition, the majority of Es are clearly single component systems, whereas the majority of S0s are clearly not and (although we do not show it here) the average B/T of S0s ($\sim 0.4$) is smaller than that of Es ($\sim 0.6$) even at smaller ellipticity.  So as to reduce the question of morphological effects as much as possible, we have chosen to separate FR Es from S0s in the analysis which follows. The next section, where we consider stellar populations, will show if this was necessary. (One can always perform a weighted sum of the E and S0 results should the separation prove to have been unnecessary.)

\begin{figure}
  \centering
  \includegraphics[width=0.99\linewidth]{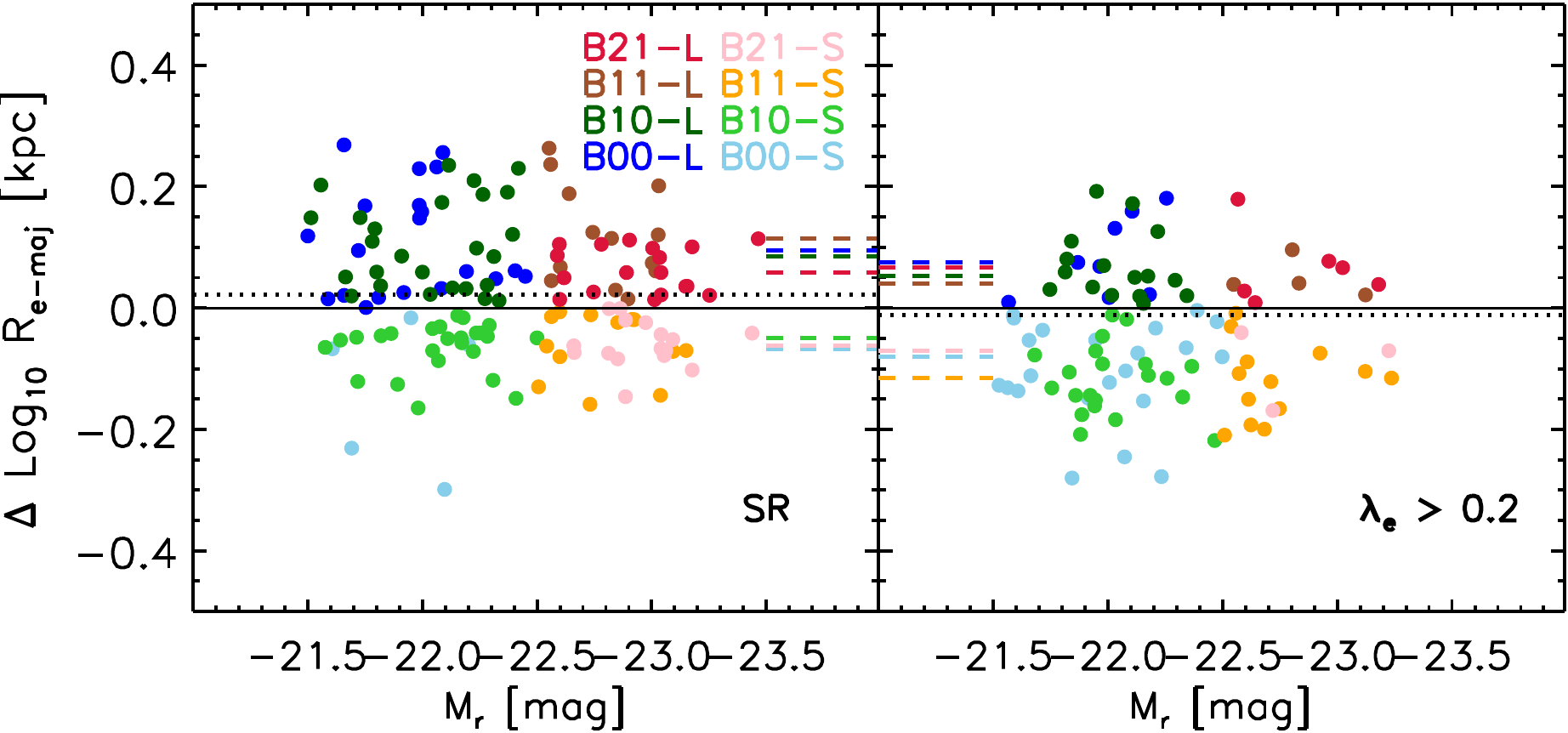}
   \vspace{-0.5cm}
  \caption{Residuals with respect the mean relation for each bin (i.e. the solid lines in Figure~\ref{fig:split}) of the size-luminosity distribution for slow (left) and fast (right) rotators. Dotted horizontal lines show the median residual for the Es in each panel, and dashed horizontal lines in each panel show the median residual in each bin: SRs are slightly larger than FRs in essentially all bins.}
  \label{fig:lambdaMR}
\end{figure}

\begin{figure}
  \centering
  \includegraphics[width=0.99\linewidth]{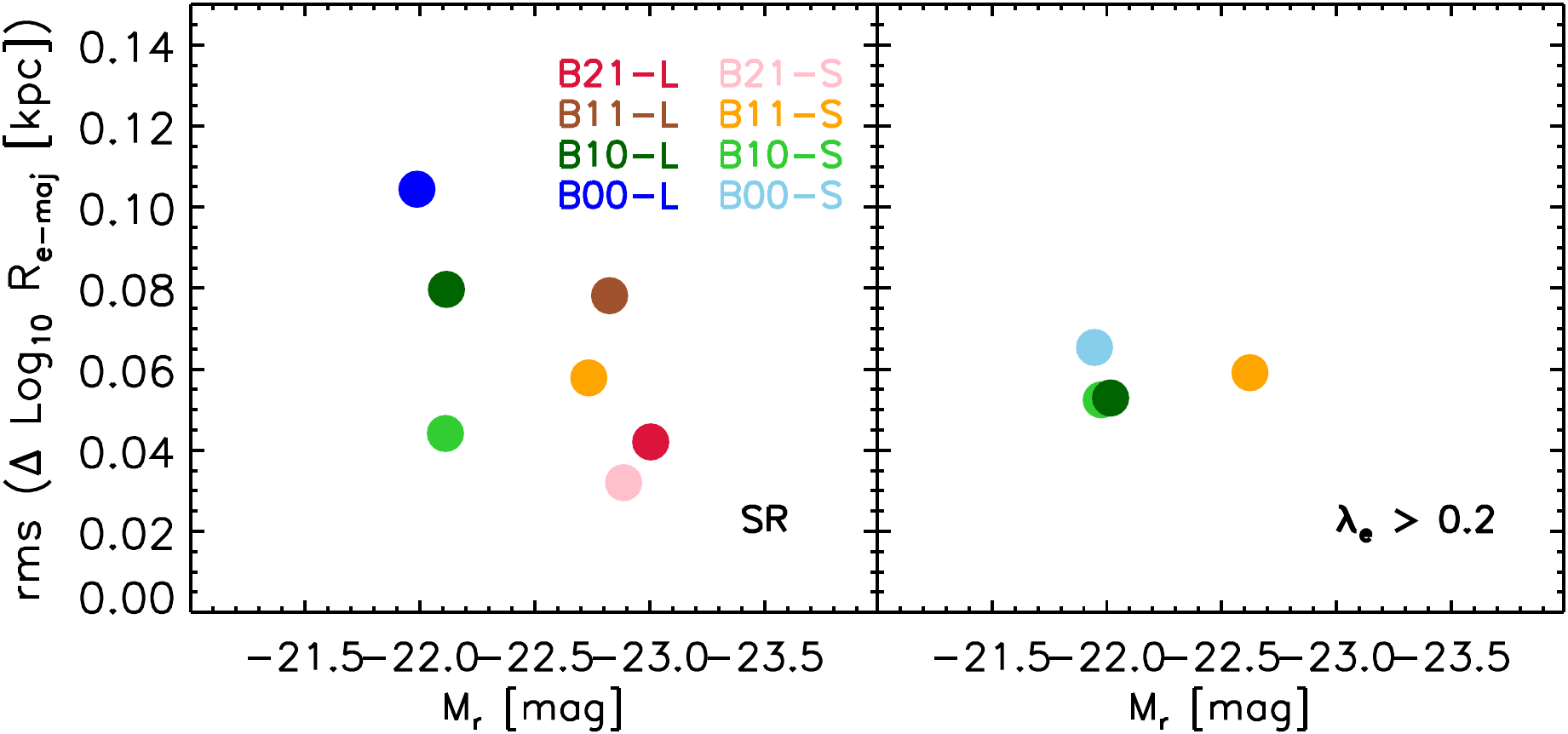}
  \vspace{-0.5cm}
  \caption{Scatter around the $R_e$-$L_r$ relation decreases as $L_r$ increases.  For slow rotators (left), this scatter is slightly smaller for the smaller objects in a given ($\sigma_0$,$L_r$) bin. Here, we only show those bins which include enough galaxies to draw statistically significant conclusions (see text).}
  \label{fig:rmsMRlambda}
\end{figure}

\subsection{Correlation between size and rotation for Es}
Since dissipationless mergers are expected to increase $n$ \cite[e.g.][]{Hilz2013}, it is tempting to view the differences between the Es in the two panels of Figure~\ref{fig:lambdanR} as further evidence that SR Es had merger-dominated histories.  Dissipationless mergers are also expected to increase galaxy sizes more dramatically than velocity dispersion.  So it is interesting that SRs tend to lie slightly above the $R_e$-$L_r$ relation defined by all the objects in the bin, whereas FRs tend to lie slightly below it (compare left and right hand panels of Figure~\ref{fig:lambdaMR}).  However, the scatter around the bin-dependent $R_e$-$L_r$ relation decreases as $L_r$ increases (Figure~\ref{fig:rmsMRlambda}) and it is not obvious that mergers can explain a decrease in this scatter.  Nevertheless, the fact that SRs tend to be larger than FRs means that one must be careful to separate correlations with size from those with rotation.  The results of the next section illustrate this nicely.

%Figure~\ref{fig:lambdanR} shows that SRs (left) tend to have slightly larger Sersic indices than FRs (right).  As minor mergers are expected to increase $n$ \citep{Hilz2013}, this small difference is consistent with the expectation that merger dominated histories involving stochastic accretion events are what keep SRs from remaining or becoming FRs.

While we have pointed out a number of circumstantial lines of evidence linking SRs to mergers, the next section provides a more careful consideration of the stellar populations. As we remarked earlier, this requires that we stack spectra, which, in turn is limited by the number of objects in a bin.  This number is provided in Table~\ref{tab:bin}.  If it is less than 10, then we do not use the bin for subsequent analysis, on the grounds that we cannot be certain of drawing statistically significant conclusions.  In addition, as we remarked earlier, we also must be certain that these objects contribute enough spaxels at each $R$ with which to make a stack of sufficiently high S/N.  Figure~\ref{fig:NspxFR}, the analog of Figure~\ref{fig:Nspx}, shows the number of spaxels in each radial bin which contribute to our SR and FR results for the eight $L_r$, $\sigma_0$ and $R_e$ bins.  The results which follow are based only on scales indicated by solid lines in Figure~\ref{fig:NspxFR}.  Dotted lines indicate too few galaxies in the bin (e.g. there are not enough objects with $\lambda_e\ge 0.2$ in bins B00-L, B11-L, B21-L and B21-S, and not enough SRs in bin B00-S), and dashed lines indicate scales which are compromised by the fact that the spaxels may not cover the entire region within $R_e$ (so the number of spaxels is decreasing rather than increasing with $R$).  This matters mainly for SRs in bins B00-L, B10-L, B21-L and especially B11-L.

\section{Single stellar population parameters}\label{sec:ssp}
We now describe differences between the stellar populations in the bins defined in Table~\ref{tab:bin}.  We first present measurements of a number of spectral features, and then the result of using single stellar population synthesis models (SSPs) to interpret these measurements in terms of IMF shape, age, metallicity [M/H], $\alpha$-element abundance [$\alpha$/Fe] and stellar mass-to-light ratio $M_*/L_r$.  We study how SSP results depend on size and rotation, both as a function of the local value of $\sigma$ and galactocentric distance $R$.

Paper~I compared a variety of single SP synthesis models, finding differences in overall values, but much smaller differences regarding relative comparisons.  E.g. whereas different SSPs might disagree on the value of the age of the stellar population, if one SSP finds that two bins differ in age by $\sim 2$ Gyrs, then the other is likely to also.\footnote{It is worth noting explicitly that all the ages we quote are H$_\beta$-luminosity weighted ages.  Recent star formation by a small fraction of the population will bias these ages younger than `mass-weighted' ages.}  Therefore, in what follows, we only show results based on the MILES-library with Padova isochrones and BiModal IMFs \citep{Vazdekis2015}. In these models, the scaled-solar spectra (i.e. [$\alpha$/Fe]=0) have abundances from \cite{Grevesse1998}, whereas the $\alpha$-enhanced spectra ([$\alpha$/Fe]$= +0.4$) assume that [X/Fe]$= +0.4$ for the elements O, Ne, Mg, Si, S, Ca and Ti, and that the other elements have solar abundances. Note that, since we only use a combination of Fe and Mg lines to constrain the $\alpha$-enhancement (see middle panel in Figure~\ref{fig:MilesBIsplit}) without taking individual element ratios into account, in practice the [$\alpha$/Fe] we report is based entirely on the [Mg/Fe] abundance \cite[e.g.][]{Johansson12}.

\subsection{Lick index diagnostics}
We measured a number of Lick indices in the stacked spectra (H$_\beta$, Mg$b$, Fe5270, Fe5335, TiO2$_{\rm SDSS}$; see Table~3 of Paper~I for details, where we also describe how the two Fe lines are combined to make what is refered to as <Fe>, and how they are combined with Mg$b$ to make [MgFe]).

\begin{figure}
 \centering
  \includegraphics[width=0.79\linewidth]{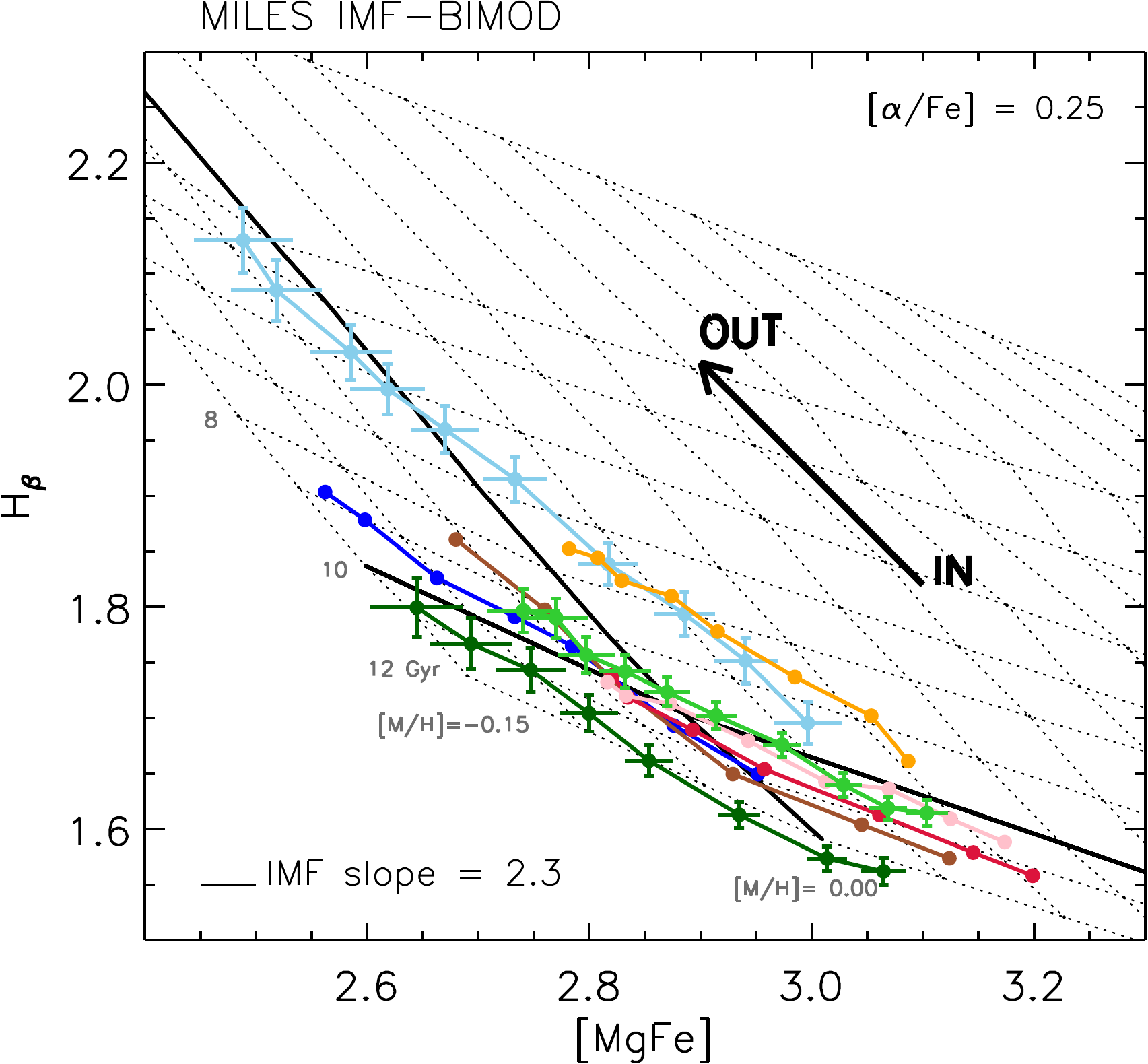}
  \includegraphics[width=0.79\linewidth]{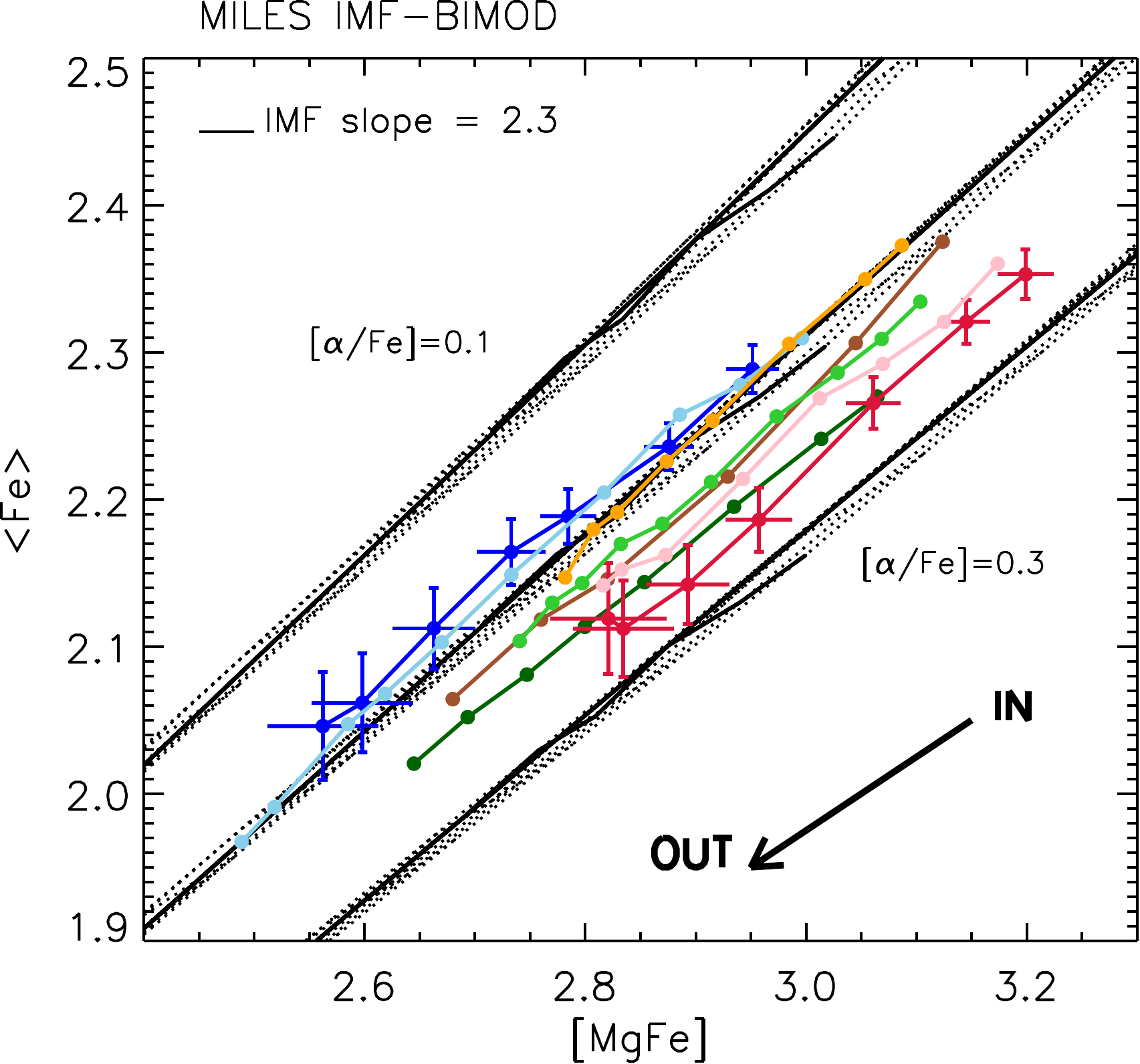}
  \includegraphics[width=0.79\linewidth]{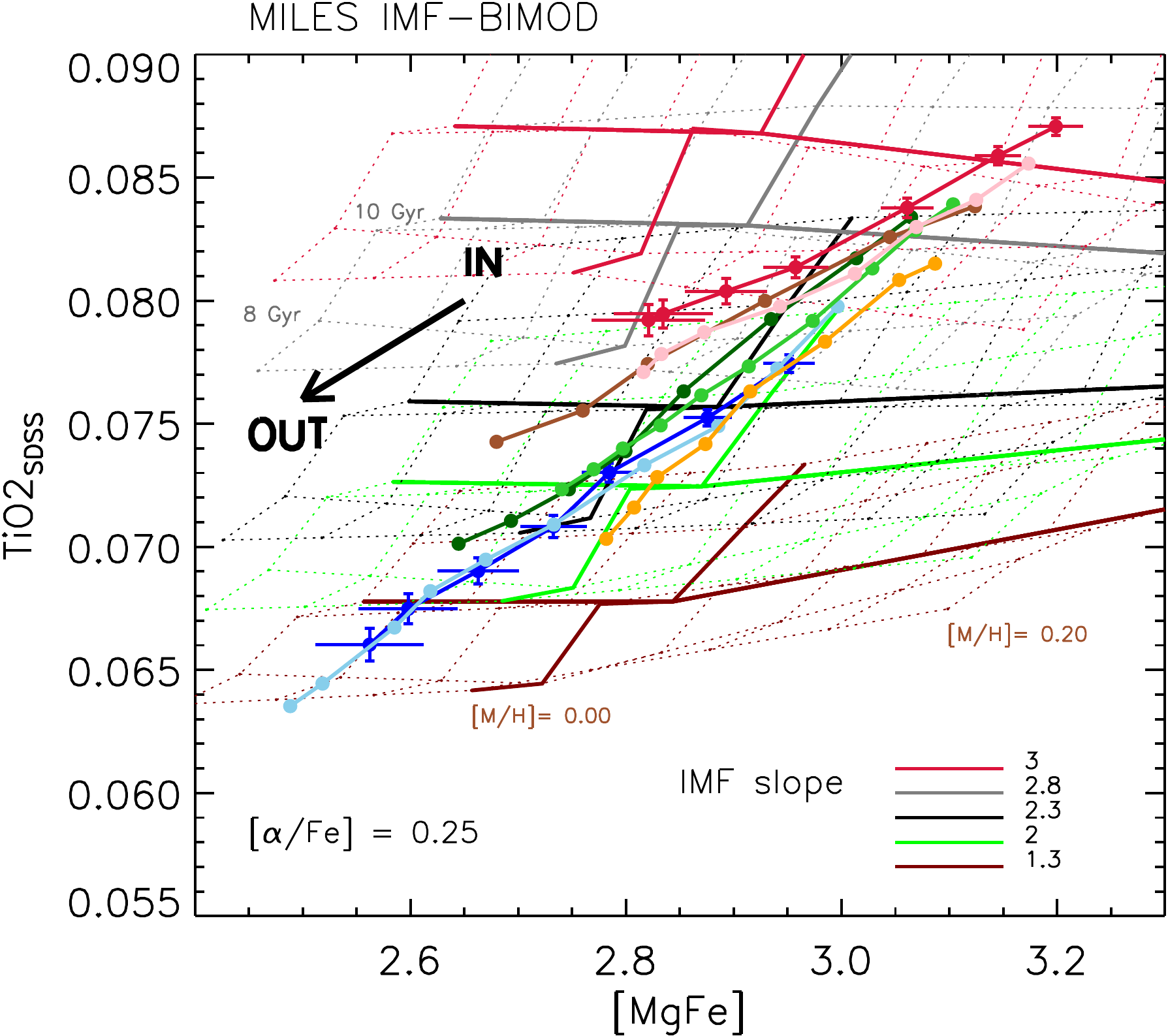}
  \caption{Lick indices H$_\beta$, <Fe> and TiO2$_{\rm SDSS}$ as a function of [MgFe] measured in the stacked spectra of the subsamples shown in Figure~\ref{fig:split}. Symbols of a fixed color show how index strength changes with distance from the center:  [MgFe] decreases from center outwards (thick bold arrow indicates the direction of increasing galactocentric distance). Lighter shades (light blue, light green, orange and pink) represent galaxies with smaller sizes given their $\sigma_0$ and $L_r$; darker shades (dark blue, dark green, brown, red) represent galaxies with larger sizes. For a fixed IMF, the top panel is an age-metallicity diagnostic and the middle panel is an indicator of $\alpha$-enhancement.  The bottom panel is used as an IMF diagnostic.  In all three panels, SSP grids for some parameters fixed as labeled are shown.  The smaller sizes, shown using lighter shades, clearly have stronger H$_\beta$;  they also have stronger <Fe> and slightly weaker TiO$_2$ (except for B00).}
  \label{fig:MilesBIsplit}
\end{figure}

Figure~\ref{fig:MilesBIsplit} shows the H$_\beta$-[MgFe], <Fe>-[MgFe] and TiO2$_{\rm SDSS}$-[MgFe] diagnostic plots.  The first is an indicator of ages and metallicities; the second of $\alpha$-element abundances; and the third is sensitive to the IMF.  The objects with smaller sizes (lighter shades:  light blue, light green, orange, pink) clearly have stronger H$_\beta$ and, except for bin B00, stronger <Fe> and weaker TiO2$_{\rm SDSS}$.  Superimposed on the measurements are model grids corresponding to a bimodal IMF with slope 2.3 and the assumed value of [$\alpha$/Fe] shown. %, which we discuss in the next subsection.  

\begin{figure}
  \centering
  \includegraphics[width=0.79\linewidth]{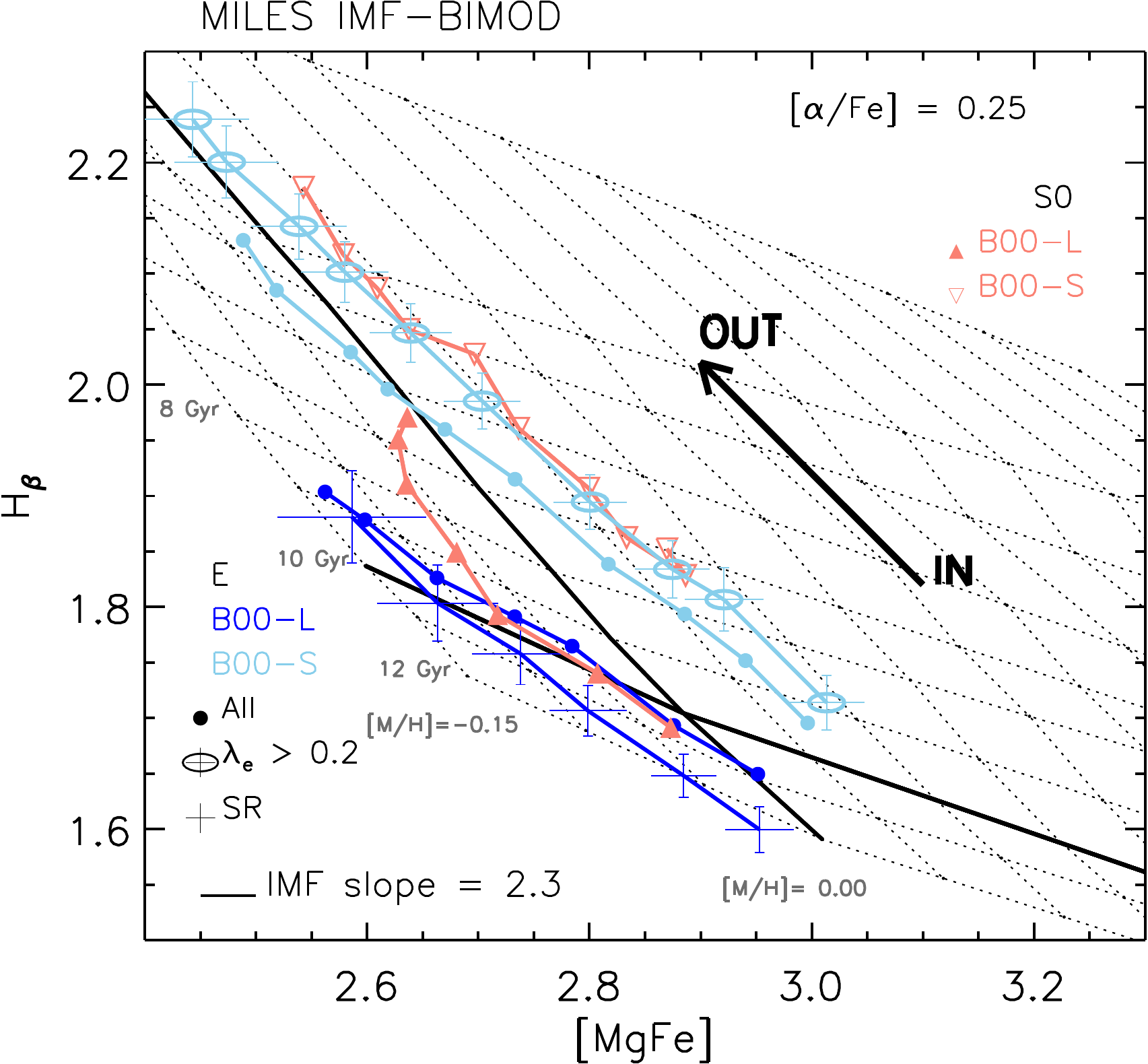}  
  \includegraphics[width=0.79\linewidth]{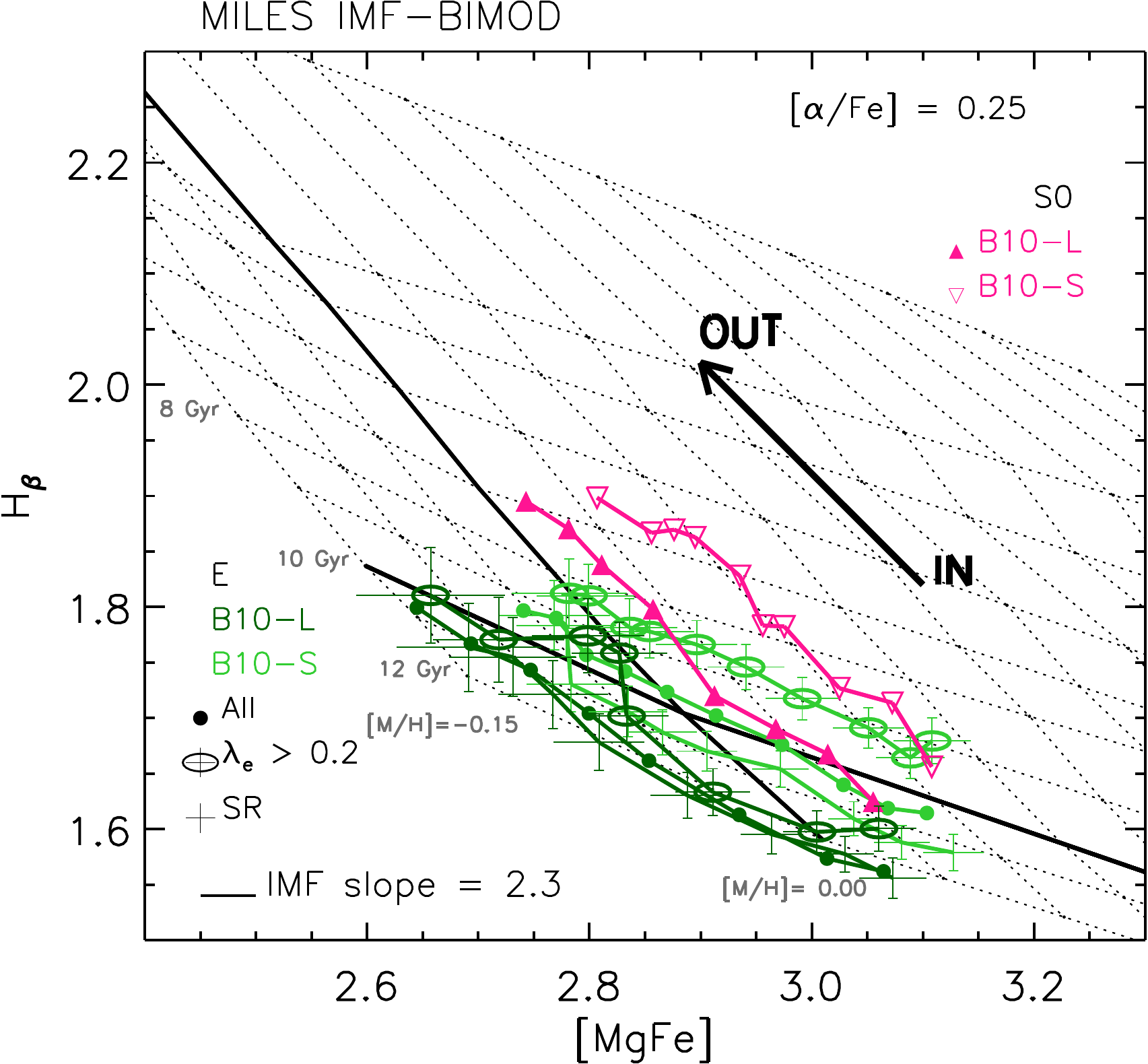}
  \includegraphics[width=0.79\linewidth]{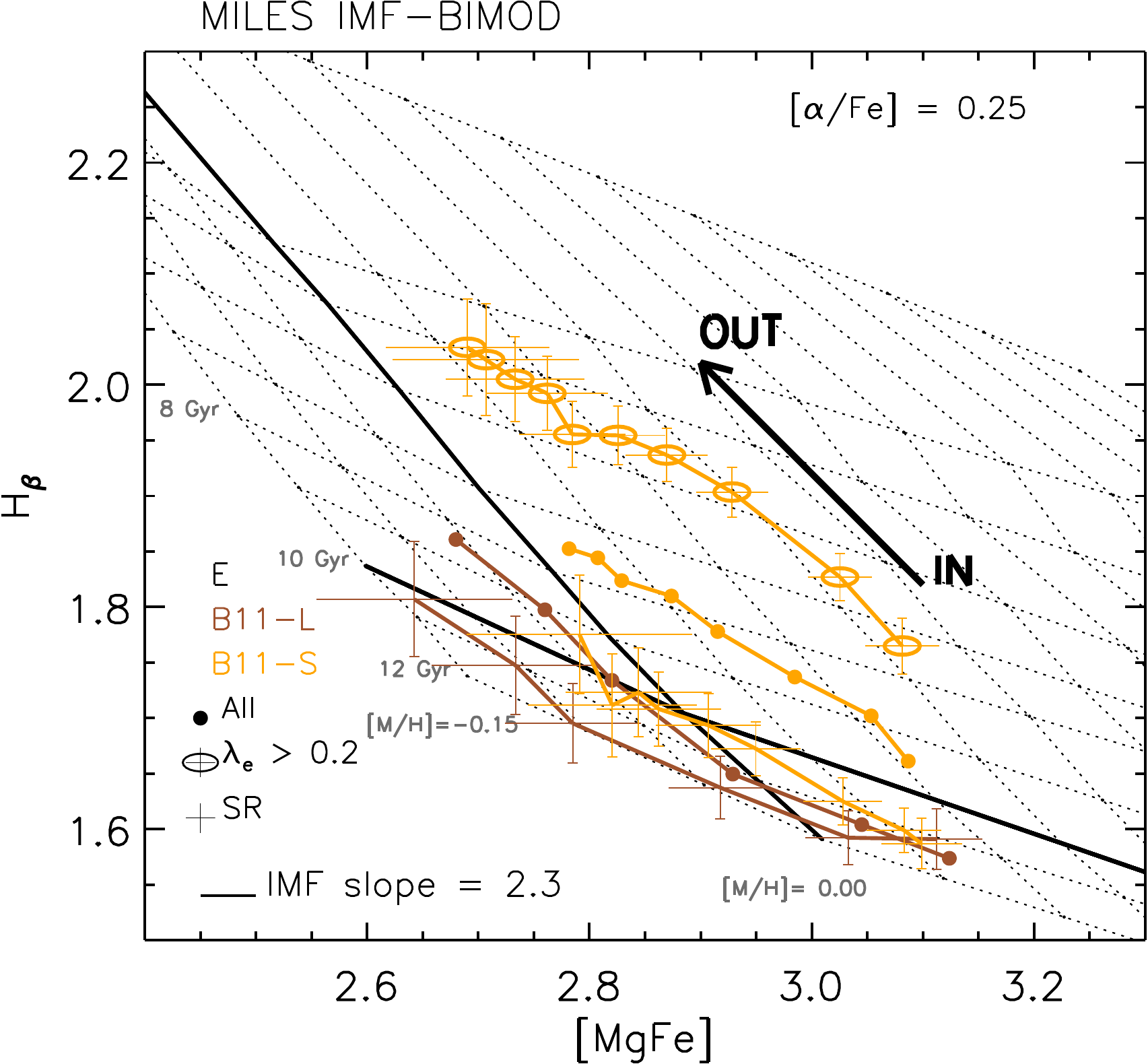}
  \caption{H$_\beta$ vs [MgFe] for Es and S0s in bins B00 (top) and B10 (middle), and Es in B11 (bottom), subdivided by size (lighter shades show the smaller sizes in a bin) and further by rotation (crosses and ellipses show SRs and FRs).  Fast rotators (ellipses) tend to have larger H$_\beta$ for their [MgFe] than do slow rotators (crosses).  SSP models interpret this as younger ages and larger metallicities for FRs. Filled circles (without error bars) show the values reported in Figure~\ref{fig:lambdasplit} before subdividing the sample in SRs and FRs. Filled and inverted open triangles in the top and middle panels show S0s which were classified as large or small using the same criteria as the Es.} 
  \label{fig:lambdasplit}
\end{figure}

\begin{figure}
  \centering
  \includegraphics[width=0.79\linewidth]{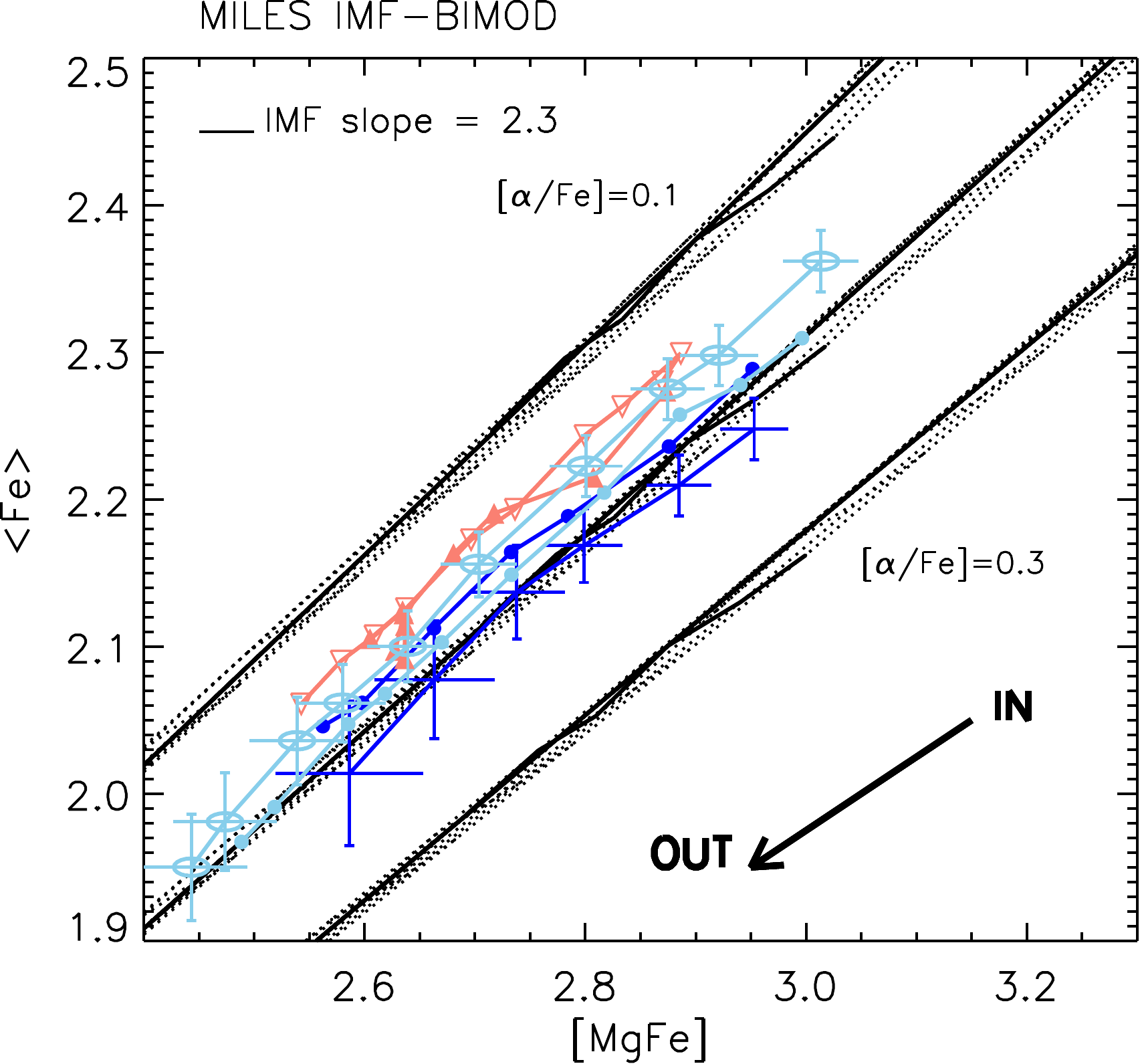}  
  \includegraphics[width=0.79\linewidth]{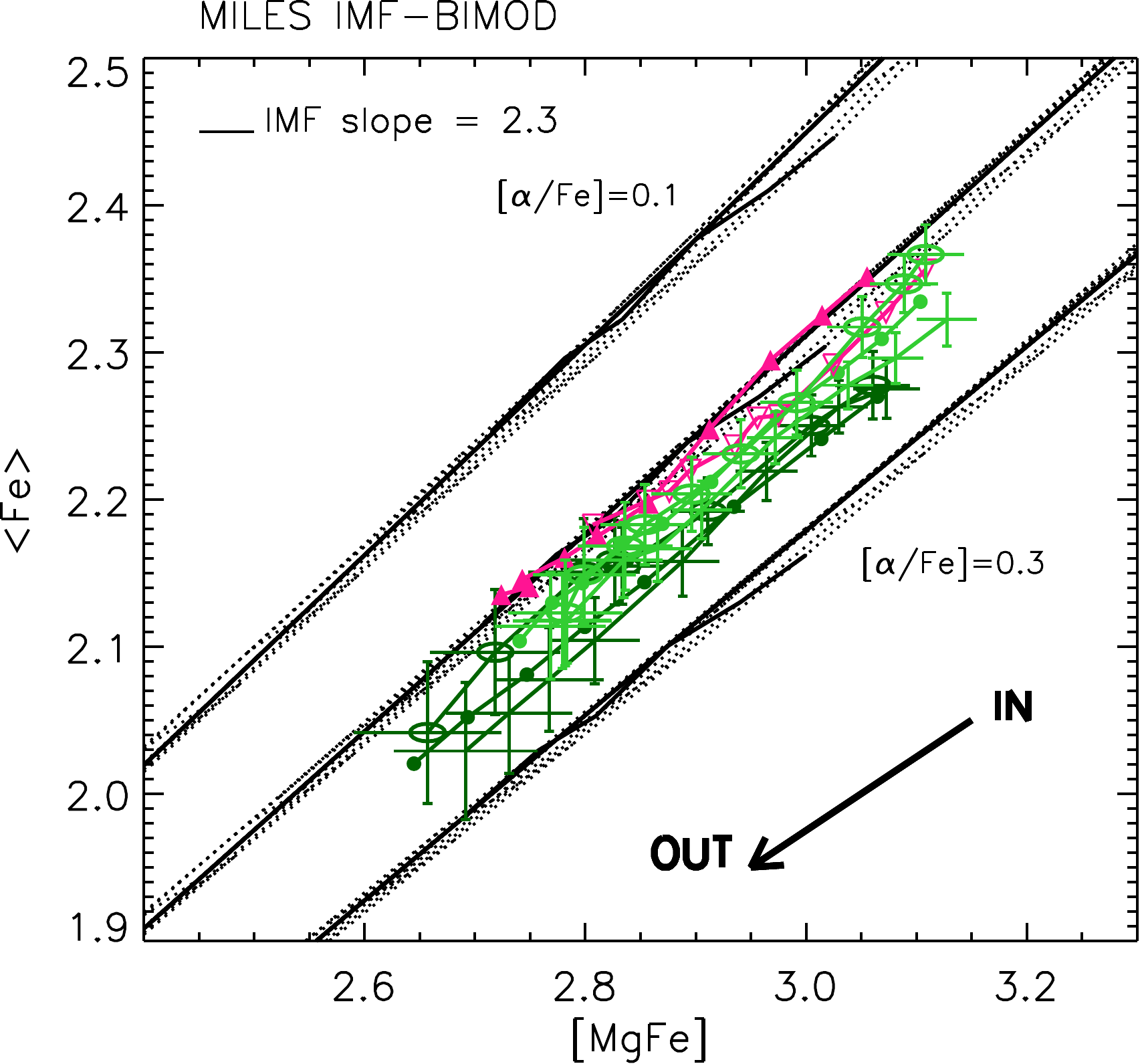}
  \includegraphics[width=0.79\linewidth]{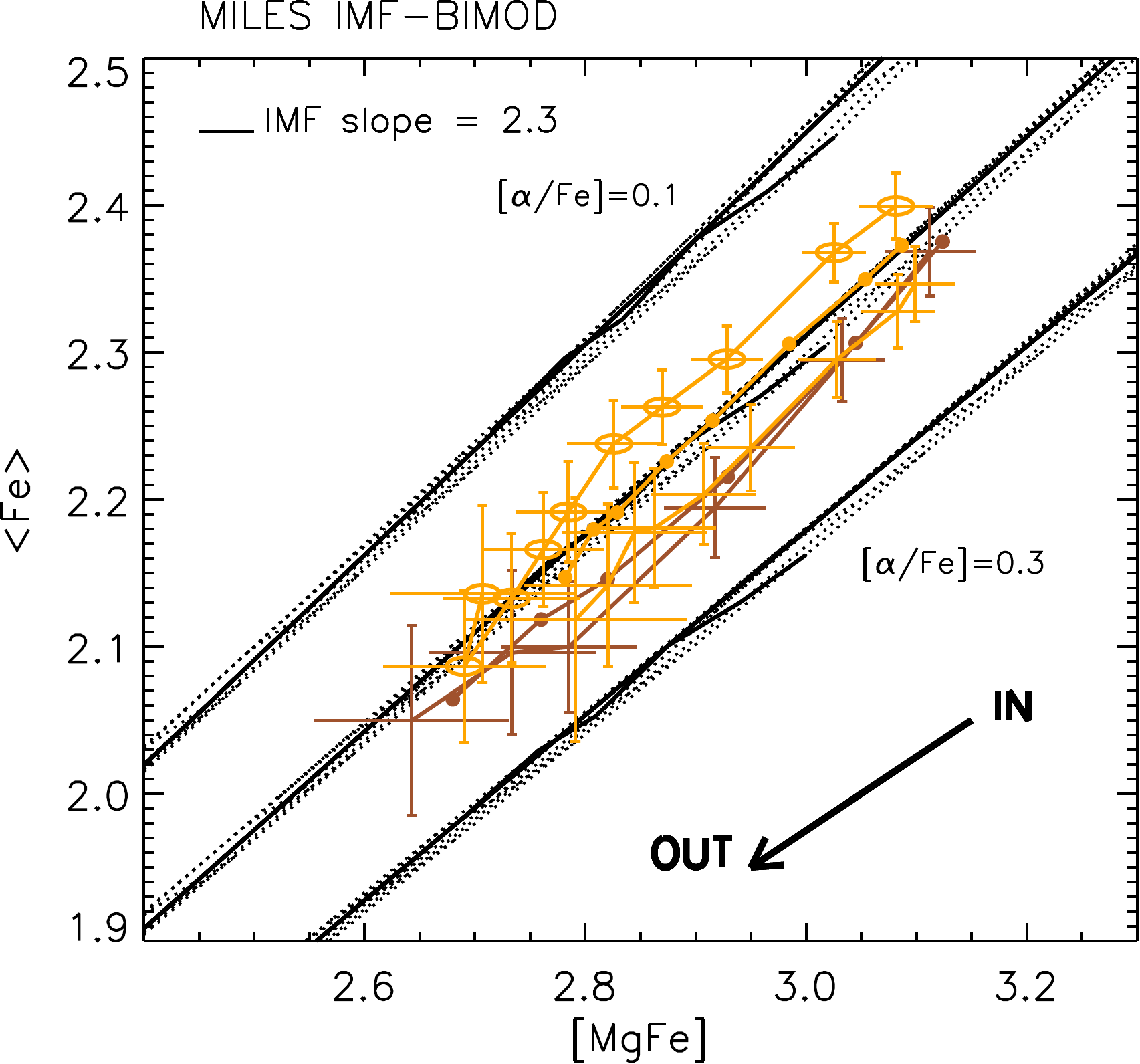}
  \caption{Same as previous figure, but now showing $\langle{\rm Fe}\rangle$ vs [MgFe]. In each bin, fast rotators (ellipses) tend to be less $\alpha$-enhanced.  }
  \label{fig:lambdasplit2}
\end{figure}

Before we use SSPs to interpret these differences, Figures~\ref{fig:lambdasplit} and~\ref{fig:lambdasplit2} show the H$_\beta$-[MgFe] and <Fe>-[MgFe] distributions for the SR and FRs in each bin.  (There are not enough FRs in bin B21, so we only show results for the other 3 bins.)  Each panel shows four sets of symbols, corresponding to the separation based on size (light vs dark shades), further subdivided into SRs and FRs (crosses and ellipses) where possible: For B00-L only SRs are shown (blue); for B00-S only FRs (light-blue); and for B11-L only SRs (brown). Filled circles (without error bars) show the values reported in Figure~\ref{fig:lambdasplit} before subdividing the sample in SRs and FRs.  In this format, Figure~\ref{fig:lambdasplit} highlights the fact that the smaller sizes have larger H$_\beta$, and at fixed size, the FRs have larger H$_\beta$.  This difference between SRs and FRs is most pronounced for bin B11-S (orange symbols in bottom panel), and least pronounced for bin B10-L (dark green in middle panel).  Similarly, Figure~\ref{fig:lambdasplit2} shows that the larger galaxies in a bin are shifted down and to the right of the smaller galaxies (i.e. towards higher [$\alpha$/Fe]), and SRs are shifted down and to the right of FRs of the same size.

It is interesting to compare these measurements for Es to those for the S0s that we had removed from bins B00 and B10.  The filled and open inverted triangles in the top and middle panels show these S0s. %(based on the same size classification as for Es).
S0s are FRs, and for bin B00-S the H$_\beta$-[MgFe] values are similar to those of FR Es (light blue ellipses). While this suggests that the distinction between FR Es and S0s is not well motivated -- consistent with \cite{Emsellem2011} and~\cite{C16} -- the <Fe>-[MgFe] plot shows that the S0s tend to be slightly offset towards larger <Fe>. Also note that the central regions of S0s in bin B00-L (filled triangles) are more similar to the B00-L SR Es. Moreover, the differences between FR Es and S0s is larger in bin B10, which is why we removed S0s from all the analysis and interpretation which follows.  It is clear that, had we included them, they would simply reinforce the differences we report between SR and FR Es.

\subsection{Stellar populations: Systematics}\label{ssp:sys}
As Paper~I discusses, to interpret these measurements, we must make assumptions about how either the IMF or the different element enhancements ([X/Fe]) depend on galaxy type. Specifically, the TiO2$_{\rm SDSS}$-[MgFe] model grids (the ones in the bottom panels of Figure~\ref{fig:MilesBIsplit}) shift vertically as the value of [X/Fe] varies (at fixed IMF). Since [X/Fe] is not known a priori, this translates into uncertainty in the best-fit IMF. To proceed, we follow Paper~I in making two rather different assumptions:  one is that all galaxies have the same IMF, whatever their $\sigma_0$ and $L_r$, and that there are no IMF gradients within galaxies (following Paper~I we refer to this as Assumption~1).  In this case, the $\Delta_{\rm [X/Fe]}$, i.e., the difference between the measured TiO2$_{\rm SDSS}$ from the stacked spectra and the value from the MILES model (with age, metallicity and [$\alpha$/Fe] given by e.g. Figure~\ref{fig:MilesBIsplit}) due to variation in [X/Fe] enhancements, shows strong gradients, which Paper~I argues are unrealistic.  The other is that $\Delta_{\rm [X/Fe]} = 0.003$ within a galaxy (Paper~I refers to this as Assumption~3).  In this case the IMF shows gradients.

One consequence of keeping $\Delta_{\rm [X/Fe]}$  constant is that metal rich objects {\em must} have bottom-heavy IMFs (steeper slopes).  In addition, there is a relatively strong degeneracy between the maximum age we allow and the IMF we infer.  Our fiducial choice requires ages to be less than 13~Gyrs (maximum redshift of formation of $z=7$).  However, if we set this upper limit to be 11~Gyrs (stars formed after redshift 2.5) then the inferred IMF is more bottom-heavy.  In turn, this impacts the value of $M_*/L_r$:  more bottom heavy IMFs have larger $M_*/L_r$.  Note that these shifts do not affect the rank ordering of the different bins.

\subsection{Stellar population trends with local velocity dispersion}\label{ssp:sig}
\begin{figure}
  \centering
  \includegraphics[width=0.9\linewidth]{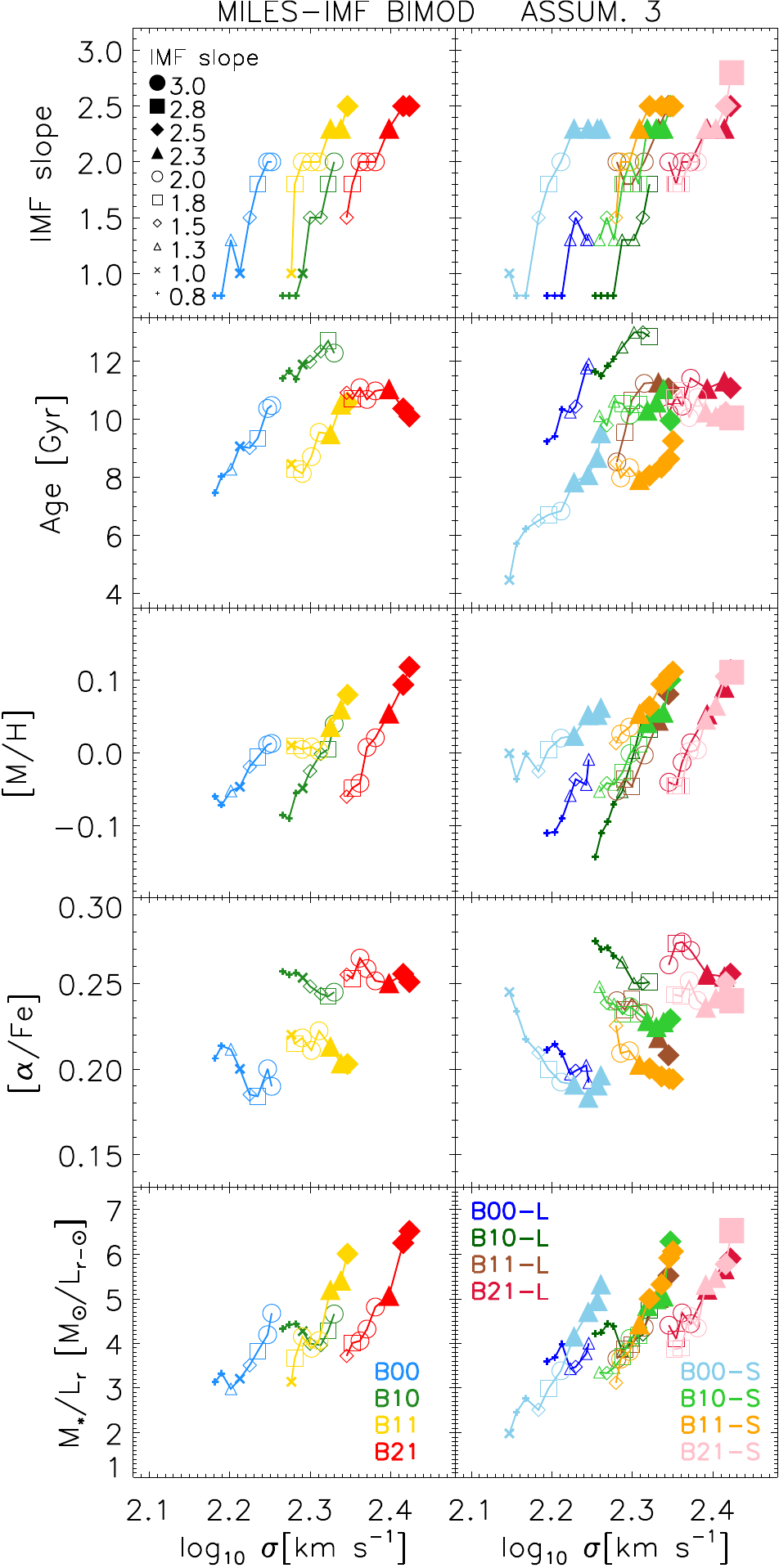}
  \caption{IMF slope, age, metallicity, [$\alpha$/Fe] and $M_*/L_r$ gradients in our four bins (left) and upon subdividing each by size (right), if we assume the IMF can differ from bin to bin, and vary within a galaxy, subject to the requirement that $\Delta_{\rm [X/Fe]}=0.003$ in a galaxy. Different symbols indicate the slope (shown in the legend) of the closest IMF model, but all the properties (including the IMF slope) shown in the panels were inferred by interpolating between the two best-fitting IMF models.}\label{option3}
\end{figure}

Figure~\ref{option3} shows the SSP-inferred stellar population parameters in our four bins (left) and when subdivided by $R_e$ (right), for our second assumption about $\Delta_{\rm [X/Fe]}$ (i.e, $\Delta_{\rm [X/Fe]}=0.003$ within a galaxy).
  %Figure~\ref{option2}, which shows what follows from assuming the same IMF for all galaxies and no IMF gradients, shows strong \textcolor{red}{$\Delta_{\rm [X/Fe]}$} gradients (top panels).\textcolor{red}{Figure~\ref{option3} shows the results for our other assumption: $\Delta_{\rm [X/Fe]}=0.003$ within a galaxy.} In this case, the top panels show the variations in the inferred IMF slope \textcolor{red}{since $\Delta_{\rm [X/Fe]}$ is constant on all scales in all bins.}
Symbols show the closest-fitting IMF model (corresponding to the legend in the top left panel), but the actual value of IMF slope (as well as all the other inferred properties) shown is got by interpolating between the two best-fitting IMF models.  The panels below them show the associated age, [M/H], [$\alpha$/Fe] and $M_*/L_r$ values which follow from the SSPs. These results are rather similar to those obtained by assuming the same IMF for all galaxies and no IMF gradients (i.e. our first assumption -- we do not show a figure because we believe the required $\Delta_{\rm [X/Fe]}$ gradients are unrealistic), except for $M_*/L_r$, which shows larger variations when IMF gradients are allowed (i.e., as in Figure~\ref{option3}).

%\begin{figure}
%  \centering
  %  \includegraphics[width=0.43\linewidth]{FIGURES/TiOenhan_sig_MILES_Padova_BI_TiO2sdss_ONEIMF_pap2.pdf}
 % \includegraphics[width=0.43\linewidth]{FIGURES_SPLIT/split_TiOenhan_sig_MILES_Padova_BI_TiO2sdss_ONEIMF_pap2.pdf}
 % \includegraphics[width=0.43\linewidth]{FIGURES/Age_sig_MILES_Padova_BI_TiO2sdss_ONEIMF_pap2.pdf}
 % \includegraphics[width=0.43\linewidth]{FIGURES_SPLIT/split_Age_sig_MILES_Padova_BI_TiO2sdss_ONEIMF_pap2.pdf}
 % \includegraphics[width=0.43\linewidth]{FIGURES/Met_sig_MILES_Padova_BI_TiO2sdss_ONEIMF_pap2.pdf}
 % \includegraphics[width=0.43\linewidth]{FIGURES_SPLIT/split_Met_sig_MILES_Padova_BI_TiO2sdss_ONEIMF_pap2.pdf}
 % \includegraphics[width=0.43\linewidth]{FIGURES/Alpha_sig_MILES_Padova_BI_TiO2sdss_ONEIMF_pap2.pdf}
 % \includegraphics[width=0.43\linewidth]{FIGURES_SPLIT/split_Alpha_sig_MILES_Padova_BI_TiO2sdss_ONEIMF_pap2.pdf}
 % \includegraphics[width=0.42\linewidth]{FIGURES/ML_sig_MILES_Padova_BI_TiO2sdss_ONEIMF_pap2.pdf}
 % \includegraphics[width=0.42\linewidth]{FIGURES_SPLIT/split_ML_sig_MILES_Padova_BI_TiO2sdss_ONEIMF_pap2.pdf}
%\includegraphics[width=0.9\linewidth]{FIGURES/results_vs_sigma_MILES_SPLIT_assum1.pdf}  
%  \caption{\textcolor{red}{$\Delta_{\rm [X/Fe]}$}, age, metallicity, [$\alpha$/Fe] and $M_*/L_r$ gradients in our four bins (left) and upon subdividing each by size (right), if we assume the IMF has slope 2.3 for all galaxies and there are no IMF gradients.  }\label{option2}
%\end{figure}

The left hand panels reproduce results from Paper~I.  They show that:
\begin{itemize}
  \item[-] at fixed $L_r$, age increases with $\sigma$ (blue and green symbols trace out one curve, and yellow and red symbols another), but the higher $L_r$ objects tend to have smaller ages for the same $\sigma$ (green is older than yellow/red). However, we find an inverted age gradient for the most massive galaxies (bin B21):  their centers are slightly younger;
  \item[-] at fixed $L_r$, [M/H] does not depend on $\sigma$ (blue and green have same [M/H], as do yellow and red), but the higher $L_r$ objects tend to have larger [M/H] for the same $\sigma$ (blue and green have smaller [M/H] than yellow and red);
  \item[-] higher $\sigma$ galaxies have higher [$\alpha$/Fe] (red lies above green and yellow, which lie above blue), but at fixed $\sigma$, if [$\alpha$/Fe] is larger then so is age, but [M/H] is smaller (compare yellow and green symbols in left hand panels: green is older and more enhanced, but less metal rich);
  \item[-] [$\alpha$/Fe] does not increase, and in some cases decreases with increasing $\sigma$ within a galaxy. 
  \item[-] IMF slope and [M/H] seem well correlated.
\end{itemize}
The trends with age, [$\alpha$/Fe] and [M/H] are well-known \cite[e.g.][]{Bernardi2005,Bernardi2006,Graves2010,McDermid2015}, and are usually taken to mean that higher $\sigma_0$ galaxies formed their stars on a shorter timescale (because ${\rm Log}_{10}$ (timescale/Gyrs) $\approx$ 1.2 - 6\,[$\alpha$/Fe], following, e.g., \citealt{Thomas2005}) and are less affected by supernovae (SN) feedback (because metallicity is low).  The anti-correlation between [M/H] and both age and [$\alpha$/Fe] at fixed $\sigma$
%\sout{applies locally within a galaxy has not been noticed or emphasized before. This anti-correlation}
is the basis for arguing that [$\alpha$/Fe] is enhanced because Fe is suppressed, and not because $\alpha$ is enhanced.  That IMF slope and [M/H] are coupled is also in qualitative agreement with previous work:  \cite{MN2015} argue that IMF slope $\approx 2.2 + 3\,$[M/H] for the BiModal models we are using here.  We find this too, but do not show it because this correlation is model dependent, as we discuss in Paper~I.
% [alpha/Fe] = 1/5 - (lg Deltat)/6 (Thomas2005)
% IMF slope = 2.2 + 3 [M/H]   (MartinNavarro2015)
The inversion of the age gradient for bin B21 is qualitatively consistent with \cite{Zibetti2019} which appeared while our paper was being refereed.  However, as we describe shortly, this inversion is not obvious when we subdivide by size. 
Finally, as we noted in Paper~I, it is surprising that the bin B21 galaxies (red), which have the largest $L_r$ {\em and} $\sigma_0$, are {\em not} the oldest.  The oldest galaxies (green) are in bin B10; they also have low [M/H] and anomalously high [$\alpha$/Fe] compared to the B11 (yellow) galaxies which have the same $\sigma_0$ but larger $L_r$.

\begin{figure*}
  \centering
 \includegraphics[width=0.75\linewidth]{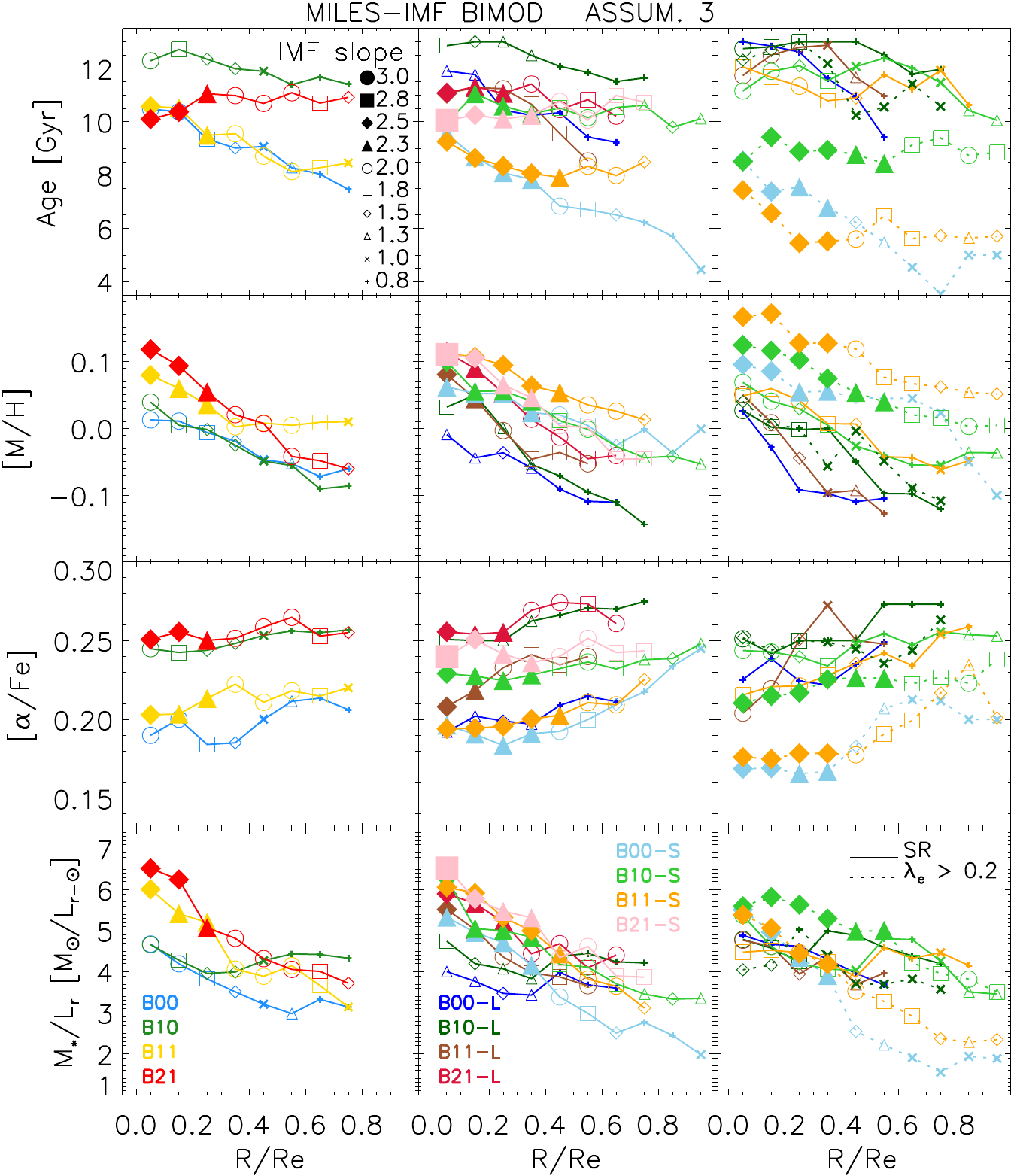}
 \caption{Age, metallicity, $\alpha$-enhancement and $M_*/L_r$ gradients when we allow IMF gradients (i.e. Assumption~3).  Left and middle panels show the results of Figure~\ref{option3} as a function of $R/R_e$ rather than $\sigma$.  Color coding is same as before (lighter shades represent bins with smaller $R_e$), and symbols indicate different IMFs, as given in the legend. Right hand panel shows gradients in subsamples further divided on the basis of rotation; solid and dashed lines show SRs and FRs, respectively, in the bins for which we have sufficient statistics (see Table~\ref{tab:bin} and related discussion). To better highlight the difference between FRs and SRs we do not show the SRs of bins B21-L (red line) and B21-S (pink line) in the third column (although we do have sufficient statistics for the SRs of both bins) because they are dominated by SRs and so are almost indistinguishable from the B21-L and B21-S bins shown in the middle panels.}   
 \label{tzR23}
\end{figure*}

%\begin{figure}
%  \centering
%  \includegraphics[width=0.45\linewidth]{FIGURES_SPLIT/split_Age_R_MILES_Padova_BI_TiO2sdss_ONEIMF_pap2.pdf}
%  \includegraphics[width=0.45\linewidth]{FIGURES_SPLIT/split_Age_R_MILES_Padova_BI_TiO2sdss_pap2.pdf}
%    \includegraphics[width=0.465\linewidth]{FIGURES_SPLIT/split_Met_R_MILES_Padova_BI_TiO2sdss_ONEIMF_pap2.pdf}
%  \includegraphics[width=0.465\linewidth]{FIGURES_SPLIT/split_Met_R_MILES_Padova_BI_TiO2sdss_pap2.pdf} 
%  \caption{Age and metallicity gradients as a function of $R/R_e$ rather than $\sigma$: (left panels) if we assume there are no IMF gradients, but that the IMF can vary with $L_r$ and $\sigma_0$ (i.e. as in Figure~\ref{option2}); (right panels) when we allow IMF gradients (i.e. as in Figure~\ref{option3}). Symbols indicate different IMFs, as given in the legend of Figure~\ref{option2}.}
%  \label{tzR23}
%\end{figure}

The right hand panels of Figure~\ref{option3} show the result of splitting the bins shown in the left hand panels in two, based on size (Table~\ref{tab:bin}).  Galaxies with smaller than average sizes (for their $\sigma_0$ and $L_r$), which we show using lighter shades, are typically about 2 Gyrs younger (except perhaps for B21 which shows a smaller difference), slightly more metal rich (except perhaps for B21), and less $\alpha$-enhanced (except for B00), but with, nevertheless, little difference in $M_*/L_r$ (the age and metallicity effects approximately cancel).  Note that the anti-correlation between either age or [$\alpha$/Fe] and [M/H] at fixed $\sigma$ persists when split by size (except perhaps in bin B21).  Bin B21, which includes galaxies with the largest $L_r$ and $\sigma_0$, shows the smallest differences. We will return to this in Section~\ref{ssp:rot}.  However, note that the inversion of the age gradient for this bin, which is obvious in the left hand panel, is no longer present when we subdivide by size.  The dependence on $R_e$ shows that there must be variations in star formation timescale and/or sensitivity to SN feedback even at fixed $\sigma_0$:  larger sizes had shorter timescales.  The tests shown in Appendix~A of Paper~I indicate that these conclusions are robust against changes in the details of the SSP models, i.e. although there are differences in overall values, the differences regarding relative comparisons are small. 

The differences between sizes (in a given $\sigma_0$ and $L_r$ bin) indicate that younger galaxies have higher surface brightnesses.  While this is qualitatively consistent with previous work, in which surface-brightness was used explicitly \cite[][see Appendix~B for more discussion]{Graves2010}, we believe that our binning in $\sigma_0$ and $L_r$ before subdividing in $R_e$ (rather than $\sigma_0$ and $R_e$ before subdividing in $I_e$) more effectively separates objects of similar masses.  Moreover, as we discuss in the next subsection, we believe that rotation plays an important role.  

It is clear that the oldest objects are those with larger than average sizes in B10-L, although the inner regions of B00-L objects are also surprisingly old: e.g. their age is similar (or even larger) than for B21 objects which have larger $L_r$, $\sigma_0$ and $R_e$.  Note also that B10-S objects (light green) have approximately the same age, [M/H] and [$\alpha$/Fe] as B11-L objects (brown), even though they have different $L_r$ and very different sizes (c.f. Figure~\ref{fig:split}).  We discuss these objects in Section~\ref{sec:B10}.

\subsection{Stellar populations of slow vs fast rotators}\label{ssp:rot}
We turn now to the gradients. The first two columns in Figure~\ref{tzR23} show the same age, [M/H], [$\alpha$/Fe] and $M_*/L_r$ values as Figure~\ref{option3}, but now as a function of $R$ rather than $\sigma(R)$.
%(We do not show a similar analysis of Figure~\ref{option2} because we believe the required $\Delta_{\rm [X/Fe]}$ gradients are unrealistic.  However, comparison of Figures~\ref{option2} and~\ref{option3} shows that allowing IMF variations slightly reduces age gradients, makes little difference to [M/H] and [$\alpha$/Fe], and increases $M_*/L_r$ gradients.)

The third column of Figure~\ref{tzR23} shows a similar analysis of the slow and fast rotators (solid and dashed lines) in all bins where we have sufficient statistics.  Whereas we do have sufficient statistics for the B21 SRs (see Section~\ref{fastSlow}), we do not show them in the third column because their populations are almost indistinguishable from the B21-L and B21-S bins shown in the middle panels (this better highlights the difference between FRs and SRs for the other bins).

Before we discuss the FRs and SRs in the right hand column, notice that plotting versus $R$ makes it obvious that:
\begin{itemize}
\item[-] In all bins, the larger galaxies are older, more metal poor and more $\alpha$-enhanced (middle panels);
 \item[-] In all bins, age and metallicity increase towards the central regions, with metallicity gradients generally being stronger; age gradients are strongest in B00 and weak otherwise;  
 \item[-] [$\alpha$/Fe] gradients are weak, with the centers being less enhanced;
\item[-] $M_*/L_r$ gradients are stronger in bins B11 and B21, but weak or non-existent in the other two bins (bottom left), in agreement with expectations from a very different analysis \citep{Bernardi2018b}; when subdivided by size (bottom middle) gradients are present in all four of the smaller bins and in bins B11-L and B21-L, but are non-existent in bins B00-L and B10-L;   
\item[-] B10 galaxies are the oldest and most metal poor (left hand columns), with B10-L galaxies being the oldest but not quite the most metal poor (middle columns); 
 \item[-] The most metal poor objects are in B00-L, and the central regions of these galaxies are remarkably old (middle panels).
\end{itemize}
In the right hand column, many of these trends are even more dramatic:
\begin{itemize}
 \item[-] SRs (solid) are significantly older (as much as $\sim 5$~Gyrs), less metal rich and more $\alpha$-enhanced than FRs in all our bins except for the B10-L FRs (dark green dashed line) which behaves more similarly to the SRs;
 \item[-] The central regions of SRs have almost the same age and metallicity in all our bins except for B21 (compare red and pink lines in middle panels with solid lines in the right hand panels); the B21 SRs are the SRs with the youngest ages and largest metallicities even though they have the largest $L_r$ and $\sigma_0$;
 \item[-] Except for bin B00, the other SR bins show rather flat age gradients but stronger [M/H] gradients;
 \item[-] FRs show slighlty stronger [$\alpha$/Fe] gradients than SRs, with the centers being less enhanced;  
 \item[-] FRs show slightly stronger $M_*/L_r$ gradients than SRs. Only SRs with large $L_r$ and $\sigma_0$ (i.e. B21-S in middle panel) show comparable $M_*/L_r$ gradients to FRs. 
 \item[-] The stellar populations in all the B10 bins are remarkably similar, with only the B10-S FRs having smaller ages and higher metallicities;  as a result, compared to the other FRs, those in bin B10-L are anomalously old, metal poor and $\alpha$-enhanced.
\end{itemize} 
As the SRs in a bin tend to be larger than FRs (c.f. Figure~\ref{fig:lambdaMR}), it appears that many of the differences between large and small sizes in the middle panels of Figure~\ref{tzR23} are more strongly correlated with differences in rotation.  In particular, SRs formed their stars rapidly, at approximately the same time, long ago, and so have sub-solar metallicities (except for bin B21).  In contrast, FRs are significantly (as much as $\sim 5$~Gyrs) younger, formed their stars over a longer timescale, and so are more metal rich (although recall that these are light-weighted conclusions that are dominated by the younger stars).
In this respect, our FR ages are rather different from those reported by the ATLAS$^{\rm 3D}$ collaboration.  Figure~11 of \cite{McDermid2015} shows that they estimate the vast majority of FRs with $\sigma\ge 220$~kms$^{-1}$ to be more than 10~Gyrs old.  Few of our FRs are this old.  We return to this in Section~\ref{sec:ageSize}.

\begin{figure}
  \centering
  \includegraphics[width=1\linewidth]{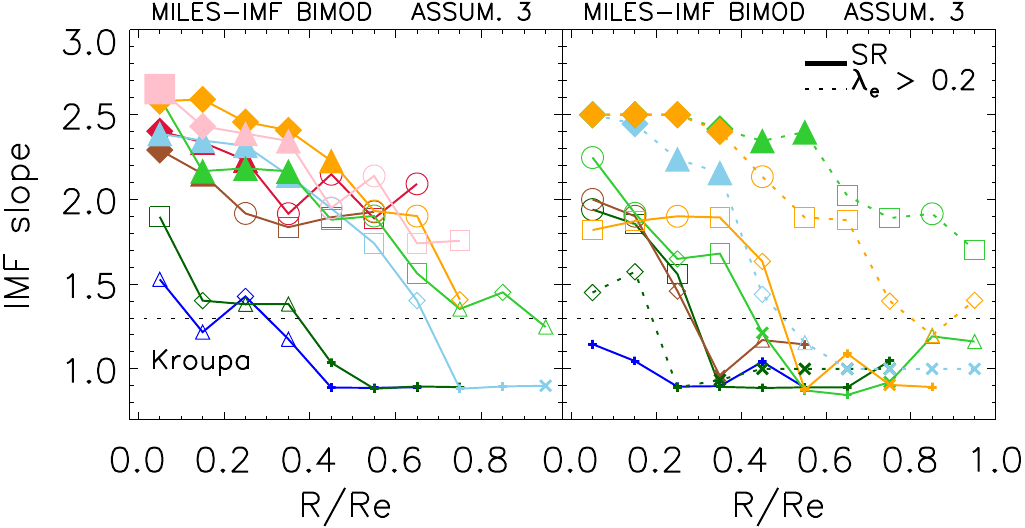}
  \caption{Inferred variation in IMF slope for samples split by size (left) and rotation (right). Symbols (same as top left panel of Figure~\ref{tzR23}) show the closest IMF model, but the IMF slope shown is got by interpolating between the two closest IMF models.  Objects with smaller than average sizes for their $L_r$ and $\sigma_0$ (lighter shades) tend to have steeper (more bottom heavy) IMFs; bins B00-L and B10-L (dark blue and green) have the smallest IMF slopes.  With the exception of the B10-L bin (dark green) FRs (dashed) tend to have steeper (more bottom heavy) IMFs than SRs.}
  \label{fig:IMF3}
\end{figure}

We remarked that it is curious that the B21 SRs are younger and more metal rich than the other SRs.  In view of our comments about SSP systematics in Section~\ref{ssp:sys}, it should come as no surprise that this bin has a different IMF than other SRs.  Figure~\ref{fig:IMF3} shows the gradient in the IMF slope parameter associated with the SSP parameters shown in the middle and right hand panels of Figure~\ref{tzR23}. (The symbols shows the closest-fitting IMF model, but the actual slope shown is got from interpolating between the two best-fitting IMF models.)   The slope is larger -- indicating a more bottom-heavy IMF -- in the central regions of all the bins.  In addition, the right panel shows that FRs (dashed lines) tend to have steeper slopes than SRs (solid).  However, the IMFs for B21 (red and pink in the left panel), which are dominated by SRs, are more like those of FRs.  In contrast, the objects in bins B00-L and B10-L -- the two oldest bins with the lowest metallicities in the middle panels of Figure~\ref{tzR23} -- have the shallowest IMFs. These results suggest that metallicity plays a major role in setting the IMF (also see Figure~17 in Paper~I). 

\subsection{Stellar and dynamical masses of slow and fast rotators}\label{sec:Mdyn}
With $M_*/L_r$ gradients in hand, we can attempt a self-consistent comparison of stellar and dynamical masses.  This is interesting because, when these gradients are ignored, then stellar population estimates of $M_*$ which assume a Kroupa or Chabrier IMF tend to lie about 0.2~dex below $M_{\rm dyn}$ estimated from a Jeans equation analysis.  Some have attributed this discrepancy to the IMF, rather than to problems with the $M_{\rm dyn}$ estimate (see Paper~I for a more detailed discussion).

\begin{figure}
 \centering
  \includegraphics[width=1\linewidth]{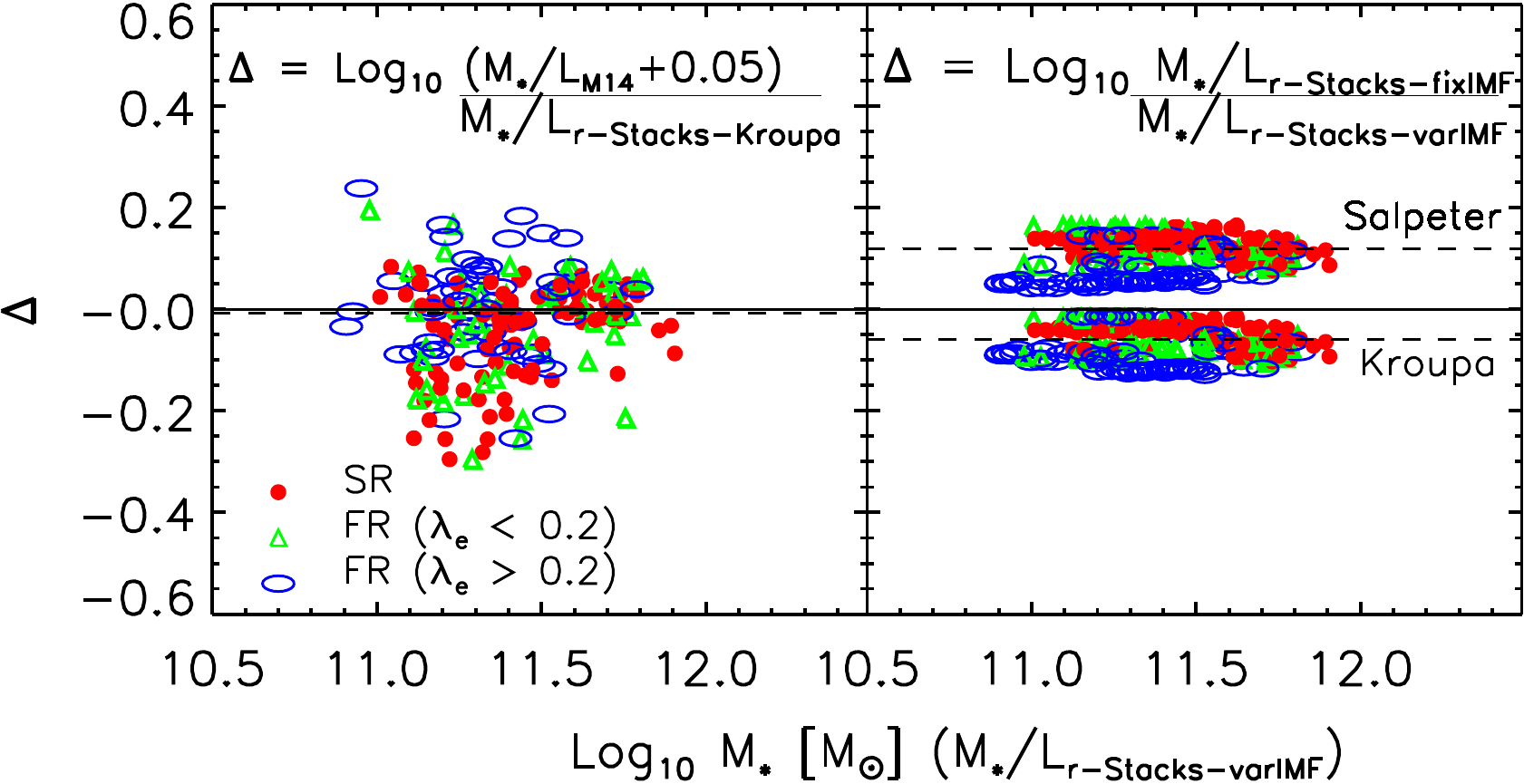}
  \caption{Comparison of integrated $M_*/L_r$ estimates. Red circles and blue ellipses show SRs and FRs with $\lambda_e>0.2$; green triangles show the other FRs; dashed lines show the median trends.  Left: Ratio of integrated $M_*/L_r$ from Mendel et al. (2014) (shifted by 0.05~dex to transform from a Chabrier to a Kroupa IMF) to our own estimate in which we assume the IMF is Kroupa on all scales for all objects. This ratio is shown as a function of the mass estimate which combines each galaxies $L_r$ with the $M_*/L_r$ profile that is appropriate for its $L, \sigma_0, R_e$ and $\lambda_e$ (shown in Figure~\ref{tzR23}).  Right: Ratio of integrated $M_*/L_r$ estimates when we assume all galaxies have the same IMF (Kroupa or Salpeter) on all scales, to that when the IMF is allowed to vary within a galaxy and across the population.}
  \label{fig:DMs}
\end{figure}

Our first step is to compare our integrated $M_*/L_r$ estimates (which are computed by dividing the total $M_*$ by the total luminosity $L_r$, see equation~3 in Paper~I and related discussion) with those from the literature.  (Comparing integrated $M_*/L_r$ rather than $M_*$ estimates themselves removes systematics associated with the total luminosity $L_r$, see \citealt[][]{Bernardi2013,Bernardi2017a,Bernardi2017b,Fischer2017}.)  The left hand panel of Figure~\ref{fig:DMs} shows the ratio of the integrated $M_*/L_r$ estimate from \cite{Mendel2014} (shifted by 0.05~dex to transform from their Chabrier IMF to a Kroupa IMF) to that returned by our analysis if we assume the IMF is Kroupa for all galaxies, and there are no IMF gradients (there may still be $M_*/L_r$ gradients).  We show this ratio as a function of our $M_*$ estimate in which we include the full $M_*/L_r$ gradient (i.e. when IMF variations are allowed).
%In practice, for each galaxy we multiply its surface brightness profile by the $M_*/L_r$ profile appropriate for its bin (in $L, \sigma_0, R_e$ and rotation) and integrate.  Since we only have $M_*/L_r$ profiles out to about $0.8\,R/R_e$, we use the $M_*/L_r$ for a Kroupa IMF to extrapolate to larger $R$.
Red circles show SRs, blue ellipses show FRs with $\lambda_e>0.2$ and green triangles show the remaining objects (i.e. FRs with small $\lambda_e$).  The agreement is rather good, thus establishing consistency with the literature.

\begin{figure*}
 \centering
  \includegraphics[width=0.9\linewidth]{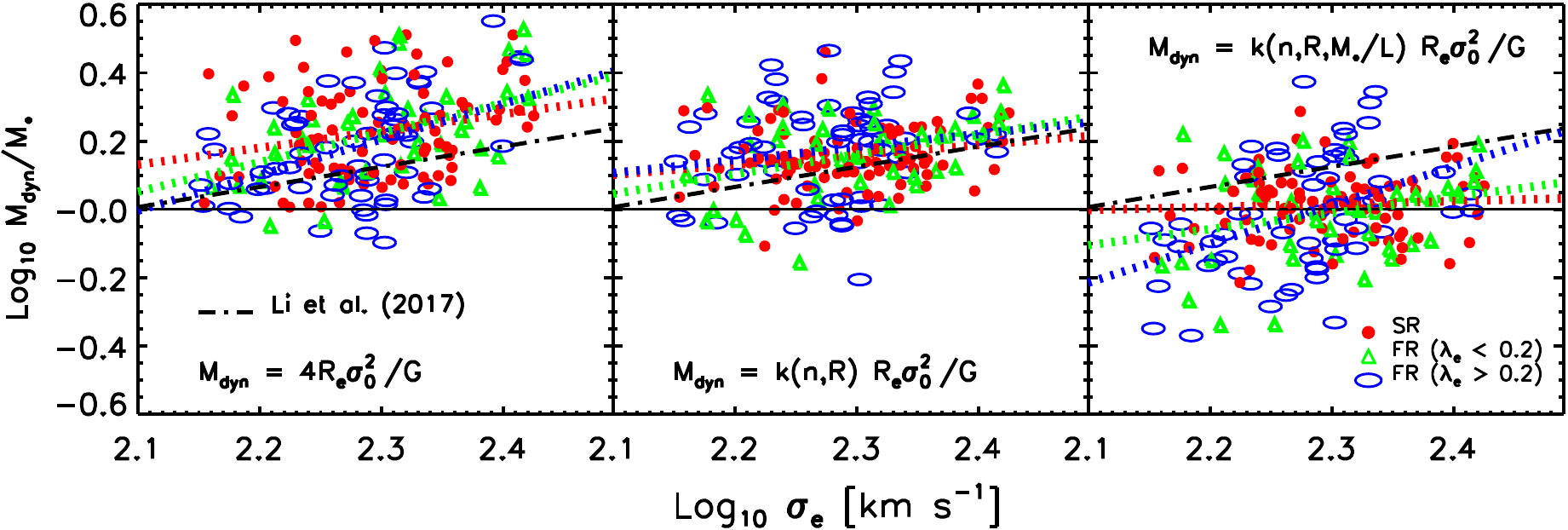}
  \caption{Comparison of three dynamical mass estimates with our variable IMF stellar mass estimate, in which IMF gradients contribute to $M_*/L_r$ gradients, shown versus $\sigma_e$.  All three panels have $M_{\rm dyn}\propto R_e\sigma_0^2/G$ with the proportionality factor: (left) constant, the same for all galaxies; (middle) dependent on the light profile shape, but ignoring $M_*/L_r$ gradients; (right) dependent on the product of the light profile and the same $M_*/L_r$ profile which was used to estimate $M_*$.  Red circles and blue ellipses show SRs and FRs with $\lambda_e>0.2$; green triangles show the other FRs; dotted colored lines show how the ratio $M_{\rm dyn}/M_*$  scales with $\sigma_e$ for each subsample; dot-dashed line shows the correlation reported by Li et al. (2017), offset slightly to account for the fact that the Salpeter IMF (their fiducial choice) has $\Delta \sim 0.13$ in the right hand panel of our Figure~\ref{fig:DMs}. In the right hand panel the offset is gone:  self-consistently accounting for $M_*/L_r$ gradients brings $M_{\rm dyn}$ into good agreement with $M_*$. In addition, the red dotted line (SRs) is significantly flatter than the blue dotted line (FRs), a consequence of the fact that our Jeans equation analysis assumes no rotation, and this is a worse approximation when $\lambda_e$ is large.}
  \label{fig:DMdMs}
\end{figure*}

The right hand panel shows the ratio of the integrated $M_*/L_r$ estimates when we fix the IMF to Kroupa or Salpeter to that when we allow the IMF to vary (both within a galaxy and between bins).  This shows that fixing the IMF to Kroupa underestimates $M_*$ by about 0.05~dex, whereas fixing it to Salpeter overestimates it by about 0.15~dex.  In addition, the fact that the red points lie slightly above the blue mainly reflects the fact that SRs have slightly larger $M_*/L$ at large $R$ (bottom right panel of Figure~\ref{tzR23}).  Note that even when the IMF in the central regions is quite different (Figure~\ref{fig:IMF3}), the net effect on the integrated $M_*/L_r$ is small.  

We now compare our (variable IMF) $M_*$ values with three different estimates of the dynamical mass, and present the results as $M_{\rm dyn}/M_*$ vs $\sigma_e$, where $\sigma_e$ is the velocity dispersion within $R_e$ (as described in Paper~I, here we show galaxies with {\tt FLAG$\_$FIT=1}, i.e. $\sim 64 \%$ of our E sample).  The left hand panel of Figure~\ref{fig:DMdMs} sets $M_{\rm dyn}=4R_e\sigma_0^2/G \approx 5R_e\sigma_e^2/G$ \cite[e.g.][]{McDermid2015}. On average, $M_{\rm dyn}$ lies about 0.2~dex above our $M_*$ estimates although the ratio $M_{\rm dyn}/M_*$ tends to increase with $\sigma_e$.  The dotted lines show linear fits to this correlation and the dot-dashed lines show the correlation reported by \cite{Li17} (here we have shifted their relation by 0.13~dex to account for the fact that the Salpeter IMF, their fiducial choice, has $\Delta \sim 0.13$ in the right hand panel of our Figure~\ref{fig:DMs}).  This correlation has been used to argue that the IMF of massive galaxies becomes more bottom-heavy (Salpeter-like) at large $\sigma_e$ \cite[e.g.][]{Cappellari2013b, Li17}.
%discrepancy is similar to that associated with assuming the IMF is Salpeter (Figure~\ref{fig:DMs}), and so has been used to argue that the IMF of massive galaxies is, in fact, Salpeter-like \cite[e.g.][]{Cappellari2013b, Li17}.
%\sout{However, in our case, this conclusion is obviously not appropriate:  the IMF is only bottom heavy in the central regions.}  
However, our analysis shows that the IMF is only bottom heavy in the central regions.

The middle panel shows $k(n,R)\,R_e\sigma_0^2/G$, where $k$ is taken from Table~1 of \cite{Bernardi2018a}.  This estimate accounts for the fact that the light profile shape (parameterized by the Sersic index $n$) differs from one galaxy to another, but assumes that $M_*/L$ is a constant that is fixed by asking that the Jeans equation estimate of the projected velocity dispersion within $R_e/10$ match the observed $\sigma_0$.  In this case the $M_{\rm dyn}/M_*$-$\sigma_e$ correlation is in even better agreement with \cite{Li17}:  if anything, the correlation is slightly tighter.   

The right hand panel of Figure~\ref{fig:DMdMs} shows the result of including the $M_*/L$ profile in our Jeans equation analysis (see Paper~I for details).  There are three remarkable differences with respect to the other two panels.  First, the mean offset is gone:  accounting for $M_*/L_r$ gradients reduces $M_{\rm dyn}$ by about 0.2~dex, bringing it into good agreement with $M_*$ estimated self-consistently using the same gradients.  Second, the slope of the $M_{\rm dyn}/M_*$-$\sigma_e$ relation is significantly flatter for SRs (red dotted) and steeper for the fastest FRs (blue dotted).  Third, the scatter is smallest for the SRs (red circles) and largest for the fastest FRs (blue ellipses).  These last two points may be a consequence of the fact that our Jeans equation analysis assumes no rotation, and this is a worse approximation when $\lambda_e$ is large. In addition, it may be that single SSPs are worse approximations to the star formation histories of FRs than they are for SRs.

To summarize: Self-consistently accounting for gradients when estimating $M_*$ and $M_{\rm dyn}$ yields good agreement between the two, confirming what we found in Paper~I.  In particular, this agreement comes because gradients reduce the $M_{\rm dyn}$ estimate (more than they increase $M_*$).  As we noted in Paper~I, this agreement implies that we can now specify the stellar mass scale identified by \cite{Bernardi2011}, $2\times 10^{11}M_\odot$ if the IMF were Chabrier, without also specifying an IMF.  The offset of $\sim 0.05$~dex from Chabrier to Kroupa combined with the $\sim 0.05$~dex offset between Kroupa and our variable IMF estimate (right hand panel of Figure~\ref{fig:DMs}), suggest that this scale is more like $3\times 10^{11}M_\odot$.

\subsection{The galaxies in bin B10}\label{sec:B10}
For completeness, we now discuss the objects in bin B10, which were identified in Paper I as being anomalously old (top left panel of Figure~\ref{tzR23}).  The top middle panel of Figure~\ref{tzR23} suggests that, in fact, it is the larger B10-L objects which are extreme -- the smaller B10-S objects are not particularly unusual (although they are unusually old compared to the smaller objects in the other bins).  Morever, the top right panel of Figure~\ref{tzR23} shows that it is rotation, not size, which seems to matter.  This is because the B10 SRs are {\em not} particularly different from the SRs in the other bins.  In this respect, the question is no longer:  `Why are the B10 Es so old?'  but
%`Why are the central regions of SRs so old?'  A separate question is:
`Why are the B10-L FRs so old?' and `Why are all the sub-bins of B10 older?' 
While we do not have good answers to these questions, we can at least address the question of why there has been no previous discussion of anomalously old galaxies with with $\sigma_0 \le 250$~km~s$^{-1}$ and $-22.5 \ge M_r \ge -23.5$ (stellar masses of $1\times 10^{11} \le M_*/M_\odot \le 3\times 10^{11}$).

One of the key differences between previous work with `early-types' and ours with Es is that we have removed S0s.  While there were essentially no S0s in bins B11 and B21 (Figure~\ref{fig:VL}), they were present in, and removed from, our lower $L_r$ bins (Figures~\ref{fig:split} and~\ref{fig:lambda}).  We have already shown some differences between S0s and Es (e.g. Figures~\ref{fig:lambdanR} and \ref{fig:lambdaBT}). In addition, Figures~\ref{fig:lambdasplit} and~\ref{fig:lambdasplit2} show that, compared to Es in bin B10, the S0s have stronger H$_\beta$ and <Fe> at a  give [MgFe].  Single stellar population model fits to S0s would return younger ages, higher metallicities, and smaller $\alpha$-enhancements than for Es.
%However, TiO2$_{\rm SDSS}$ is not very different (right).
Hence, adding the S0s to our sample of Es would reduce the anomalously large ages for bin B10 and, to some extent, the SRs in bin B00-L as well.  As there are very few S0s in bins B11 and B21, the question of what happens if they are included is moot.  We conclude that it would have been more difficult to notice that bin B10 is anomalous if we had not removed its S0s.

\begin{figure*}
 \centering
  \includegraphics[width=0.9\linewidth]{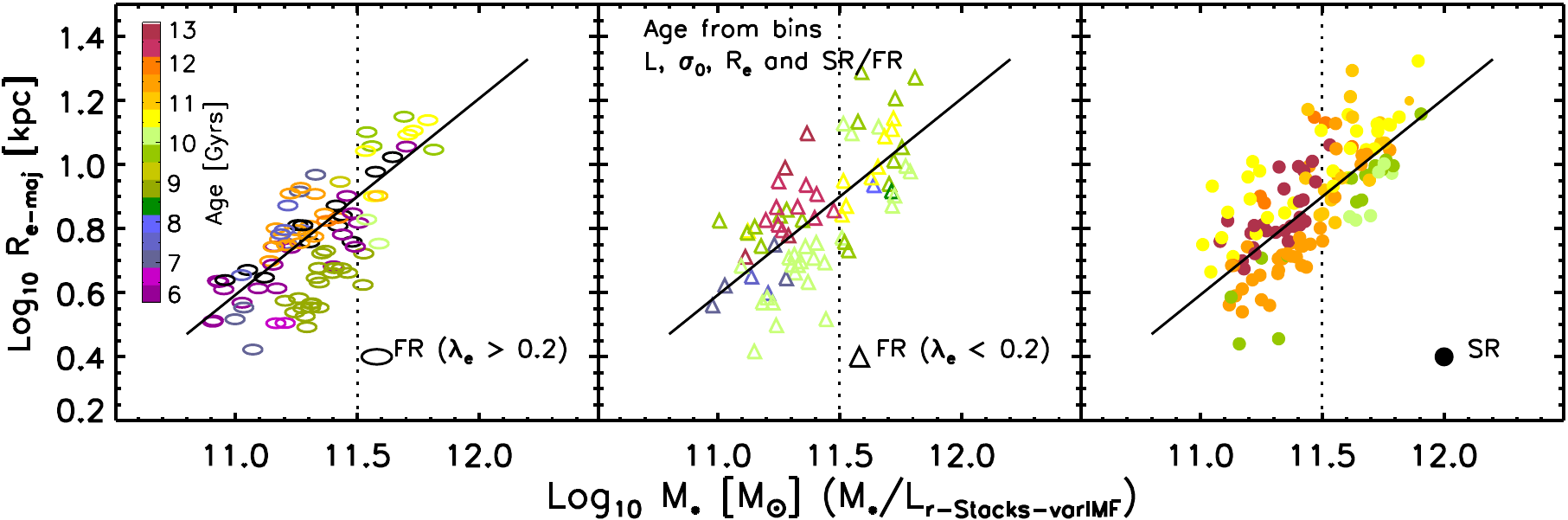}
  \caption{Size-stellar mass correlation for FRs with $\lambda_e > 0.2$ (left), FRs with $\lambda_e < 0.2$ (middle) and SRs (right), with symbol color indicating light-weighted age.  FRs are younger than SRs of the same mass. Solid line shows the best linear fit of the size-mass relation to the full sample. Dotted vertical line shows the mass scale identified by \citealt{Bernardi2011} (after accounting for the fact that the IMF is variable).  Above this mass, the number of objects below the solid line drops, indicating that the slope of the $R_e$-$M_*$ correlation steepens. In addition, above this mass the SR/FR ratio increases.}
  \label{fig:RMsFR}
\end{figure*}

\begin{figure*}
 \centering
  \includegraphics[width=0.9\linewidth]{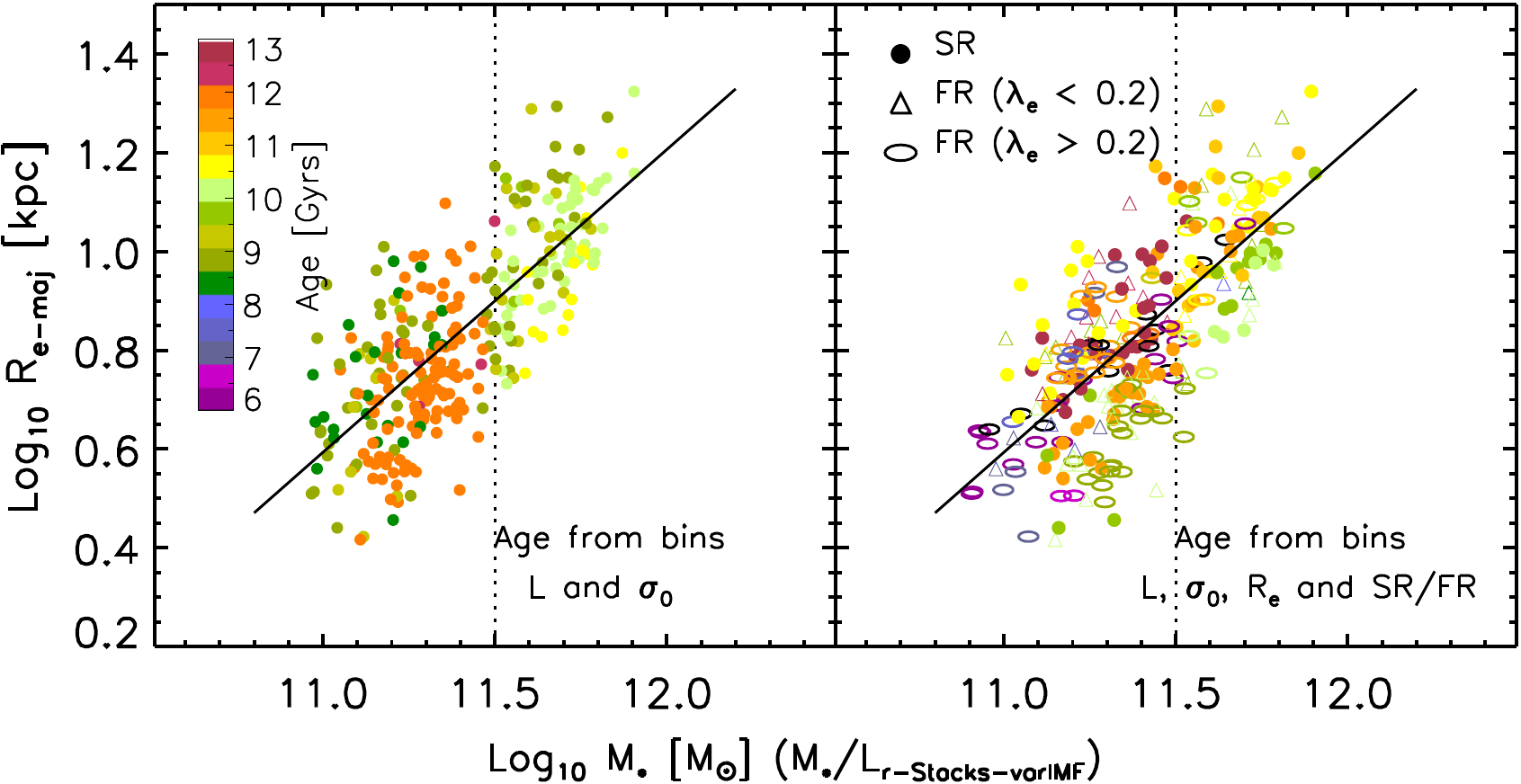}
  \caption{Correlation between size and stellar mass with symbols colored by light-weighted age, where age in left and right panels is estimated from the profiles shown in the top left and top right panels of Figure~\ref{tzR23}, respectively.  (Note that right panel simply shows all the symbols from all the panels of Figure~\ref{fig:RMsFR}.)  Left panel suggests that more compact galaxies are older; right panel suggests almost the opposite.}
  \label{fig:RMs}
\end{figure*}

\subsection{Ages, sizes and the SR/FR dichotomy}\label{sec:ageSize}
The SR/FR dichotomy has important implications for galaxy scaling relations.  To illustrate, Figure~\ref{fig:RMsFR} shows the correlation between size and stellar mass, with colors indicating age.  (The age of each galaxy is obtained by weighting the age profile shown in the top right panel of Figure~\ref{tzR23} that is appropriate for its $L_r$ and $\sigma_0$ with its surface brightness profile.)  The three panels show FRs with $\lambda_e > 0.2$ (left), FRs with $\lambda_e < 0.2$ (middle) and SRs (right).  Comparison of the left and right panels shows clearly that FRs are younger than SRs of the same mass, as might have been expected from the dichotomy in the top right panel of Figure~\ref{tzR23}.

The vertical dotted line in Figure~\ref{fig:RMsFR} shows the mass scale identified by \cite{Bernardi2011}, which we noted is $3\times 10^{11}M_\odot$ after accounting for the fact that the IMF is variable.  Above this mass, most SRs and FRs lie above the solid line shown, indicating that the $R_e$-$M_*$ relation steepens.  In addition, there are few FRs above this mass.  That these mass scales coincide is consistent with ATLAS$^{\rm 3D}$ \citep{Cappellari2013b}, although the value of this mass scale differs from what they report, since our $M_*$ estimates are different (e.g. Figure~\ref{fig:DMdMs}).

It is interesting to contrast this with Figure~\ref{fig:RMs}, in the right hand panel of which we have simply stacked together all three bottom panels of Figure~\ref{fig:RMsFR}.  While this shows that a wide variety of ages contributes to each $R_e$ and $M_*$ bin, it appears that if one averages the SRs and FRs together, then one would still find that the smaller objects are younger (mainly because smaller galaxies tend to be FRs).  Now consider the left hand panel.  In this case, we have colored objects by the ages estimated from the four $(L,\sigma_0)$ stacks without subdividing by size or rotation.  (I.e., in this case the age estimates use the profiles in the top left rather than top right panel of Figure~\ref{tzR23}.)  This shows that, at fixed $M_*$, smaller galaxies are older, not younger.  Evidently, how one does the averaging matters.  The age from the average stack (left) is not the same as averaging the ages from many stacks (right).  This difference should be borne in mind in future analyses.  

Before ending this subsection, it is worth noting that smaller galaxies being older is consistent with one of the major results of Paper XXX of the ATLAS$^{\rm 3D}$ collaboration \cite[e.g. Figure~6 of][but note that our sample only probes the largest masses and velocity dispersions]{McDermid2015}.  While the left hand panel of our Figure~\ref{fig:RMs} appears to be consistent with this, their analysis, based on averaging the ages of individual galaxies, should be closer to the trend observed in the right hand panel, which indicates that compact galaxies are younger, not older.  For us, especially at smaller $M_*$, this is driven by the fact that compact galaxies tend to be FRs, and FRs are younger.  As we noted earlier, \cite{McDermid2015} report much older FRs.  This difference in the ages of FRs is the main reason our conclusion about the size-mass-age correlation differs from theirs.  We provide more discussion of how our results compare with those of ATLAS$^{\rm 3D}$ in Appendix~\ref{sec:compare}.

\section{Does environment matter?}\label{Sec:env}
We now check if the SR/FR dichotomy is correlated with environment.   

\subsection{Environment: Large-scale density and tidal fields}
\begin{figure}
  \centering
  \includegraphics[width=0.99\linewidth]{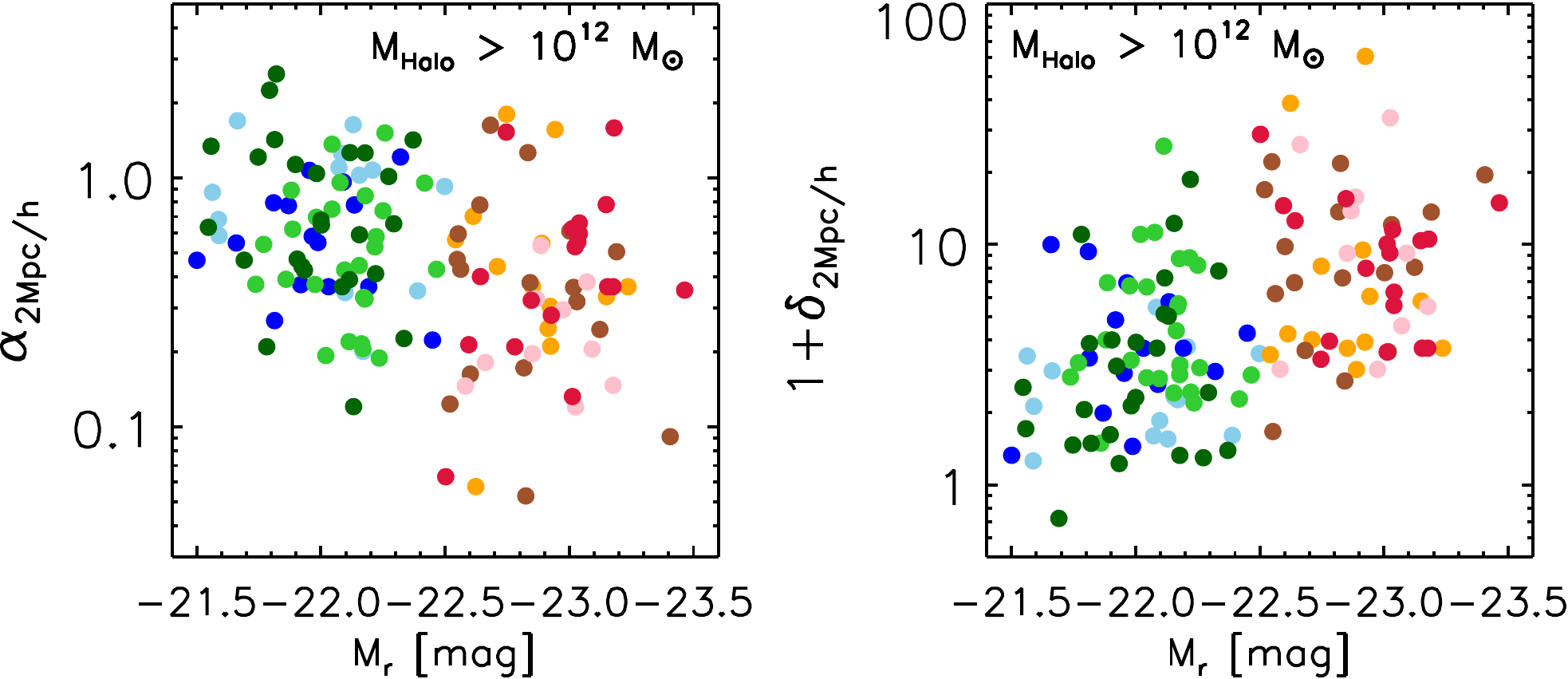}
   \vspace{-0.5cm}
  \caption{Fainter galaxies in our sample are in lower density regions (right) with stronger tidal fields (left).} \label{alphad}
\end{figure}

In this subsection, we use two measures of the environment, both provided by \cite{Paranjape2018}:  one is a measure of the density smoothed over a few Mpc, and the other is a measure of the tidal field strength smoothed over the same scale. (These measures are only available for approximately half of our sample.)  Figure~\ref{alphad} shows that, as expected, the lower $L_r$ bins are in less dense environments where the tidal field is slightly stronger.  However, we find no striking difference between the environments of the two low-$L_r$ bins (B00 and B10).   

\subsection{Environment:  Central vs satellite}
\begin{figure}
  \centering
  \includegraphics[width=0.7\linewidth]{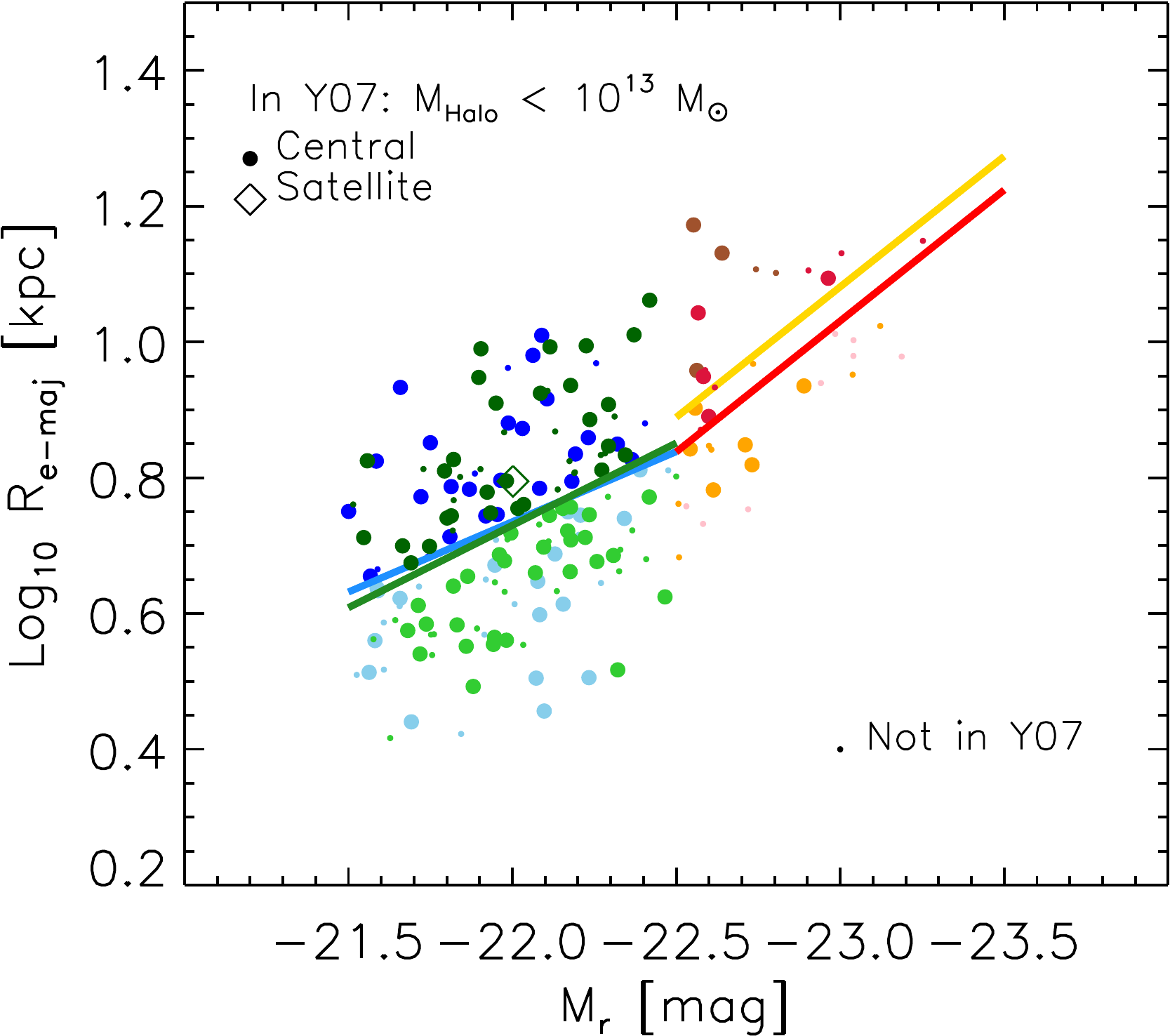}
  \includegraphics[width=0.7\linewidth]{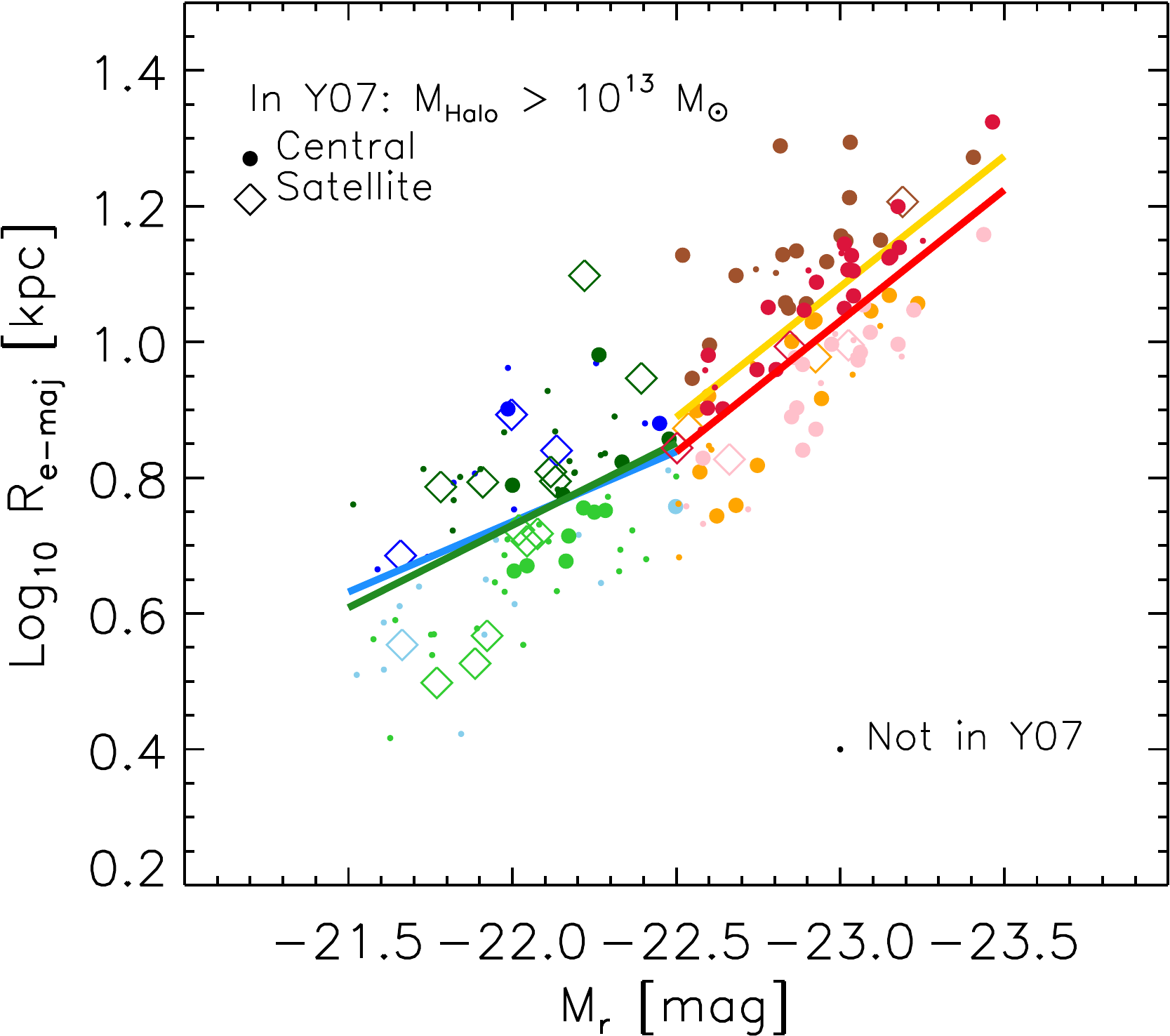}
   %\vspace{-0.5cm}
  \caption{Size-luminosity relation of the MaNGA E galaxies in groups identified by Yang+ as being less (top) or more (bottom) massive than $10^{13}M_\odot$. Filled circles and open diamonds show MaNGA Es identified as central and satellite galaxies in the groups; small dots show MaNGA Es that are not in the group catalog.} \label{fig:yangAll}
\end{figure}

\begin{figure}
  \centering
  \includegraphics[width=0.99\linewidth]{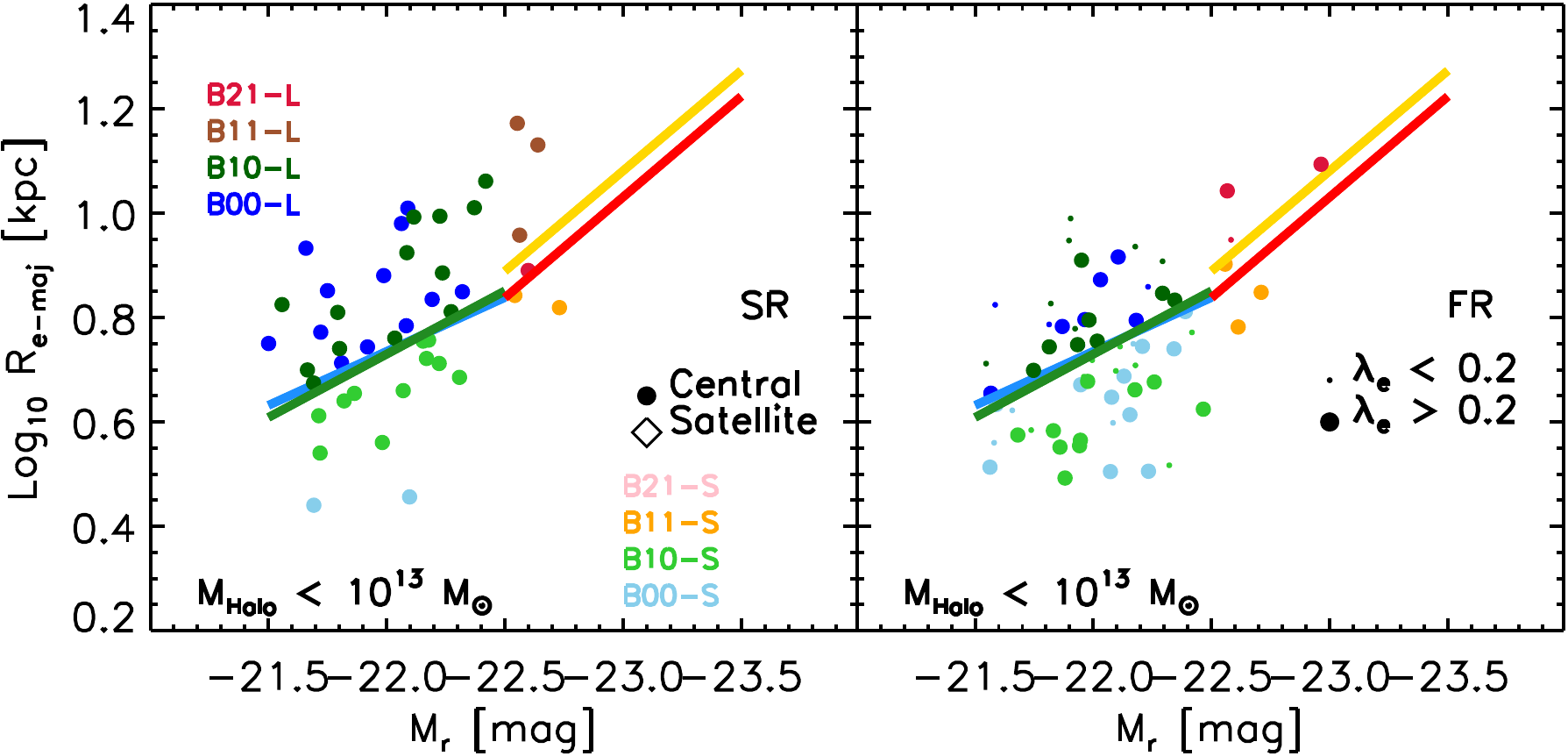}
  \includegraphics[width=0.99\linewidth]{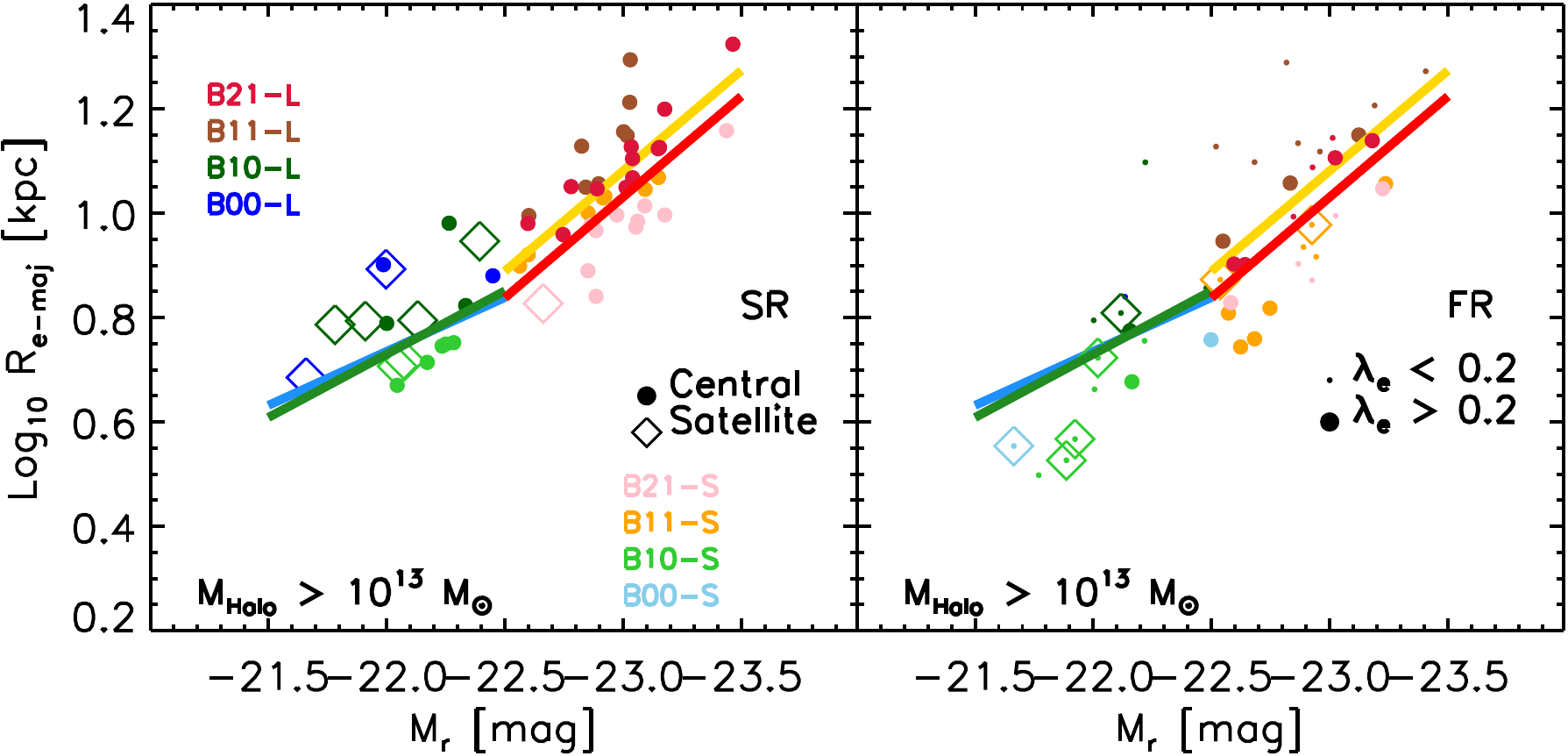}
   \vspace{-0.5cm}
  \caption{Same as previous figure, but now subdivided into SRs (left) and FRs (right). } \label{fig:yang}
\end{figure}

Figures~\ref{fig:yangAll} and~\ref{fig:yang} show the result of a different test.  Here, we have matched our galaxies to the group catalog of \citet[hereafter Yang+]{Yang2007}, and classified galaxies by whether they are central galaxies in their group or not.  Thus, here, the `environment' is defined on a smaller scale than before.

In the groups which Yang+ estimate to have halo masses in excess of $10^{13}M_\odot$, there are 1550 central Es with $-22.5\le M_r\le -23.5$ and only 450 satellite Es (morphological classification for the SDSS galaxies in the Yang+ sample comes from \citealt{DS2018}).  In this luminosity range, most Es are centrals; one in three or four is a satellite.  
However, for $-21.5\le M_r\le -22.5$, there are 500 central Es but 1750 satellites: there are more than three times as many satellite Es as centrals.  Evidently, at $M_r \approx -22.5$ ($M_* \sim 3\times 10^{11} M_\odot$), there is a significant change in the E population:  brighter than this, most Es are centrals; fainter than this, they tend to be satellites.  Mergers are expected to have played a role in the assembly and star formation histories of central galaxies in clusters, so it is probably no coincidence that this is the luminosity scale at which early-type galaxy scaling relations change \citep{Bernardi2011}.  Since bins B10 and B11 are separated by this critical luminosity ($M_r \sim -22.5$) but have same $\sigma_0$, it is interesting to check how their group environments compare.  

Before we do so, there is an interesting fact about MaNGA Es which has not been highlighted before:  
In haloes more massive than $10^{13}M_\odot$, MaNGA has 15 centrals and 16 satellites for $M_r$ between $-21.5$ and $-22.5$ and 64 centrals and 7 satellites for $M_r$ between $-22.5$ and $-23.5$ (205 centrals and 14 satellites if we do not limit the sample to $z < 0.08$).  Thus, satellite Es are very under-represented in MaNGA compared to in the full Yang catalog.  Therefore, it is a good approximation to think of MaNGA Es as being dominated by centrals in all bins.

Figure~\ref{fig:yangAll} shows that low luminosity Es (B00 and B10) tend to be centrals of low mass halos (top panel), whereas high luminosity Es (B11 and B21) are centrals of more massive halos (bottom panel).  B00 and B10 objects in massive halos tend to be satellites. These are not surprising trends, since group mass and central galaxy luminosity are known to be well-correlated.  (A number of MaNGA galaxies are not in the group catalog:  we have shown them as small dots in both panels.)

The process of becoming a central galaxy is generally believed to involve mergers \cite[e.g.][]{Mo2010}, though whether these are major or minor and if they are between similar morphological types or not is still debated \cite[e.g.][]{Bundy2009,Hilz2013,Behroozi2015,Man2016}. Presumably this has affected the Es in bins B11 and B21. In contrast, the B00 and B10 Es are primarily centrals of lower mass halos. These halos are more isolated, suggesting they have not undergone significant recent merger activity.  With this in mind, it is interesting to ask if there is any correlation between SR or FRs and halo mass.  Figure~\ref{fig:yang} shows the same $R_e$-$L_r$ correlation as before, but now for SRs (left) and FRs (right) in low (top) and high (bottom) mass groups.  Unfortunately, the numbers are too small to allow clear conclusions. For example, while the ratio of FRs to SRs in each bin is slightly reduced in the more massive groups, the sample is too small to draw firm conclusions, especially given that MaNGA tends to pick centrals, and not all MaNGA objects are present in the group catalog.

Nevertheless, our results are consistent with the statements that both the SRs and the FRs in B10 are older and less metal rich than objects in B11 because they formed in lower mass halos.  In addition, our results showing that the FRs in bin B11 appear younger and more metal rich than those in bin B10 support a senario in which these FRs may be the result from a relatively recent major merger between a fast rotator (S or S0) and an E.

\section{Discussion}
Appendix~\ref{sec:compare} discusses how our findings compare with previous work.  This section provides a brief summary. We describe global quantities before describing gradients.

We find that at fixed $L_r$ and $\sigma_0$, the objects with larger $R_e$, in addition to being older, are less metal rich and more [$\alpha$/Fe] enhanced.  Although \cite{Graves2010} did not present their findings in this way -- their stacks contain a mix of objects from different bins in our scheme -- we show in Appendix~\ref{sec:compare} that our results are qualitatively consistent, even though they did not account for IMF gradients whereas we do.  However, there are important differences.  For example, their binning scheme hides the fact that objects in our bin B10-L are extremely old.  In addition, their analysis extends to much smaller $\sigma_0$ and $L_r$, in part because they do not distinguish between Es and S0s.  Indeed, they did not consider the FR/SR dichotomy at all, whereas our results indicate this separation is useful.

While the FR/SR dichotomy was addressed by the ATLAS$^{\rm 3D}$ collaboration (\citealt{McDermid2015}), our results differ from theirs in a number of respects.  Our analysis is confined to a much smaller range of $\sigma_0$ then theirs.  However, at the highest $\sigma_0$ where we can compare results, they report that the vast majority of FRs are more than 10 Gyrs old.  Few of our FRs are this old.  As a result, while they report that smaller FRs are older, we find the opposite.  Instead, our size-age scaling agrees with the age-rotation anti-correlation for early-types reported by SAMI (\citealt{vdSande2018}).

We find strong metallicity gradients but weaker age and [$\alpha$/Fe] gradients, in qualitative agreement with previous work (see Introduction).  In addition, the IMF gradients we find are qualitatively consistent, though slightly weaker than, the many recent IFU-based studies cited in the Introduction.  However, we find that self-consistently accounting for gradients when estimating $M_*$ and $M_{\rm dyn}$ yields good agreement between the two, especially for SRs, because gradients reduce the $M_{\rm dyn}$ estimate (more than they increase $M_*$).  Thus, our resolution of the $M_*$-$M_{\rm dyn}$ discrepancy is opposite to that which has been advocated in the recent literature (\citealt{C16,Li17}). 

\section{Conclusions} \label{sec:conclusions}
We used the methodology of Paper I to estimate single stellar population parameters using stacked spectra of elliptical galaxies binned in $\sigma_0$ and $L_r$ which were further subdivided based on $R_e$ in each bin (Figure~\ref{fig:split}) as well as on rotation (Figure~\ref{fig:lambda} and Table~\ref{tab:bin}).  Our analysis has uncovered a number of trends.  We first summarize our findings having to do with global quantities before describing gradients.  

\subsection{Global quantities}

\begin{itemize}
  \item Absorption line-strengths show clear differences between $L_r$ and $\sigma_0$ bins, and a clear dependence on size in each bin (Figure~\ref{fig:MilesBIsplit}).  The Es with larger sizes in a bin are older, less metal-rich and more $\alpha$-enhanced (Figure~\ref{option3}).
  \item Many of the trends with size are driven by the fact that Es with above average sizes for their $L_r$ and $\sigma_0$ tend to be slow rotators, whereas smaller than average Es are fast rotators (Figure~\ref{fig:lambdaMR}).  
  \item Absorption line strengths (Figures~\ref{fig:lambdasplit} and~\ref{fig:lambdasplit2}) and SSP inferred stellar populations show a strong dichotomy between fast and slow rotators (Figure~\ref{tzR23}), which drives many of the trends with size:  FRs are younger, more metal rich, and less $\alpha$-enhanced.  This confirms the age-rotation anti-correlation for early-types reported by SAMI \citep{vdSande2018}, shows that rotation matters for the other stellar population parameters as well, and shows that these trends are present even when S0s have been removed from the early-type population.  This agreement is reassuring, as few of our FRs are more than 10~Gyrs old, and this is in contrast to ATLAS$^{\rm 3D}$ \cite[see Figure~11 of][]{McDermid2015}.  
  \item Ignoring the SR/FR dichotomy can lead to puzzling results.  Although more compact Es are younger than their larger counterparts of the same mass (Figure~\ref{fig:RMsFR}), if one estimates ages from stacked spectra which ignore the SR/FR dichotomy, one finds that more compact galaxies are older rather than younger (Figure~\ref{fig:RMs}).  This strongly suggests one should be cognizant of the SR/FR dichotomy when studying stellar populations and assembly histories.
  \item The stellar populations of S0s and fast rotating Es are similar at lower $\sigma_0$, consistent with the suggestion that FR Es could be misclassified S0s \citep{Emsellem2011, C16}.  However, above $\sigma_0\sim 200$~km~s$^{-1}$, S0s tend to have larger rotation $\lambda_e$ (Figure~\ref{fig:lambda}), smaller $n$ (Figure~\ref{fig:lambdanR}), smaller B/T (Figure~\ref{fig:lambdaBT}), younger ages, smaller $\alpha$-enhancements, and larger metallicities (Figures~\ref{fig:lambdasplit} and~\ref{fig:lambdasplit2}) than FR Es.  Therefore, had we combined S0s with Es in our analyses, then this would have made the differences between FRs and SRs even more pronounced.  
  \item The metallicity of the FRs in bin B11 is significantly higher than in other bins (Figures~\ref{fig:lambdasplit}, \ref{fig:lambdasplit2} and~\ref{tzR23}). The high central metallicity and lower [$\alpha$/Fe] of FRs in this bin suggest that they may have formed from a relatively recent major merger involving at least one FR (spiral or S0) -- they can not be FRs whose disks have faded or been stripped away. This is consistent with the fact that the FR population basically disappears at $M_* \ge 3\times 10^{11} M_\odot$. 
 \item The FRs in bin B10-L are unusual: their properties are more similar to those of slow rotators than to S0s (of the same $L_r$ and $\sigma_0$). The majority of these galaxies are centrals of halos less massive than $10^{13} M_\odot$ (Figure~\ref{fig:yang}). This suggests that fast rotators at low $L_r$ (i.e. in bin B10) are systems with larger dissipation than their slow rotator counterparts with either more quiescent merger histories, or histories in which the assembly happened much earlier compared to the FRs at higher masses (i.e. bin B11). 
  \item There are two distinct types of SRs.  Although both are $\alpha$-enhanced, at smaller $\sigma_0$ and/or $L_r$ (bins B00, B10 and B11) SRs are old and metal poor, whereas at large $\sigma_0$ and $L_r$ (bin B21) they are not quite as old and are more metal rich (compare red/pink lines in the middle panels of Figure~\ref{tzR23} with the solid lines in the left panels).  From this it appears that the general belief that age, metallicity and $\alpha$-enhancement increase monotonically with $\sigma$ only applies if one has averaged over SRs and FRs (or over all sizes at a given $\sigma$). 
\end{itemize}

\subsection{Gradients}
Regarding gradients, we found that:
\begin{itemize}
\item Gradients preserve the global correlation of $\sigma$ with age, metallicity and $M_*/L_r$ while we observe a weak local anti-correlation with [$\alpha$/Fe] (Figure~\ref{option3}).  
\item At fixed $L_r$, ages track the local value of velocity dispersion. When we split our sample in objects which are larger/smaller than average for their bin they define different tracks. However, the ages of each subsample (i.e. small or large sizes) of a given $L_r$ still track the local value of velocity dispersion (in panels which are second from top in Figure~\ref{option3} dark/light blue connect to dark/light green symbols, and brown/orange connect to red/pink symbols). 
\item The local metallicity and IMF slope are also strongly correlated (compare top and third from top panels in Figure~\ref{option3}).
\item The anti-correlation between metallicity and age or $\alpha$-enhancement also applies locally. This is more evident for bins B00 and B10 (e.g. in right hand panels of Figure~\ref{option3} compare dark/light blue lines in the second through fourth panels).  
  \item Inside $\sim R_e$ SRs are remarkably homogeneous, being uniformly old, metal poor and $\alpha$-enhanced for bins B00, B10 and B11 (right hand panels of Figure~\ref{tzR23}). SRs in bin B21 are slightly younger and more metal rich (pink and red symbols in Figure~\ref{tzR23}).  In this context, it is worth noting that \cite{Gu2018} have reported that early-type galaxies in clusters show no correlation between [$\alpha$/Fe] and galaxy mass, nor any gradient in [$\alpha$/Fe], whereas [$\alpha$/Fe] decreases from the center towards the outskirts in field galaxies.  They argue that their observations indicate a coordinated assembly of the stellar mass in cluster galaxies.  However, this cannot be the full story because the majority of B00 and B10 objects in our sample do not inhabit massive clusters (Figures~\ref{fig:yangAll}), yet they are remarkably homogeneous.  Moreover, despite being in relatively isolated enviroments, they have [$\alpha$/Fe] increasing towards the outskirts.  (Unfortunately, as we discuss in Paper~I, [$\alpha$/Fe] gradients are model dependent.)  
  \item Age gradients in SRs tend to be stronger in the less massive objects.  Metallicity gradients are generally stronger than age gradients.  While it is often argued that stronger gradients imply quiescent merger histories at least in the recent past, some recent work suggests that minor mergers tend to create positive age gradients, bring in $\alpha$-enhanced stars, and steepen metallicity gradients \citep{Hirschmann2015}.  Our measurements of the B21 SRs do indeed show such gradients. On the other hand the properties of SRs with smaller $\sigma_0$ and/or $L_r$ (bins B00, B10 and B11) can be explained by either a scenario in which both star formation and assembly happened at very early times, or the assembly at later times involved minor mergers with old, low metallicity and highly $\alpha$-enhanced systems. The lower [M/H] of SRs in bins B00, B10 and B11 imply that they could not have been formed from minor mergers of FRs having similar properties to the ones we observe (i.e. they must have been fast rotators with low [M/H] and [$\alpha$/Fe]).        
  \item FRs are slightly less $\alpha$-enhanced in their centers, suggesting more extended star formation histories in the central regions (though we again caution that [$\alpha$/Fe] gradients are model dependent).   [$\alpha$/Fe] increasing slightly towards the outskirts has also been reported by \cite{Boardman2017} in a sample of 12 intermediate mass ETGs (E+S0s with $M_* < 2\times 10^{11} M_\odot$), and by \cite{MN2018}.  Such gradients are in qualitative agreement with the outside-in models of \cite{Pipino2006}, although more recent work suggests that the gradients imply mostly quiet evolutionary histories in which many objects have experienced some kind of gaseous interaction in recent times.
  \item Gradients in $M_*/L_r$ are stronger for FRs. Only SRs with large $L_r$ and $\sigma_0$ (i.e. B21) show comparable $M_*/L_r$ gradients (bottom middle and right panels of Figure~\ref{tzR23}).
  \item  $M_*/L$ gradients matter when estimating the dynamical mass, confirming what we found in Paper~I.  Even when we subdivide each $L_r$ and $\sigma_0$ bin on the basis of half-light radius $R_e$ and rotation, self-consistently accounting for gradients when estimating $M_*$ and $M_{\rm dyn}$ yields good agreement between the two (Figure~\ref{fig:DMdMs}).  This is because accounting for $M_*/L$ gradients reduces $M_{\rm dyn}$ by $\sim 0.2$ dex so it agrees with the stellar population estimate of $M_*$ (rather than the other way around).  Moreover, because only the mass in the central regions may have bottom-heavy IMFs (Figure~\ref{fig:IMF3}), this $M_*$ is more consistent with using a Kroupa IMF than a Salpeter IMF, even for massive galaxies.  
  \item This agreement between $M_*$ and $M_{\rm dyn}$ means we no longer need to specify an IMF when specifying the mass scale at which scaling relations change and the population becomes dominated by SRs:  This scale is $3\times 10^{11}M_\odot$ (e.g. Figure~\ref{fig:RMsFR}).  
\end{itemize}

In his review of the field \cite{C16} states that early-type galaxies form via two main channels: FRs start as star-forming discs and grow their bulges via dissipative processes, followed by quenching, while the more massive SRs form as in the two-phase scenario, with an early rapid dissipative formation followed by repeated dry merger events.  Our results suggest that, within $\sim R_e$, SR Es were indeed formed in gas-rich, rapid star formation events at $z\sim 4$. This will have led to relatively steep radial metallicity variations and positive [$\alpha$/Fe] gradients which were weakened by the subsequent assembly history, in broad agreement with \cite{MN2018}.  

As most of our analysis was confined to scales smaller than the half-light radius, the next natural observational step is to push to larger scales, to see how gradients change (do they flatten out? invert?).  From the theory side we are left with a number of open questions which we hope will stimulate further work. Why are FR Es so much younger than SR Es? Are FR Es just S0s seen face-on? In this case, why are the properties of FRs in bin B10-L more similar to those of slow rotators than to S0s?  Are FR Es with large $L_r$ and $\sigma_0$ (i.e. in bin B11, where metallicity is significantly higher than in other bins) formed from a relatively recent major merger, perhaps involving a fast rotator galaxy (spiral or S0), while SR Es are the result of many dry minor mergers?  Or did SR Es form through major dry mergers which happened at very early times? Could major versus minor dry mergers explain why there are two types of SRs, one metal rich and the other metal poor?

\section*{Acknowledgements}
We are grateful to E. Emsellem for helpful correspondence about whether or not FRs are really S0s, to K. Westfall and T. Parikh for many helpful discussions about the MaNGA data, to C. Maraston, D. Thomas, A. Vazdekis and G. Worthey for correspondence about their SSP models, and to the referee for a helpful report. This work was supported in part by NSF grant AST-1816330.

Funding for the Sloan Digital Sky Survey IV has been provided by the Alfred P. Sloan Foundation, the U.S. Department of Energy Office of Science, and the Participating Institutions. SDSS acknowledges support and resources from the Center for High-Performance Computing at the University of Utah. The SDSS web site is www.sdss.org.

SDSS is managed by the Astrophysical Research Consortium for the Participating Institutions of the SDSS Collaboration including the Brazilian Participation Group, the Carnegie Institution for Science, Carnegie Mellon University, the Chilean Participation Group, the French Participation Group, Harvard-Smithsonian Center for Astrophysics, Instituto de Astrof{\'i}sica de Canarias, The Johns Hopkins University, Kavli Institute for the Physics and Mathematics of the Universe (IPMU) / University of Tokyo, Lawrence Berkeley National Laboratory, Leibniz Institut f{\"u}r Astrophysik Potsdam (AIP), Max-Planck-Institut f{\"u}r Astronomie (MPIA Heidelberg), Max-Planck-Institut f{\"u}r Astrophysik (MPA Garching), Max-Planck-Institut f{\"u}r Extraterrestrische Physik (MPE), National Astronomical Observatories of China, New Mexico State University, New York University, University of Notre Dame, Observat{\'o}rio Nacional / MCTI, The Ohio State University, Pennsylvania State University, Shanghai Astronomical Observatory, United Kingdom Participation Group, Universidad Nacional Aut{\'o}noma de M{\'e}xico, University of Arizona, University of Colorado Boulder, University of Oxford, University of Portsmouth, University of Utah, University of Virginia, University of Washington, University of Wisconsin, Vanderbilt University, and Yale University.

%%%%%%%%%%%%%%%%%%%%%%%%%%%%%%%%%%%%%%%%%%%%%%%%%%



%\bibliographystyle{mnras}

\bibliographystyle{mnras}
\bibliography{biblio} % if your bibtex file is called example.bib

% \begin{thebibliography}{99}
% \bibitem[\protect\citeauthoryear{Author}{2012}]{Author2012}
% Author A.~N., 2013, Journal of Improbable Astronomy, 1, 1
% \bibitem[\protect\citeauthoryear{Others}{2013}]{Others2013}
% Others S., 2012, Journal of Interesting Stuff, 17, 198
% \end{thebibliography}

%%%%%%%%%%%%%%%%%%%%%%%%%%%%%%%%%%%%%%%%%%%%%%%%%%

%%%%%%%%%%%%%%%%% APPENDICES %%%%%%%%%%%%%%%%%%%%%

%%%%%%%%%%%%%%%%%%%%%%%%%%%%%%%%%%%%%%%%%%%%%%%%%%

\appendix

\section{Signal-to-noise of stacked spectra}\label{sec:SN}

\begin{figure}
  \centering
  \includegraphics[width=0.9\linewidth]{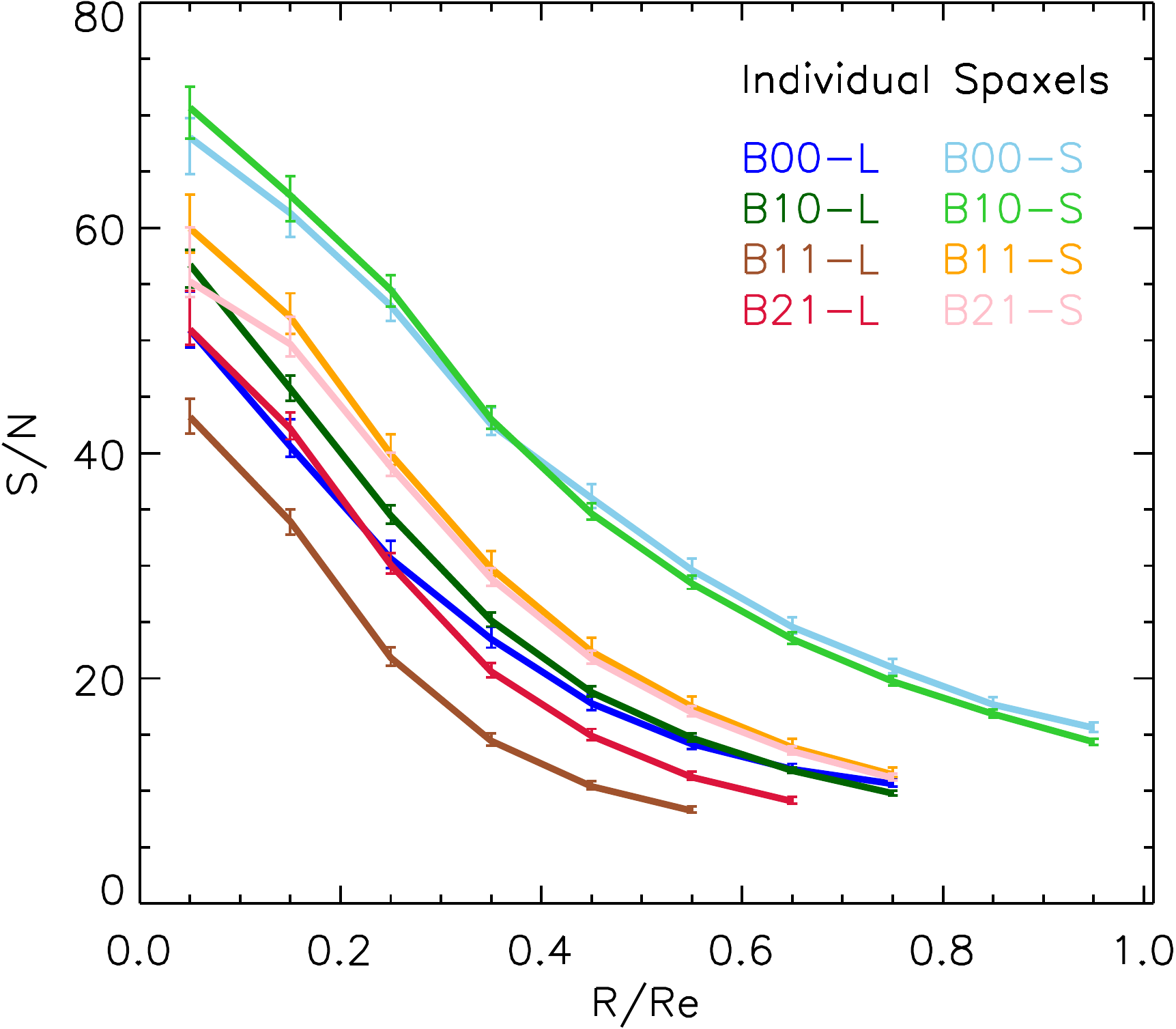}
  \caption{The median signal-to-noise per spaxel decreases monotonically with distance from the center, for the eight bins defined in Table~\ref{tab:bin}.  The smaller objects of each type (lighter shade of each color) have higher S/N.  This represents independent confirmation from the spaxels that the photometric analysis has correctly separated each $\sigma$ and $L_r$ bin into high- and low-surface brightness (small and large sizes) objects.  The typical S/N per spaxel is less than 100, which is why a stacking analysis is necessary.}
  \label{fig:SNsplitindiv}
\end{figure}

The analysis in the main text uses stellar population synthesis models to interpret a number of absorption line strengths.  It is well known that spectra with S/N$>100$ are required to make sufficiently reliable measurements.  Here we show that our stacked spectra do indeed have sufficient signal-to-noise.

As Figure~\ref{fig:SNsplitindiv} shows, the typical S/N in a spaxel lies well-below this desired value of S/N = 100. (We note in passing that the S/N is clearly larger for the more compact galaxies.  This is reassuring, since the $R_e$ determinations were made from the photometry, without regard to the spectroscopy.  As we noted, smaller $R_e$ implies larger surface brightness at fixed $L_r$, hence higher S/N in a spaxel.)  

\begin{figure}
  \centering
  \includegraphics[width=0.9\linewidth]{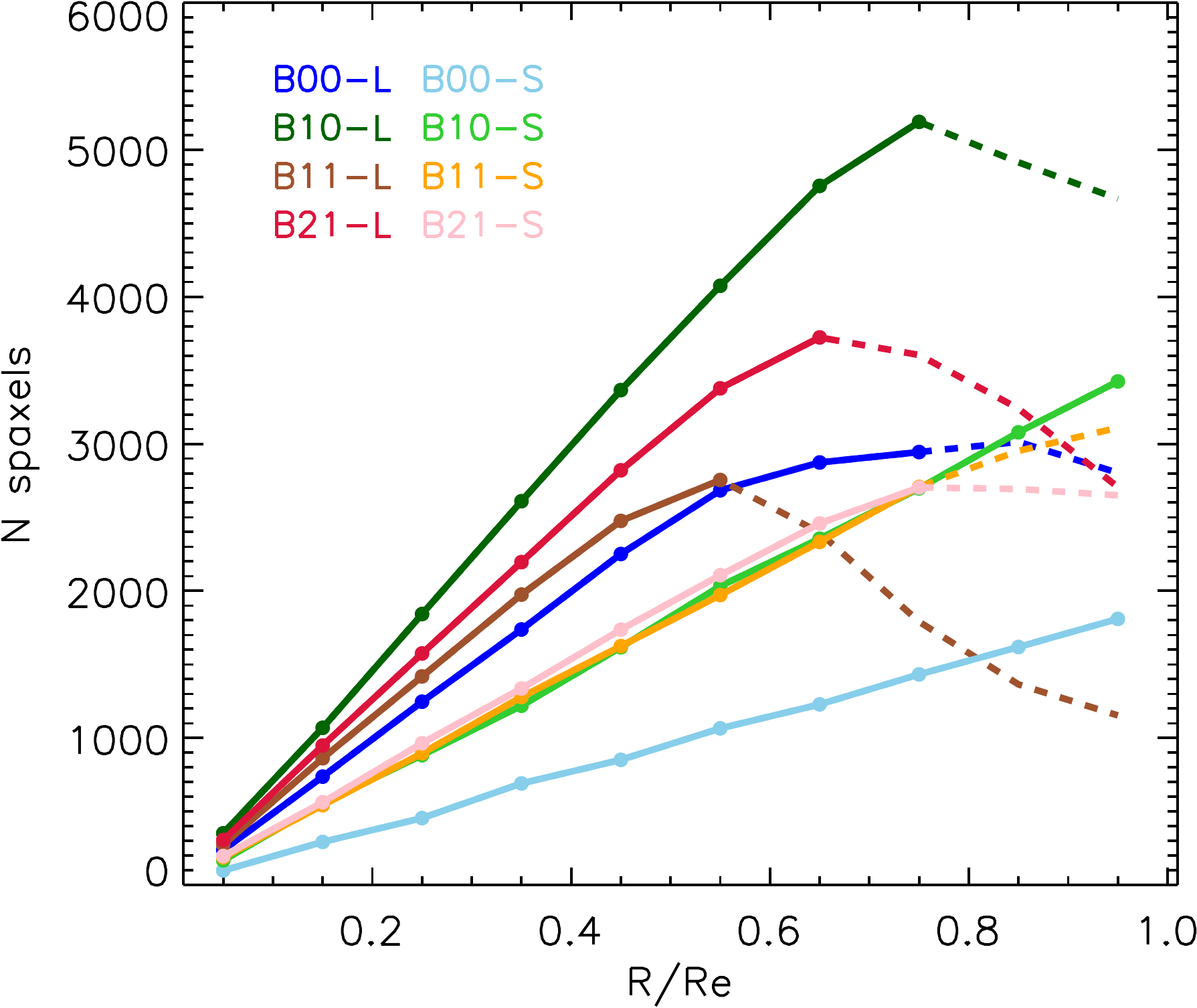}
  \caption{Number of spaxels in each radial bin which contribute to our results for the eight bins in $L_r$, $\sigma_0$ and $R_e$ defined in Table~\ref{tab:bin} (lighter shades show smaller sizes). Dashed curves show scales which are compromised by incomplete IFU coverage, so they are not used when studying gradients.}
  \label{fig:Nspx}
\end{figure}
\begin{figure}
  \centering
  \includegraphics[width=0.9\linewidth]{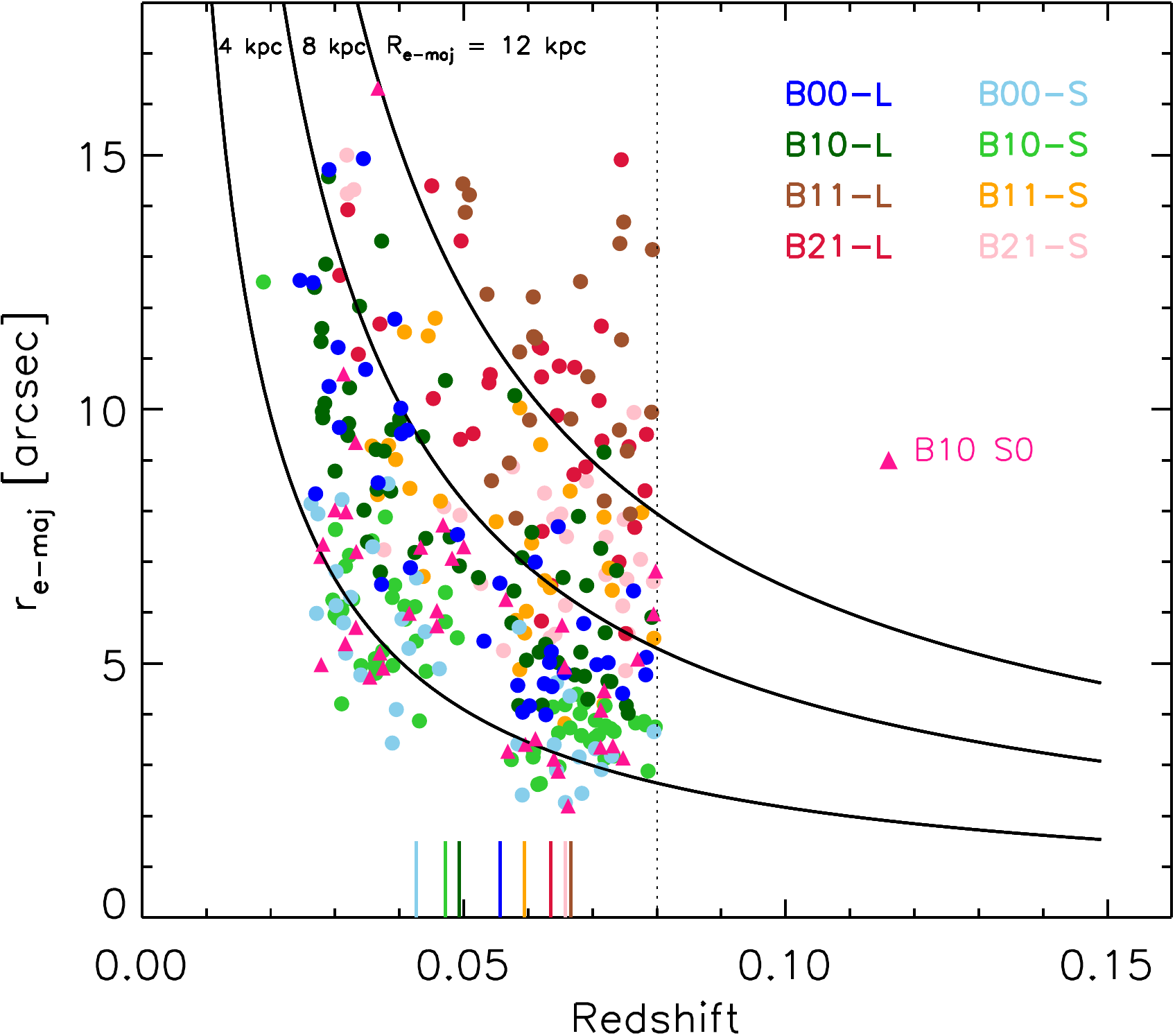}
  \caption{Apparent size - redshift relation for the objects in our sample.  Objects with large apparent sizes are covered by more fibers; at fixed $R_e$, lower redshift objects dominate the spaxel count. Small vertical lines show the median $z$ for each bin.}
  \label{fig:rz}
\end{figure}

Figure~\ref{fig:Nspx} shows the number of spaxels in each radial bin which contribute to our results for the eight bins in $L_r$, $\sigma_0$ and $R_e$ defined in Table~\ref{tab:bin}.
Some of the curves decrease at large $R$ because, for the largest galaxies, the spaxels may not cover the entire region within $R_e$.
To appreciate why, Figure~\ref{fig:rz} shows the distribution of apparent size and redshift for these Es.  Since the spaxel size is fixed, for fixed $R_e$, nearby objects are sampled by more spaxels.  Conversely, at fixed $z$, the objects with larger $R_e$ (darker shade of each color) are sampled by more spaxels, meaning they contribute more spaxels to a stack on the scale $R/R_e$. Figure~\ref{fig:Nspx} shows that this bias is stronger for bin B11-L. Dashed curves show scales which are compromised by incomplete IFU coverage, so they are not used when studying gradients.

\begin{figure}
  \centering
  \includegraphics[width=0.9\linewidth]{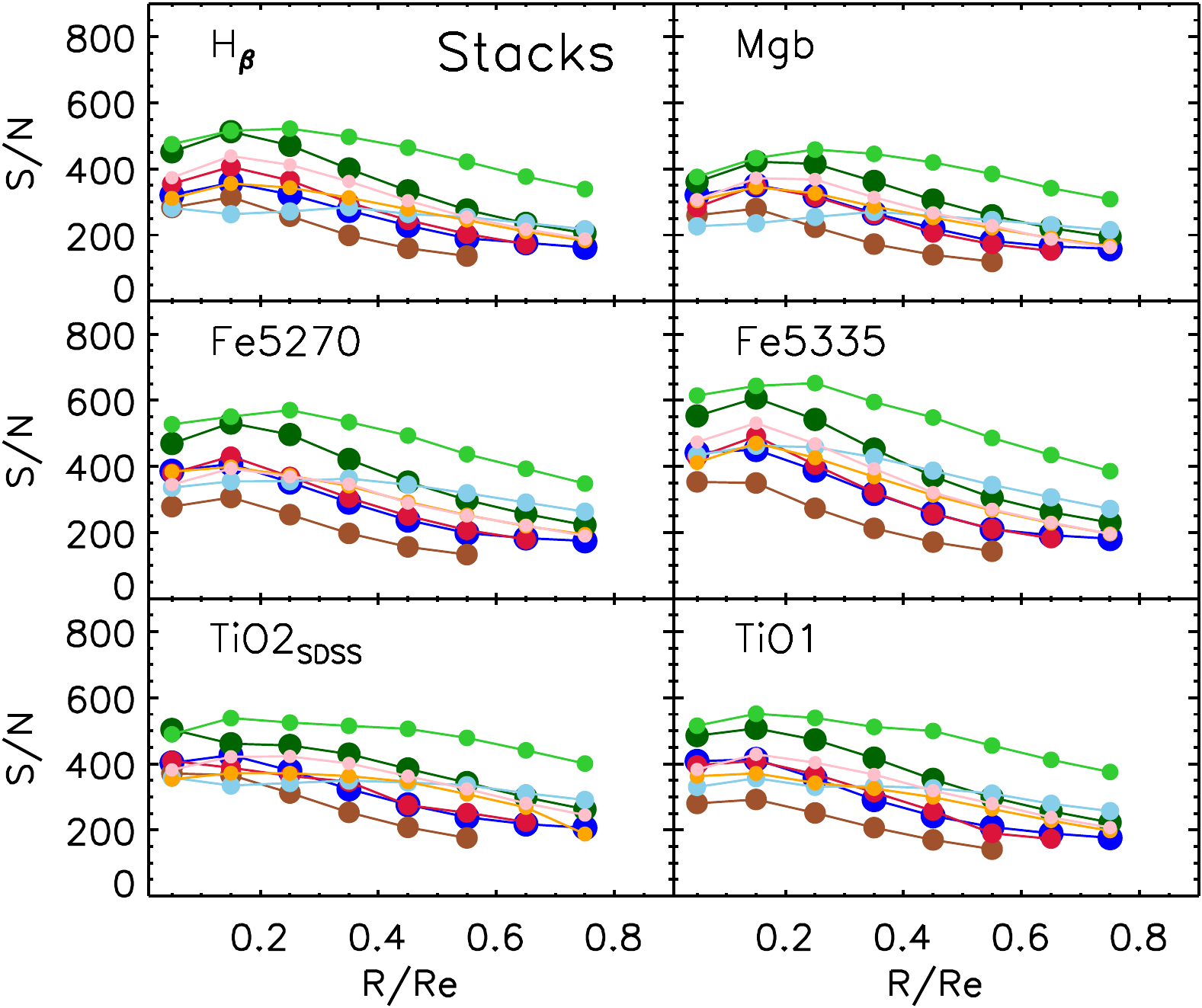}
  \caption{Signal-to-noise profiles for the Lick indices measured from the stacked spectra of galaxies in the bins defined in Table~\ref{tab:bin}.  Our stacks have S/N $> 100$ on all scales we explore in this paper.}
  \label{fig:SNsplit}
\end{figure}

Figure~\ref{fig:SNsplit} shows that our stacks have S/N $> 100$ on all scales we explore in this paper.
  
\begin{figure}
  \centering
  \includegraphics[width=0.9\linewidth]{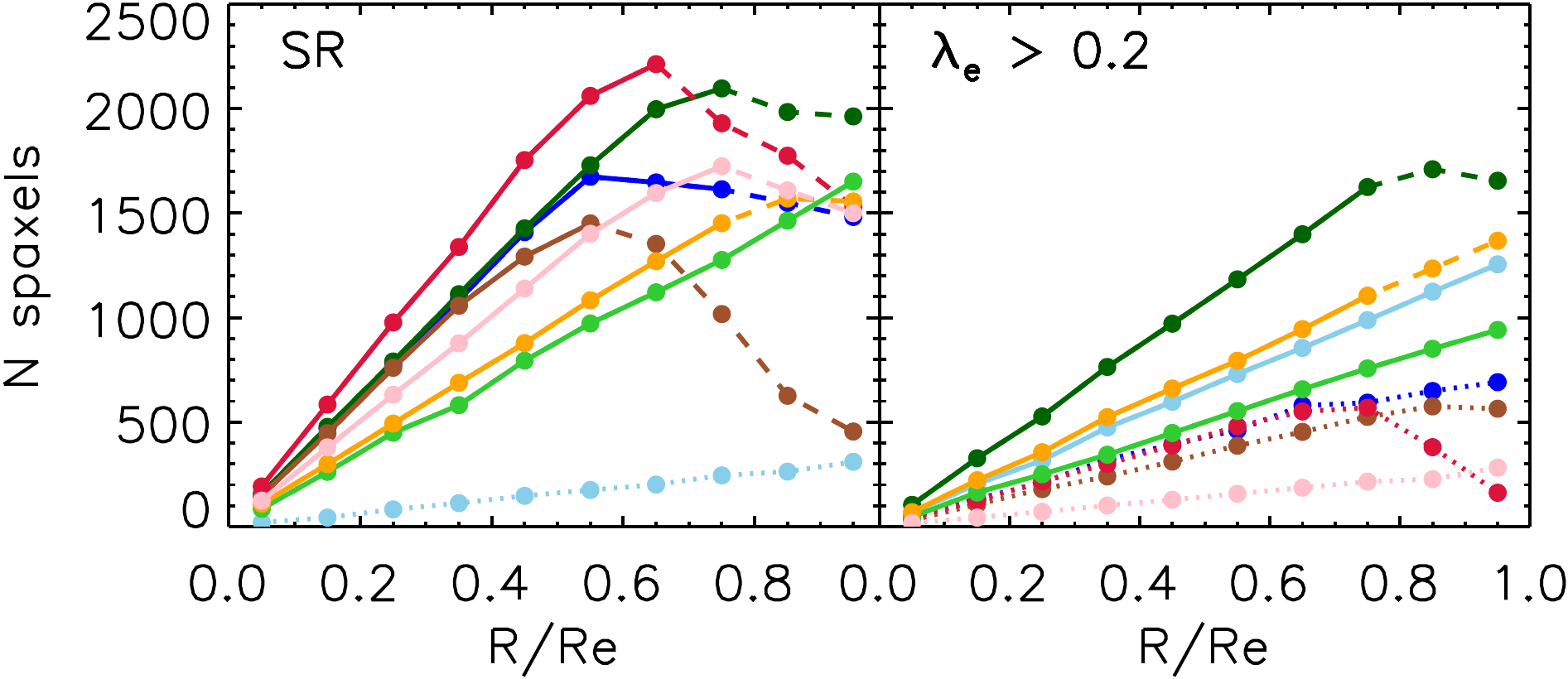}  
  \caption{Number of spaxels which contribute to our slow (left) and fast rotator (right) stacks.  Dotted lines connect bins in which there are too few spaxels to make reliable Lick index measurements:   these are either slowly rotating compact low luminosity B00 galaxies (light blue), fast-rotating luminous galaxies from bin B21 (pink and red), or fast-rotating large B11 galaxies (brown).  Dashed curves show scales which are compromised by incomplete IFU coverage, so they are not used when studying gradients. }
  \label{fig:NspxFR}
\end{figure}

In Section~\ref{fastSlow} we study the difference between slow and fast rotators.  This means we must further subdivide our sample, reducing the number of spaxels which contribute to each stack.  Dotted lines in Figure~\ref{fig:NspxFR} show when there are too few spaxels to make reliable Lick index measurements:   these are either slowly rotating compact low luminosity B00-S galaxies (light blue), fast-rotating large low luminosity B00-L galaxies (dark blue), fast-rotating luminous galaxies from bin B21 (pink and red), and fast-rotating large B11 galaxies (brown).  Dashed curves show scales which are compromised by incomplete IFU coverage, so they are not used when studying gradients.

\section{Comparison with previous work}\label{sec:compare}

\subsection{Comparison with Graves et al. (2010)}
\cite{Graves2010} estimated ages, metallicities and $\alpha$-enhancements from stacked spectra of early-type galaxies in the SDSS, assuming the same (Chabrier) IMF for all objects.  Converting their Table~2 scalings in $\sigma$, $R_e$ and $\Delta I_e\propto (L/R_e^2)R_e^{1.21}\sigma^{-1.16}$ into our $L, R_e$ and $\sigma_0$ yields  
\begin{align}
 {\rm age} &\propto R_e^{0.02} \sigma_0^{0.73} \Delta I_e^{-0.92}
             \propto R_e^{0.75} \sigma_0^{1.8} L^{-0.9} \label{ageRVI}\\
 {\rm [Fe/H]} &\propto  R_e^{-0.46} \sigma_0^{-0.47} L^{0.64} \label{metRVI}\\
 %{\rm [Mg/H]} &\propto \sigma_0^{0.18} R_e^{-0.29}L^{0.35}\\
 {\rm [Mg/Fe]} &\propto R_e^{0.24} \sigma_0^{0.58} L^{-0.27} \propto {\rm age}^{0.31}
 \label{alphaRVI}
\end{align}
(if we ignore the curvature in their age relation).  
%      dI = I - (1.16V - 1.21 R + 0.55)
%         = L - 2R - 1.16V + 1.21 R - 0.55
%         = L - 0.79R - 1.16V - 0.55
% lg(age) = 0.85 + 0.73 V + 0.02 Re - 0.92 (dI) - 2.9 dI^2
%         = 0.8 V + 0.75 Re - 0.92 L
% [Fe/H]  = 0.27 V + 0.04 Re + 0.64 (L - 0.79R - 1.16V - 0.55)
%         = -0.47 V - 0.46 Re + 0.64L
% [Mg/H]  = 0.58 V + 0.02 Re + 0.35 (L - 0.79R - 1.16V - 0.55)
%         = 0.18 V - 0.26 Re + 0.35 L
% [Mg/Fe] = 0.27 V - 0.27 (I - 1.16V + 1.21 Re - 0.55)
%         = 0.27 V - 0.27 (L - 0.79R - 1.16V)
%         = 0.58 V - 0.27 L + 0.21 Re
% Note that
%  [Mg/H]/[Fe/H] ~ 0.18 V - 0.26 Re + 0.35 L - (-0.47 V - 0.46 Re + 0.64L)
%                ~ 0.65 V + 0.2 Re - 0.29 L
% Also
%  [M/H] = [Fe/H] + lg(0.694 10^[alpha/Fe] + 0.306)

Thus, they find that, at fixed $L_r$ and $\sigma_0$, objects with larger $R_e$ are older:  a 0.1~dex difference in $R_e$ corresponds to an 18\% change in age.  This is consistent with our finding of a $\sim 2$~Gyr age difference in the main text.  These scalings are also consistent with our finding that, at fixed $L_r$ and $\sigma_0$, the objects with larger $R_e$, in addition to being older, are less metal rich and more [Mg/Fe] enhanced.  Thus, our analysis shows that these qualitative scalings persist even when one allows for IMF gradients.

\begin{figure}
 \centering
  \includegraphics[width=0.9\linewidth]{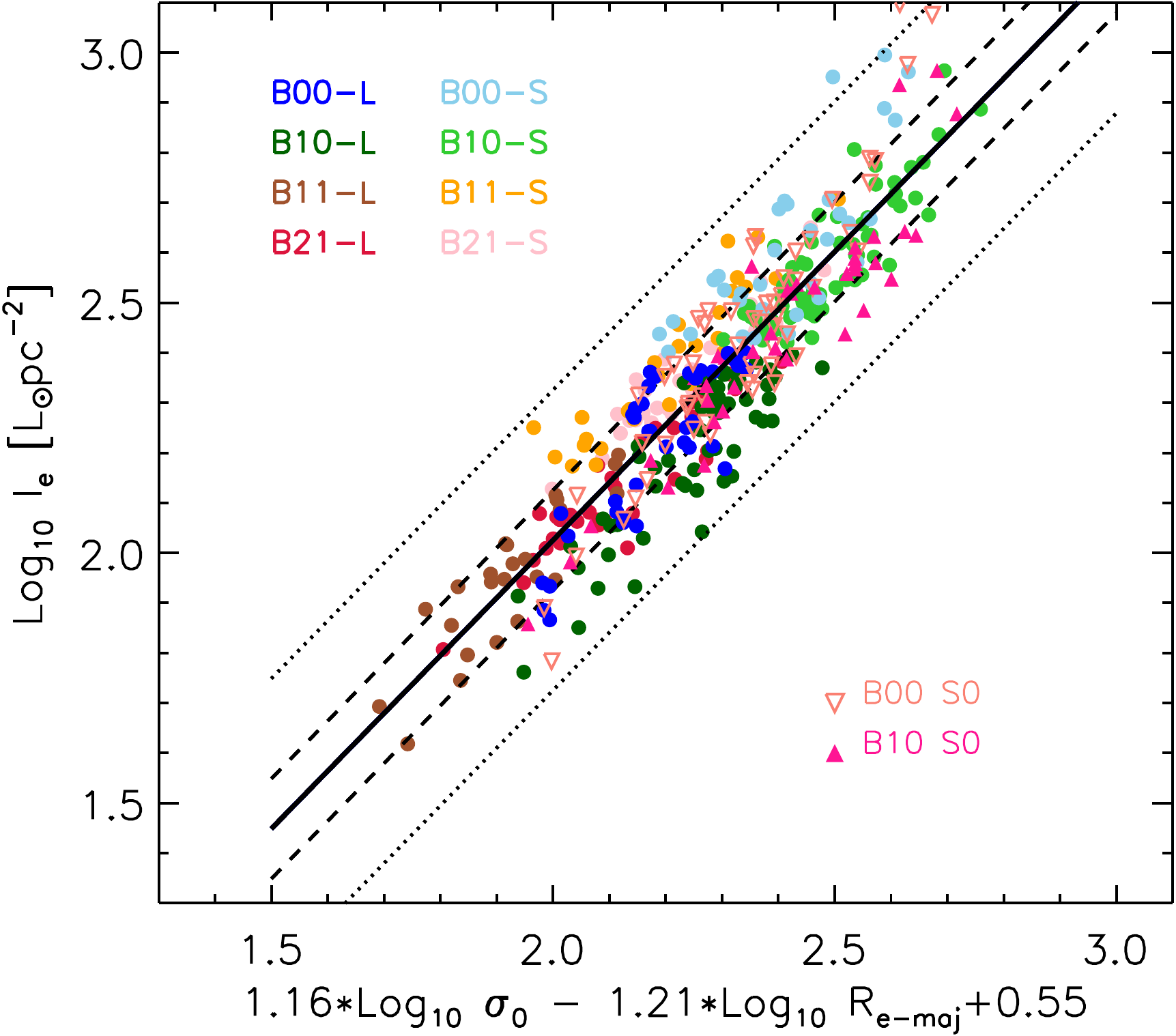}
  \includegraphics[width=0.9\linewidth]{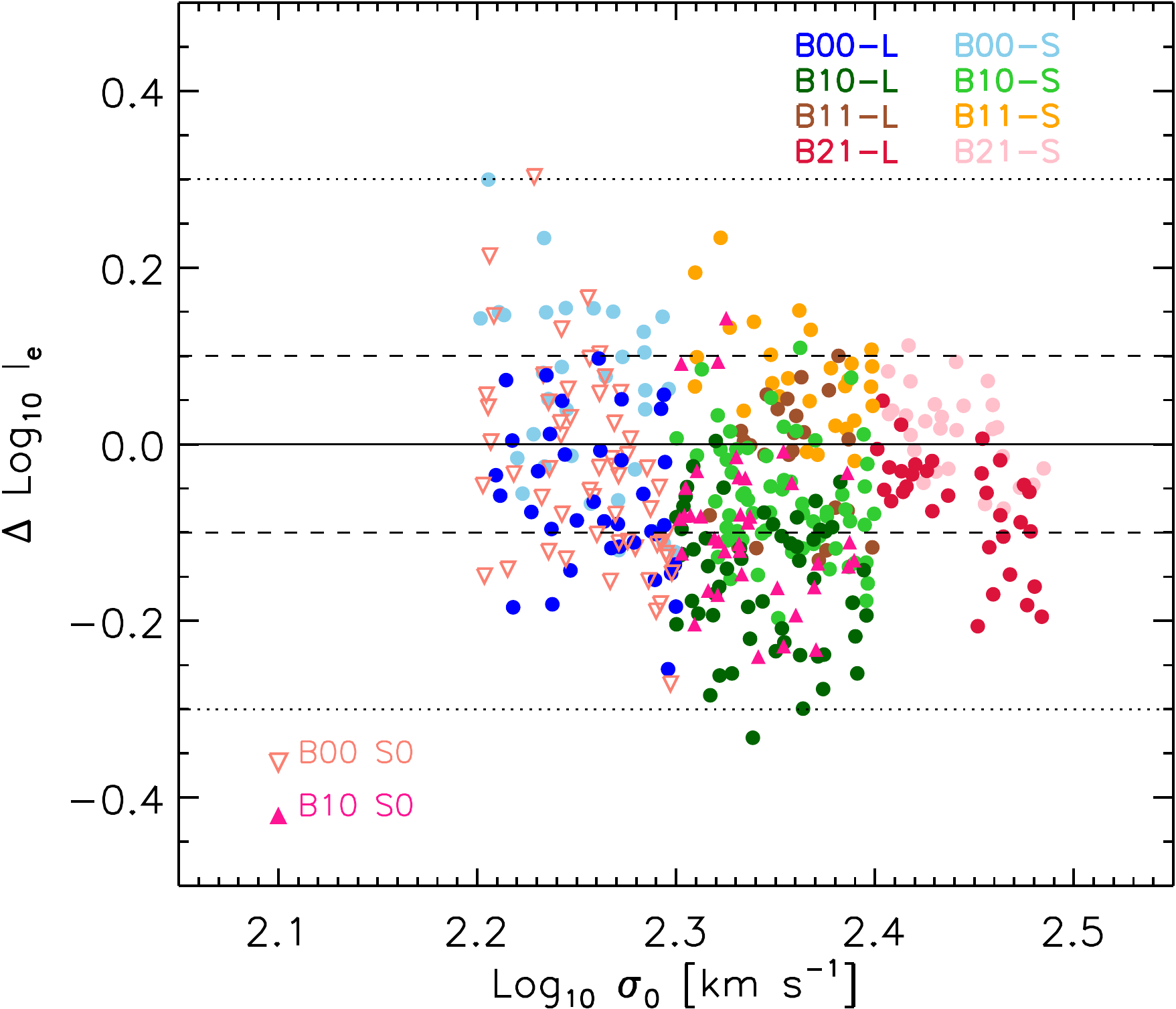}
  \caption{Top:  Surface-brightness as predicted by the Fundamental Plane combination of $\sigma_0$ and $R_e$.  A wide range of $L_r$ and $R_e$ have the same $I_e$.  Bottom: Surface brightness residuals $\Delta I_e\propto (L/R_e^2)/(R_e^{1.21}\sigma_0^{-1.16})$ from the Fundamental Plane vs $\sigma_0$. Inclusion of S0s in the B10 bin would reduce the estimated ages in that bin.  Equation~(\ref{ageRVI}) implies that lines of constant age slope up and to the right.}
  \label{fig:FP}
\end{figure}

Although these trends with size at fixed $\sigma_0$ and $L_r$ are in qualitative agreement with our findings, there are important differences when it comes to the scaling of, e.g., age with $\sigma_0$ and $L_r$ for Es.  Figure~\ref{fig:FP} illustrates why.  The colored symbols show the objects in the bins we define, plotted in the variables \cite{Graves2010} used for determining which spectra to stack.  Comparison with their Figure~1 shows that their bins correspond to a horizontal division in the bottom panel of Figure~\ref{fig:FP}.  Their analysis extends to much smaller $\sigma_0$ and $L_r$.  This is in part because they do not distinguish between Es and S0s.  But over the $\sigma$ range where our samples overlap, it is clear that their stacks contain a mix of objects from different bins in our scheme.  While our scheme could be crudely approximated by drawing parallelograms in this coordinate system, we believe that by working directly with $L_r$ and $\sigma_0$ our binning scheme more effectively isolates objects of similar mass, so trends in it are simpler to interpret.

As a result, while we have qualitative agreement on some points -- e.g. the age af bin B10-L Es, which are the oldest galaxies in our sample, have the lowest $\Delta\log_{10} I_e$ -- there are important differences in detail. In particular, 
the distribution of B10 S0s in the bottom panel of Figure~\ref{fig:FP} is more similar to that of bin B10-L while their ages, metallicities and $\alpha$-enhancements are more similar to those of bin B10-S (see Figures~\ref{fig:lambdasplit} and~\ref{fig:lambdasplit2}). In addition, the expressions above do not accurately describe how, in the main text, age, metallicity and [Mg/Fe] depend on $\sigma_0$ and $L_r$.  E.g., their bins hide the fact that B10-L is extremely old.  While the dark green symbols which show this bin do have amongst the largest values of $\Delta I_e$, equation~(\ref{ageRVI}) would not make them significantly older than the objects in bin B21-L (dark red).  

% age = 0.73 V - 0.92 I - 2.9 I^2
% 0 = 2.9 I^2 + 0.92 I + age-0.73 V
% I = [sqrt(1 + 4*3*(0.73V-age)/0.9^2) - 1]*0.3/2

% dI = -age/0.9 + 0.73/0.9 V
% 

Finally, we note that our anomalous bin B10-L (dark green symbols) is rather different from the three outliers identified in \cite{Graves2010}:  theirs had $\sigma\le 150$~km~s$^{-1}$, which is lower than any of the B10 Es, and they had low [Mg/Fe] for their otherwise normal ages.  Our B10-L Es have anomalously high ages, but [Mg/Fe], while high, is not particularly unusual given the estimated age.

\begin{figure}
 \centering
  \includegraphics[width=0.9\linewidth]{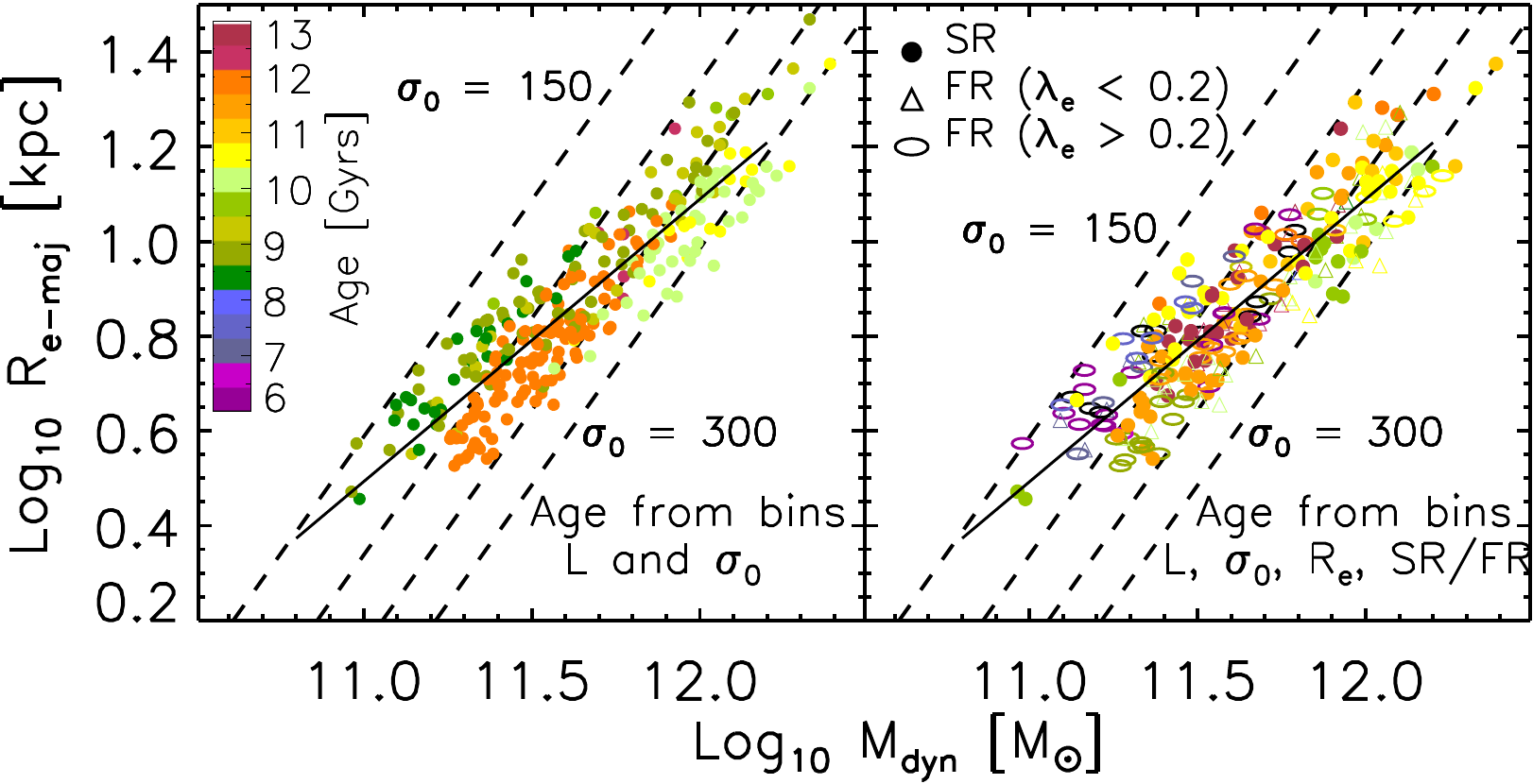}
  \caption{Same as Figure~\ref{fig:RMsFR} of the main text, but with $M_*$ replaced by $4R_e\sigma_0^2/G$. Dashed lines show locii of fixed $\sigma_0$ as labeled.  In both panels, the symbol which represents a galaxy is colored by the age estimate for its bin, but the age estimates in the two panels differ.  In the left hand panel, the smaller objects are older, but this does not hold in the right hand panel. }
  \label{fig:McD}
\end{figure}

\subsection{Comparison with McDermid et al. (2015)}\label{sec:atlas}
The main text showed that the stellar populations of FRs and SRs are very different.  The FR/SR dichotomy was not considered at all by \cite{Graves2010}.  However, \cite{McDermid2015} (Paper XXX of the ATLAS$^{\rm 3D}$ collaboration) do study this issue.  They state that compact early-type galaxies tend to be older, more metal-rich, and more $\alpha$-enhanced than larger ETGs of the same mass.  To compare directly with them,  Figure~\ref{fig:McD} shows the size-dynamical mass correlation in our sample.  The dashed lines shows locii of fixed $\sigma_0$; lines for larger $\sigma_0$ are displaced down and to the right.  In both panels, the symbol which represents a galaxy is colored by the age estimate for its bin.

The first point to make is that their age-size trend is most dramatic when the full range of $\sigma_0$ is shown; our sample spans a much smaller range in $\sigma$ and $M_{\rm dyn}$ (it corresponds to the top right corner of their plot) -- so the question is if the age-size correlation (at fixed mass) remains well-defined at the highest $\sigma_0$.  The left hand panel suggests that, at fixed mass, the smaller objects are indeed older.  While this agrees with \cite{McDermid2015}, note that our most massive objects ($\log_{10}(M_{\rm dyn}/M_\odot) \ge 12$) are not the oldest: objects with $\log_{10}(M_{\rm dyn}/M_\odot) \sim 11.5$ are older.  This is inconsistent with their Figure~6.

Moreover, their Figure~11 shows that the vast majority of FRs with $\sigma\ge 220$~kms$^{-1}$ are more than 10~Gyrs old.  As we discuss in the main text, few of our FRs are this old, and this is one reason why we find that, even at high masses, smaller objects are younger whereas \cite{McDermid2015} report the opposite.

% Don't change these lines
\bsp	% typesetting comment
\label{lastpage}
\end{document}